\RequirePackage{fix-cm}
\documentclass[twocolumn,epjc3]{svjour3}       
\smartqed  
\RequirePackage{graphicx}
\RequirePackage{mathptmx}      
\RequirePackage{latexsym}
\RequirePackage[numbers,sort&compress]{natbib}
\RequirePackage[colorlinks,citecolor=blue,urlcolor=blue,linkcolor=blue]{hyperref}
\usepackage{amsmath} 

\usepackage{amssymb}
\usepackage{multirow}
\usepackage{multicol}
\usepackage{graphicx}
\usepackage{amsmath}
\usepackage{lineno}
\usepackage{setspace}
\usepackage{subcaption}
\usepackage{listings}
\usepackage{xspace}
\usepackage{xcolor}
\usepackage{float}
\usepackage{comment}
\usepackage{soul}
\usepackage{wasysym}
\usepackage{isotope}
\usepackage{siunitx}
\newcommand{\orphee}{\textsc{ORPHEE}\xspace}
\newcommand{\icone}{\textsc{ICONE}\xspace}
\newcommand{\iphi}{\textsc{IPHI}\xspace}
\newcommand{\satelit}{\textsc{SATELIT}\xspace}
\newcommand{\geant}{\textsc{Geant4}\xspace}
\newcommand{\ncrystal}{\textsc{NCrystal}\xspace}
\newcommand{\toucans}{\textsc{TOUCANS}\xspace}
\newcommand{\mcnp}{\textsc{MCNP6}\xspace} 

\newcommand{\tripoli}{{\textsc{Tripoli-4}}\textsuperscript{\textregistered}\xspace}
\newcommand{\armco}{ARMCO\textsuperscript{\textregistered}\xspace}
\newcommand{\lego}{{\textsc{LEGO}}\textsuperscript{\textregistered}\xspace}
\newcommand{\degree}{$^\circ\text{C}$\xspace}
\newcommand{\cadmesh}{\textsc{CADMesh}\xspace}
\newcommand{\promethee}{\textsc{PROMETHEE}\xspace}
\newcommand{\srim}{\textsc{SRIM-2013}\xspace}
\newcommand{\copright}{\textsuperscript{\textcopyright}\xspace}

\definecolor{darkred}{rgb}{0.75, 0.0, 0.2}

\journalname{Eur. Phys. J. A}
\begin{document}

\title{Neutron production with a 10~kW HiCANS based on SATELIT, a CEA-Saclay target with liquid lithium }


\author{
    L.~Thulliez\thanksref{IRFU,e1}  
    \and N.~Berton\thanksref{IRFU}   
    \and R.~Boudouin\thanksref{IRFU}   
    \and N.~Cavalière\thanksref{ISAS}   
    \and S.~Cazaux\thanksref{IRFU,e1}   
    \and T.~Chaminade\thanksref{IRFU}   
    \and N.~Chauvin\thanksref{IRFU,e1} 
    \and D.~Chirpaz\thanksref{IRFU}   
    \and J.-L.~Courouau\thanksref{ISAS}   
    \and Q.~Cridling\thanksref{ISAS}   
    \and P.~Daniel-Thomas\thanksref{IRFU}      
    \and J.~Darpentigny\thanksref{LLB}
    \and G.~Debras\thanksref{IRFU}   
    \and G.~Disset\thanksref{IRFU}   
    \and A.~Drouart\thanksref{IRFU,e1}   
    \and E.~Dumonteil\thanksref{IRFU} 
    \and R.~Duperrier\thanksref{IRFU} 
    \and R.~Ferdinand\thanksref{IRFU}  
    \and E.~Giner Demange\thanksref{IRFU}   
    \and J.-C.~Guillard\thanksref{IRFU}   
    \and N.~Jonqueres\thanksref{IRFU}   
    \and T.~Lebrun\thanksref{IRFU}      
    \and D.~Loiseau\thanksref{IRFU}   
    \and J.~Mendes\thanksref{IRFU}   
    \and B.~Mom\thanksref{IRFU}   
     \and F.~Ott\thanksref{LLB}  
    \and C.~Péron\thanksref{IRFU}   
    \and J.~Phocas\thanksref{ISAS}   
    \and Y.~Reinert\thanksref{IRFU}   
    \and A.~Roger\thanksref{IRFU}   
    \and Y.~Sauce\thanksref{IRFU}   
    \and F.~Senée\thanksref{IRFU} 
    \and M.~Trocm\'e\thanksref{list} 
    \and C.~Veyssière\thanksref{IRFU}   
    \and D.~Vurpillot\thanksref{IRFU}   
    \and X.~Wohleber\thanksref{IRFU}
    %
}


\thankstext{e1}{emails: loic.thulliez@cea.fr, antoine.drouart@cea.fr, sandrine.cazaux@cea.fr, nicolas.chauvin@cea.fr}

\institute{IRFU, CEA, Universit\'e Paris-Saclay, 91191 Gif-sur-Yvette, France \label{IRFU}
    \and
    Universit\'e Paris-Saclay, CEA, Service de recherche en Corrosion et Comportement des Mat\'eriaux, 91191 Gif-sur-Yvette, France \label{ISAS}
    \and
    Universit\'e Paris-Saclay, CEA, CNRS, IRAMIS, LLB, 91191, Gif-sur-Yvette, France \label{LLB}
    \and 
    Universit\'e Paris-Saclay, CEA, List, F-91120, Palaiseau, France \label{list}
}

\date{Received: date / Accepted: date}

\maketitle


\begin{abstract}

High-Current Accelerator-driven Neutron Sources (HiCANS) are currently under development across Europe to address the shortage of medium-scale neutron sources, as many research nuclear reactors have been decommissioned over the past several years.
At CEA-Saclay, a HiCANS has been developed utilizing the \iphi accelerator, which delivers a 3~MeV proton beam with a current of up to 100~mA, and the CEA-Saclay liquid lithium target named \satelit.
In 2024-2025, a successful experimental campaign was conducted, during which a 10~kW proton beam was directed at the liquid lithium target for nearly 100~h to generate neutrons \textit{via} the \isotope[7][]{Li}(p,n)\isotope[7][]{Be} nuclear reaction.
Throughout the experimental campaign, a total deposited beam power of 840~kW.h was accumulated, including two continuous operational days exceeding 11~h each.
A polyethylene moderator coupled with \satelit enabled the extraction of a thermal neutron beam, with a flux measured at 1.4~m from the extraction point exceeding $10^6$~n.cm$^{-2}$·s$^{-1}$, which is sufficient for numerous neutron applications.
The next step for the long-term operation of this facility involves developing strategies to mitigate the radiological concerns associated with the accumulation of \isotope[7][]{Be} within the system.
Overall, this work demonstrates that such facilities can play a significant role in the future of medium-scale neutron sources in Europe.

\end{abstract}

\section{Introduction}
Facilities providing thermal neutron beams are of primary importance for condensed matter and nuclear fundamental research, as well as for applications such as neutron-imaging or medical therapy. However, in Europe, many small and medium-sized neutron facilities, predominantly research nuclear reactors, have been shut down in favor of constructing large-scale neutron sources, mainly based on spallation reactions, such as the European Spallation Source (ESS). While these large facilities are well-suited for applications requiring intense neutron beams, they cannot replace local facilities that can be geographically situated close to end-users, such as laboratories, universities, industries, or hospitals. Such facilities are, by definition, more conveniently and directly available, and thus more adequate for fostering the development of new experimental techniques and detectors, and for supporting educational initiatives essential for training the next generation of physicists. Therefore, since few years, several research institutions across Europe have been developing High-Current Accelerator-driven Neutron Sources (HiCANS), which are Compact Accelerator-driven Neutron Sources (CANS) with a beam power of 10~kW or higher, to fill this gap~\cite{ZAKALEK2025104163,ELENA}. For instance, in Germany, there is the HBS facility at Jülich~\cite{HBS,BAGGEMANN2024169912}, in Spain, the ARGITU project at ESS Bilbao~\cite{Perez01102020}, and in France, the \icone (Innovative COmpact NEutron) facility at CEA-Saclay~\cite{ICONE2023}.

The context of these developments in France is rooted in the shutdown of the \orphee research nuclear reactor at CEA-Saclay, which had served the French neutron scattering community for 40~years until its closure in 2019. To find an alternative to reactor-based neutron sources, CEA-Saclay has been developing and testing CANS and HiCANS since 2016, utilizing the IPHI (High Intensity Proton Injector) accelerator. This accelerator delivers a 3~MeV proton beam with an intensity of up to 100~mA in continuous mode. This initiative paves the way for the \icone facility, which is being jointly developed by CEA and CNRS. The \icone facility will generate neutron beams for scattering experiments using a 25~MeV proton beam directed at a solid beryllium target (melting temperature of 1280~\degree), with an average deposited power of 80~kW~\cite{ICONE_website}. The main challenges include the operation of a high-intensity accelerator, the development of a target able to withstand such power, and the optimization of the target-moderator-reflector-shielding assembly (TMRS) to maximize the extracted neutron flux. To pursue this goal, initial experiments were conducted in 2016 using a beryllium target and a 10~W proton beam. These experiments allowed for the verification of the angle and energy of the produced neutrons and the generation of the first thermal neutrons~\cite{Tran2018,Tran2020}. In 2019, the power on the beryllium target was increased to 3~kW, thanks to a new target design~\cite{Thulliez2020}. Then, in 2022, a 30~kW beam was directed at an updated beryllium target, which was tilted at a 20° angle relative to the proton beam axis to maintain a deposited power density below 500~W.cm$^{-2}$~\cite{Schwindling2022}. In addition to the characterization of the thermal neutron flux, this experiment enabled neutron diffraction studies using the DIoGENE setup~\cite{Darpentigny2022}.
\newline
\indent
In parallel with the development of solid beryllium targets, a lithium target named \satelit (Saclay Lithium Target) has been under development at CEA-Saclay since 2020 to maximize neutron production. Indeed, with 3~MeV protons, neutron production with lithium is approximately five times higher than that with beryllium~\cite{Porges1970}. However, developing a liquid lithium target able to sustain a beam power exceeding 10~kW and minimize lithium evaporation is challenging (the lithium melting temperature is 180~\degree). The liquid lithium target presented in this work is an improvement over the design of the liquid lithium target LiLiT~\cite{Paul2015,Silverman2020}. The ultimate objective of the \satelit project is to maximize the thermal neutron flux using the capabilities of the 3~MeV proton beam from the \iphi accelerator, which can provide several tens of kilowatts of beam power.

The development of this lithium loop for a HiCANS facility has been carried out in two phases. Phase 0, initiated in 2020, focused on circulating liquid lithium without the proton beam. Then, in 2024-2025, Phase 1 coupled the lithium loop to the \iphi accelerator and a moderator to produce and efficiently extract thermal neutrons (see Figure \ref{fig:IPHI_CAD}). The primary objective was to characterize the interaction between the proton beam and lithium, validate sustainable neutron production, and thermal neutron extraction obtained by coupling a polyethylene moderator with the lithium target, while also characterizing the properties of the extracted thermal neutron beam. Overall, this experiment aimed to identify the weak points of such a HiCANS to propose a design able to deliver thermal neutron beams to research experiments on a long-term basis.

This work details the different components of this new HiCANS and the main results obtained during this experimental campaign. Section~\ref{sec:iphi} provides a detailed overview of the \iphi accelerator and its operation during the experimental campaign. Following this, Section~\ref{sec:satelit} describes the \satelit liquid lithium target, while Section~\ref{sec:cmrbDesign} elaborates on the integration of the neutron moderator and shielding around the target. Section~\ref{sec:simulation} presents the simulation package used to develop and optimize the neutron performance of the HiCANS. Section~\ref{sec:7Be_buildup} focuses on the monitoring of the \isotope[7][]{Be} build-up in the lithium loop, which is essential for radiological safety assessments. Finally, the experimental characterization of the extracted neutron beam is presented in Section~\ref{sec:measurements}.


\section{The IPHI facility}
\label{sec:iphi}

\subsection{IPHI facility experimental setup}  
\label{IPHI_facility}

\subsubsection{Accelerator}
\begin{figure}[h!]
    \centering \includegraphics[scale=0.4]{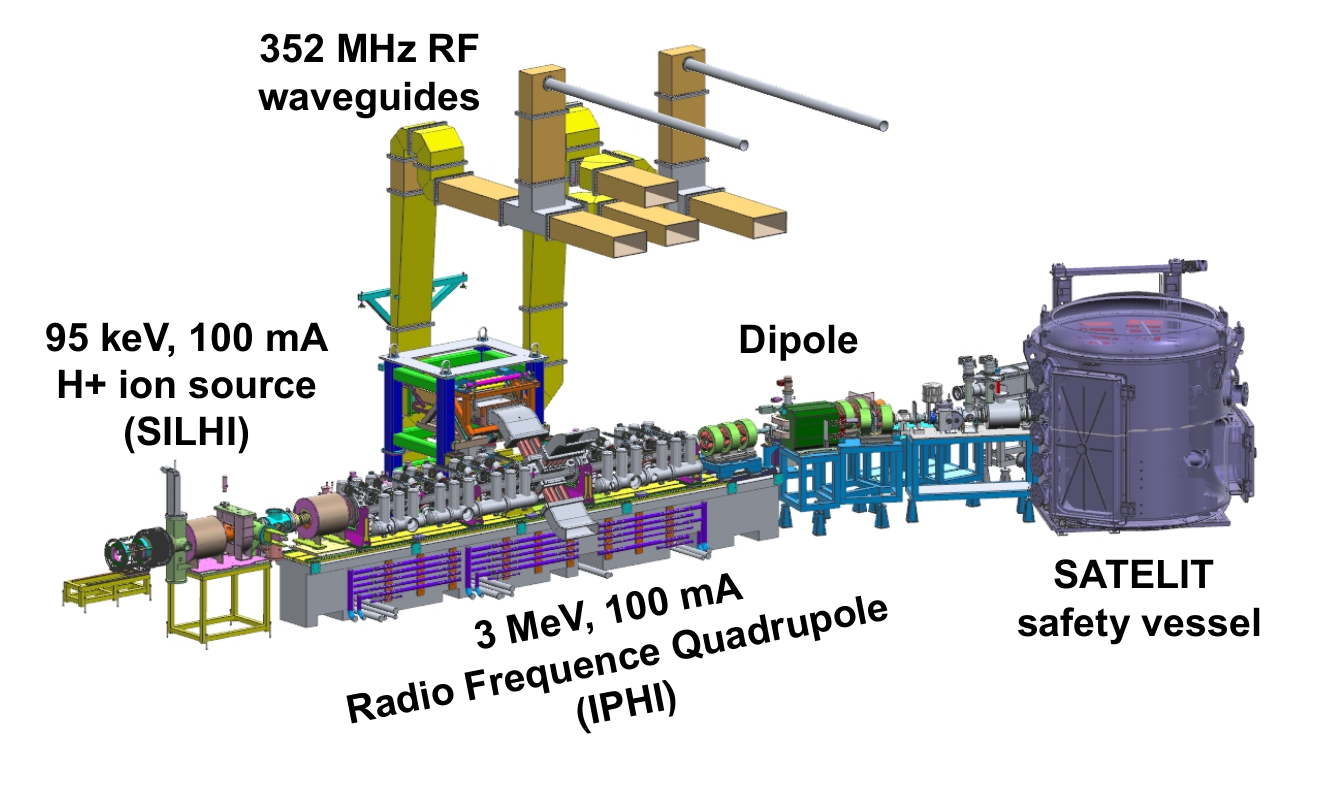}
    \caption{3D CAD model of \satelit integrated within the IPHI facility.}
    \label{fig:IPHI_CAD}
\end{figure}

\begin{figure*}[h!]
    \centering \includegraphics[width=1.\linewidth]{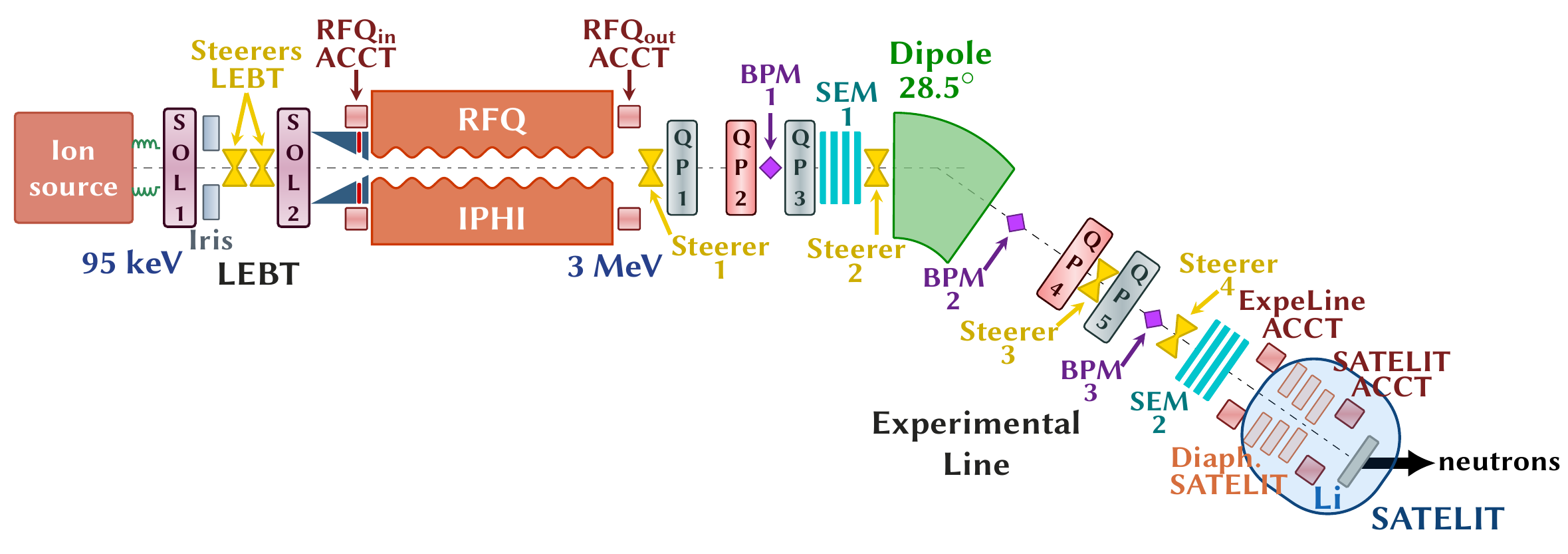}
    \caption{Scheme of the IPHI facility.}
    \label{fig:IPHI_Scheme}
\end{figure*}

The IPHI (Injecteur de Protons Haute Intensit\'e) facility~\cite{Senee2018}, developed at CEA-Saclay, is designed to deliver a high current \SI{3}{MeV} proton beam for accelerator research and development and various applications.
Figures~\ref{fig:IPHI_CAD} and~\ref{fig:IPHI_Scheme} present the IPHI accelerator as configured for the SATELIT experiment.
It comprises an ion source and a Low-Energy Beam Transport (LEBT) line, followed by a radio-frequency quadrupole (RFQ) accelerating cavity, and finally a High Energy Beam Transport (HEBT) line, known as the "Experimental Line", which transports the beam to the SATELIT target.

The IPHI ion source is the Source of Light Ions with High Intensities (SILHI)~\cite{Gobin_RSI2002}, an electron cyclotron resonance (ECR) source operated at \SI{2.45}{GHz}. It delivers a \SI{95}{keV}, \SI{80}{mA} continuous-wave (CW) proton beam. During the experiments with the liquid lithium target, the source was operated in pulsed mode, providing beam pulses with durations ranging from \SIrange{0.1}{8}{ms} at frequencies between \SIrange{1}{50}{Hz}, depending on the required beam power.

In the LEBT line, two solenoids, SOL1 and SOL2, are used to transport and adapt the beam, optimizing its injection into the Radio-Frequency Quadrupole (RFQ). Additionally, two magnetic steerers are employed to correct the beam trajectory in case of misalignments.
Following SOL1, a variable aperture diaphragm, known as the 'Iris,' is installed in the beamline. This diaphragm limits the beam size and, consequently, its intensity before injection into the IPHI RFQ, without requiring adjustments to the ion source settings.
An AC Current Transformer (ACCT), named RFQ$_{in}$ ACCT, is installed at the end of the LEBT line to measure the beam current injected into the RFQ.
The IPHI RFQ~\cite{Piquet_IPAC2016}, designed in the 1990s for continuous-wave (CW) operation, is a \emph{4-vane}-type accelerating cavity powered by a \SI{352}{\mega\hertz} RF wave. It bunches the beam and accelerates it from \SI{95}{\kilo\electronvolt} to \SI{3}{\mega\electronvolt}, while also providing transverse focusing.

The HEBT line guides the \SI{3}{\mega\electronvolt} proton beam from the Radio-Frequency Quadrupole (RFQ) to the SATELIT target.
Beam focusing at the RFQ exit is performed by a magnetic quadrupole triplet (QP1, QP2, and QP3), which ensures its transport to the subsequent section of the beamline.
A \SI{28.5}{\degree} bending magnet deflects the beam toward the Experimental Line up to the SATELIT target. After this dipole, a quadrupole doublet (QP4 and QP5) focuses the beam and shapes it to the desired transverse sizes at the target position.
The correction scheme for the beam centroid position along the HEBT relies on four magnetic steerers, with position measurements provided by three Beam Position Monitors (BPM1, BPM2, and BPM3). Secondary Emission Monitors (SEMs) have been installed at locations SEM1 and SEM2 to measure horizontal and vertical beam profiles in short pulse mode.
Three ACCTs are distributed along the HEBT to measure beam intensity: one at the RFQ exit (RFQ$_{out}$ ACCT), another at the end of the experimental line (ExpeLine ACCT), and the third just before the SATELIT target (SATELIT ACCT).
In a beam section delimited by two consecutive ACCTs, ACCT$_1$ and ACCT$_2$, measuring currents $\text{I}_{\text{ACCT}_1}$ and $\text{I}_{\text{ACCT}_2}$ respectively, the beam transmission T$_{1 \rightarrow 2}$ through this section is defined by:

\begin{equation} \label{eq:transmission}
    \text{T}_{1 \rightarrow 2 } = \frac{\text{I}_{\text{ACCT}_2}}{\text{I}_{\text{ACCT}_1}}
\end{equation}
\noindent
Correspondingly, the beam loss within this section is measured online as $\text{I}_{\text{ACCT}_1} - \text{I}_{\text{ACCT}_2}$.

\subsubsection{Diaphragms before the lithium target} 
\label{subsub:diaph}

\begin{figure}[h!]
    \centering    
    \includegraphics[width=1.\linewidth]{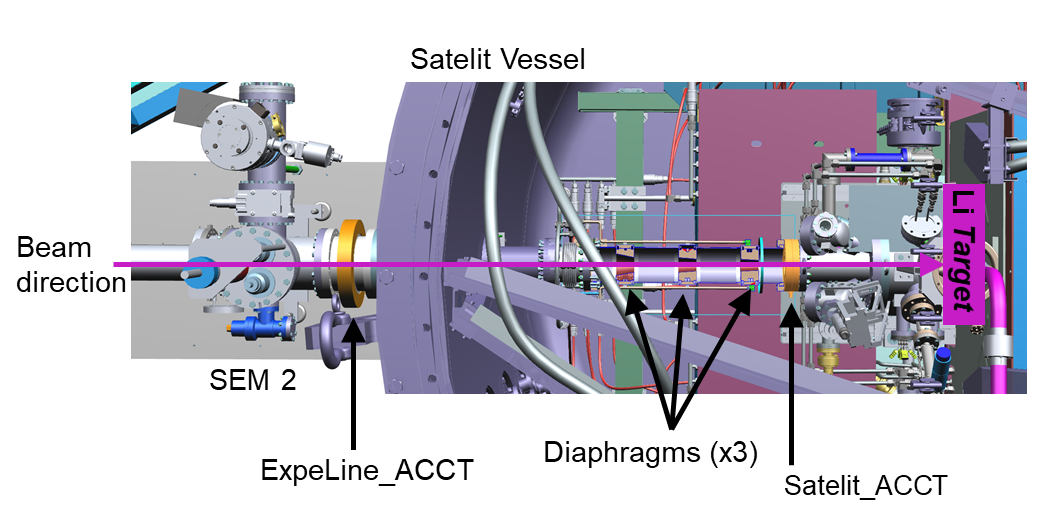}
    \caption{Simplified 3D CAD model of the last 2 meters of the experimental proton beam line up to the target.}
    \label{fig:Target_Setup}
\end{figure}

\noindent
Figure~\ref{fig:Target_Setup} illustrates the final section of the beamline—the last 2~m leading to the target—where three diaphragms are positioned between the ExpeLine ACCT and the SATELIT ACCT. These diaphragms are designed to protect the extremities of the \satelit nozzle (see Section~\ref{subsec:Satelit_Components}) from any power deposition that may occur due to potential beam misalignment.
Each diaphragm is engineered to dissipate a maximum average power of \SI{1.5}{kW}.

\subsection{Beam tuning}
\label{subsec:beamTuning}

The objective of the experimental campaign regarding the accelerator was to deliver a \SI{10}{kW} proton beam that matches the operational capabilities of the lithium target.
The beam intensity on the target was chosen to be \SI{11}{mA} at an energy of \SI{3}{MeV}. To achieve this, the IPHI accelerator had to be operated in pulsed mode, with beam pulses of \SI{6.1}{ms} at a repetition rate of \SI{50}{Hz}, corresponding to a duty cycle of 30.5\%. To account for potential losses (around 10\%) in the beamline and diaphragms described in Section~\ref{subsub:diaph}, the beam current extracted from the RFQ and transported to the target had to be around \SI{12}{mA}.
Before accelerating a \SI{10}{kW} proton beam through IPHI, the beam optics were carefully tuned to ensure safe and reliable transport up to the target. During these beam tuning phases, the duty cycle was reduced to $2 \times 10^{-4}$ (\textit{i.e.} \SI{200}{\micro\second} pulses at \SI{1}{Hz}) to limit the beam power to a few watts and prevent potential damage to the accelerator diagnostics or in case of beam losses. However, the beam current was kept at its nominal value of \SI{12}{mA} to ensure that the beam dynamics remained consistent with the full-duty-cycle experimental conditions.

Beam transport optimization in the LEBT line was carried out by adjusting the solenoid fields to maximize beam transmission through the RFQ. Transmission was calculated using Eq.~\ref{eq:transmission}, based on measurements from the RFQ\textsubscript{in} and RFQ\textsubscript{out} ACCTs. During all neutron production experiments reported in this paper, RFQ transmission consistently exceeded 97\%, approaching the 99\% theoretical value. The aperture of the Iris in the LEBT was adjusted to measure a beam current of \SI{12}{mA} after the RFQ.

\begin{figure}[h!]
    \centering    
    \includegraphics[width=.8\linewidth]{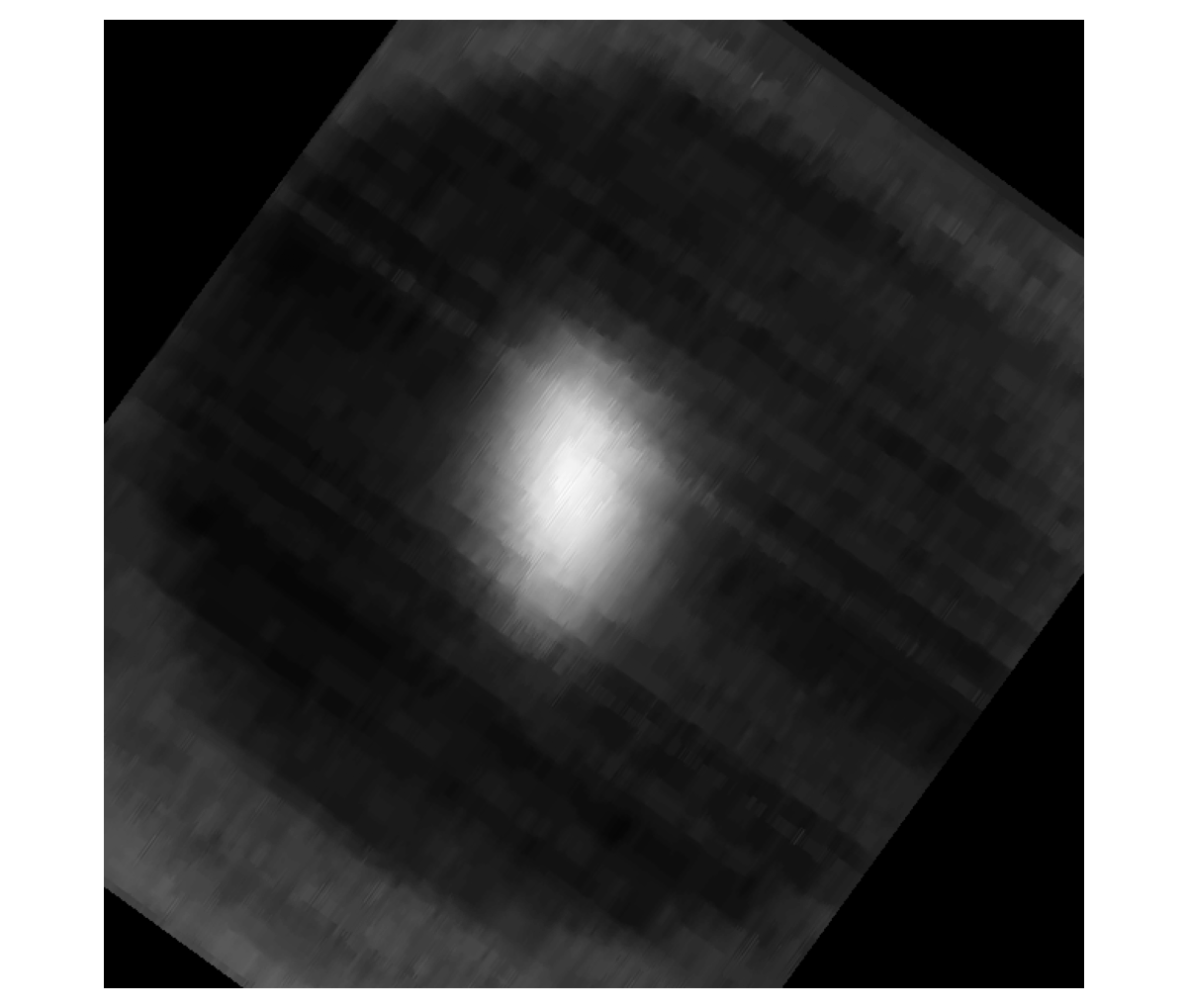}
    \caption{Image of the beam spot on the "beam blocker" made of a tantalum disk of \SI{42}{mm} diameter seen as a black circle.}
    \label{fig:Spot_on_Beam_Blocker}
\end{figure}

The HEBT line was optimized to ensure minimal beam losses along the beam pipe and to shape the beam to meet the experimental spot size specifications at the lithium target. The goal was to achieve a root-mean-square (rms) beam size of $8 \times 12\ \si{mm}$ (horizontal $\times$ vertical dimension). Initial quadrupole settings were determined using beam dynamics simulations performed with the TraceWin software~\cite{Uriot_IPAC2015} to satisfy these requirements.
The final beam tuning was performed experimentally by inserting a tantalum plate, referred to as the "beam blocker," just upstream of the lithium target. This plate was designed to withstand an average beam power of \SI{10}{W} without active cooling.
An optical camera is used to monitor the tantalum plate and observe the luminescence induced when the beam impacts it. An image of the beam striking the beam blocker is shown in Figure~\ref{fig:Spot_on_Beam_Blocker}.
It is important to note that the image of the beam spot cannot be directly considered an accurate measurement of the beam size, since the "transfer function" linking the observed luminescence to the beam density is unknown.
While this diagnostic does not provide precise beam profile measurements, it serves as a valuable tool for monitoring and adjusting beam centering on the target in close proximity to the lithium flow. Additionally, beam size reproducibility was regularly verified after the restart of the accelerator.

\subsection{Power ramp-up and beam time}\label{sub:beam_time}

\begin{figure}[h!]
    \centering    
    \includegraphics[width=1.\linewidth]{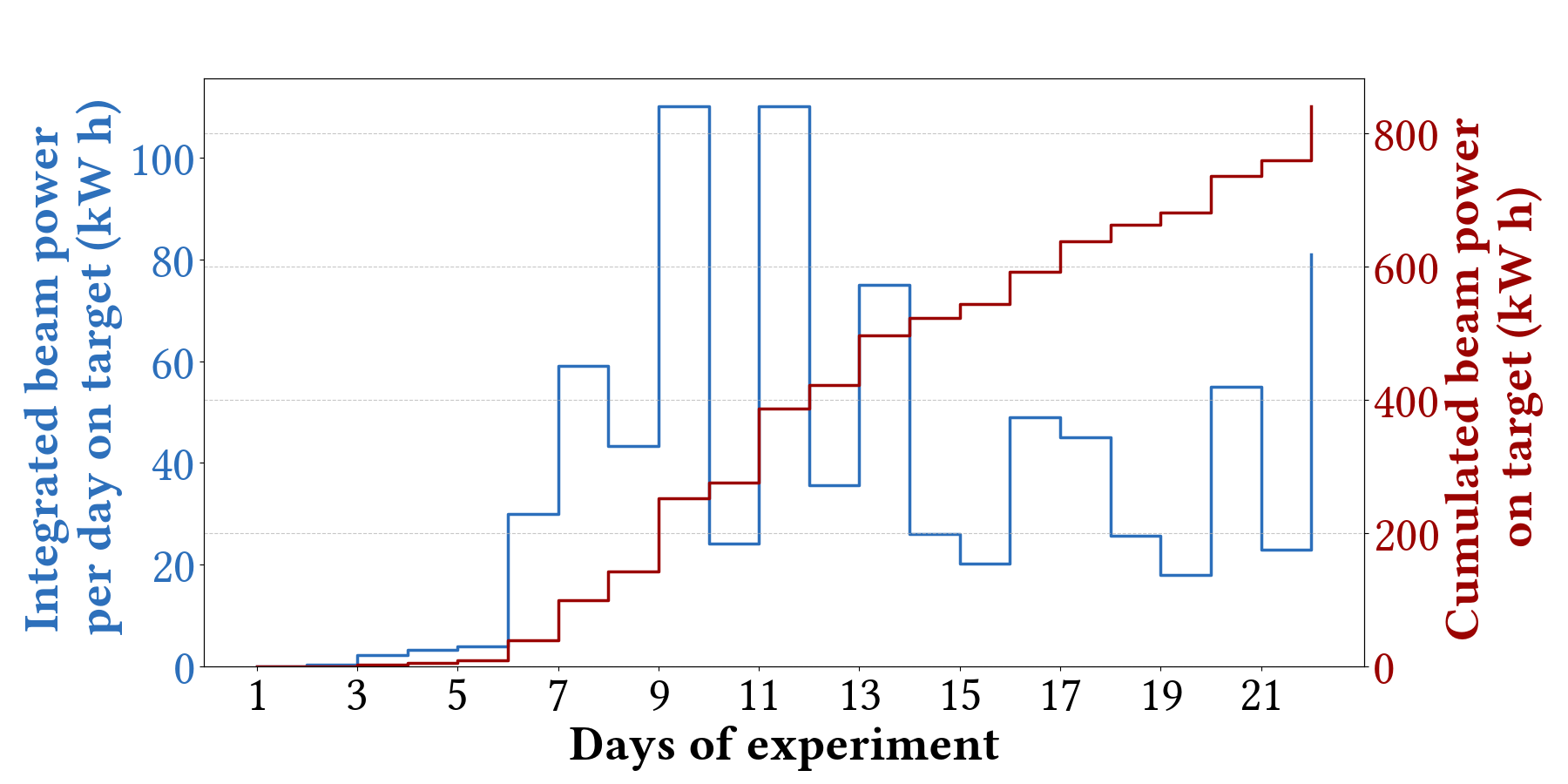}
    \caption{Integrated beam power on target per day of operation (in blue, left axis) and cumulated beam power on target (in red, right axis).}
    \label{fig:Cumulated_Power}
\end{figure}

During the first five days of operation, the beam power delivered to the target was carefully and progressively increased by extending the pulse length and raising the repetition rate, as the behavior of the lithium target needed to be closely monitored. On the sixth day, the nominal beam power of \SI{10}{kW} was successfully reached on SATELIT without any difficulties. Subsequently, several neutron production experiments were conducted at the nominal power.
It is worth noting that, following the initial cautious ramp-ups during the first few days, the proton beam could be switched on and off from 0 to \SI{10}{kW} without any additional precautions, whether at the start of an experimental session or immediately after a beam trip. This was possible because the lithium demonstrated excellent capability to reliably dissipate the full beam power.
Figure~\ref{fig:Cumulated_Power} shows the integrated beam power delivered to the target each day, along with the total accumulated power over the 22-day experimental campaign. While minor interruptions did occur, such as those related to the ion source, RF amplifier, or target interlock system, most were resolved within several tens of minutes. Moreover, the reliability improved over the course of the campaign: whereas one or two beam interruptions occurred per day in the early stages, the later stages, including the final day, experienced uninterrupted operation.

In total, we accumulated 840~kW.h, which is a world premiere at this proton beam energy.
The overall reliability of the experimental setup, comprising the accelerator and \satelit, proved to be very high. On two separate occasions, following the test program, we were able to deliver up to 110~kW.h of beam power to the target over 11 hours of continuous operation.

\subsection{Beam losses}\label{sub:beam_Losses}

The power losses in the SATELIT diaphragms were initially the limiting factor for ramping up to \SI{10}{kW} on the target. Therefore, these losses were minimized by adjusting the beam center on the target, particularly using steerer~4.
More generally, the objective was to minimize beam losses throughout the accelerator, particularly within the HEBT line (see Section~\ref{subsec:beamTuning}), to optimize beam transmission and minimize beam pipe material activation.
Losses within the Experimental Line (ExpeLine\_Losses) are determined using current measurements taken at the RFQ exit (RFQ\textsubscript{out} ACCT) and at the end of the Experimental Line (ExpeLine ACCT), just before the target. Similarly, the beam losses that occur in the SATELIT diaphragms are assessed by a differential measurement between the ExpeLine ACCT and SATELIT ACCT.

\begin{figure}[h!]
    \centering    
    \includegraphics[width=1.\linewidth]{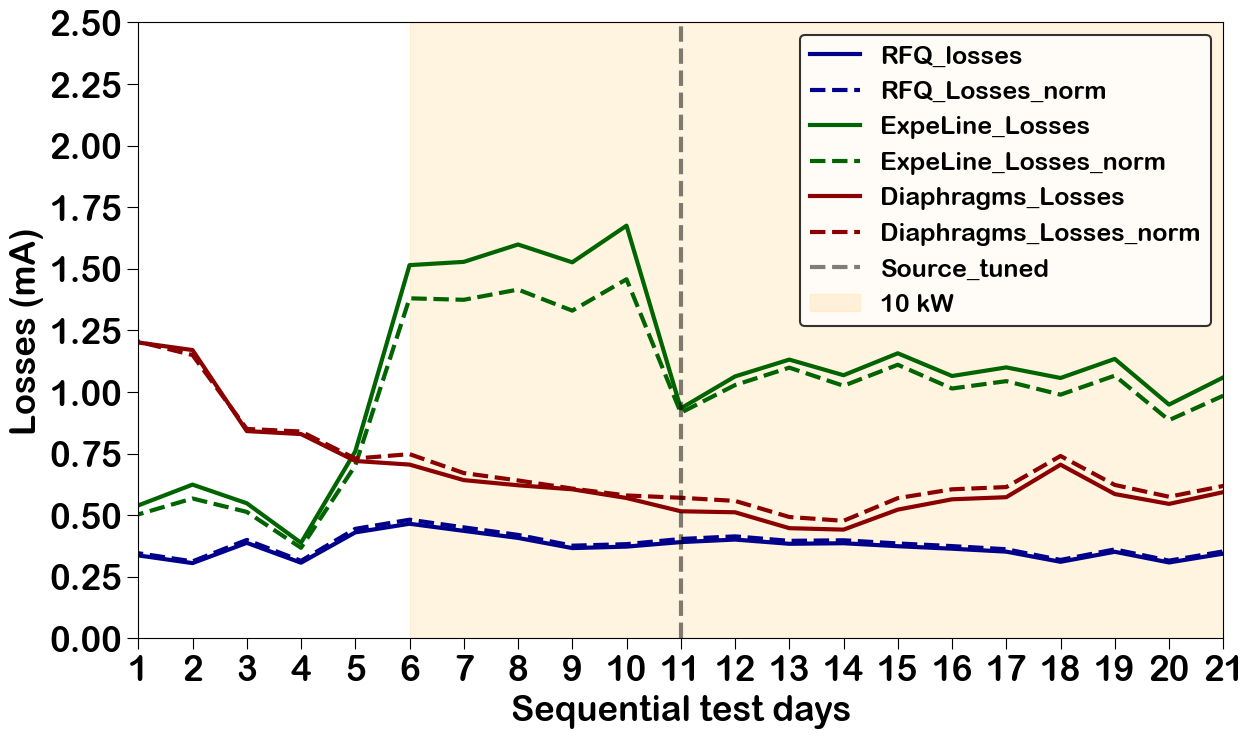}
    \caption{Evolution of average losses along the accelerator during SATELIT phase 1 tests. The vertical grey line indicated the moment of the ion source optimization.}
    \label{fig:Losses_Tests}
\end{figure}

Figure~\ref{fig:Losses_Tests} shows the day-by-day evolution of the average losses measured in the different sections of the beamline throughout the experimental campaign. These average losses were normalized to a reference current of \SI{12}{mA}, which corresponds to the overall mean of the four ACCT measurements over the entire experimental period. This reference value was chosen to smooth out random fluctuations and to enable a global assessment of beam losses under comparable operating conditions. The analysis is based on these normalized losses (namely RFQ\_Losses\_norm, ExpeLine\_Losses\_norm, and Diaphragms\_Losses\_norm in Figure~\ref{fig:Losses_Tests}).
In particular, Figure~\ref{fig:Losses_Tests} highlights the optimization of steerer~4 during the power ramp-up phases (the first five days), which led to a significant reduction in beam losses in the \satelit diaphragms (shown in dark red). When the accelerated beam power is at \SI{10}{kW} (yellow-highlighted region of interest, starting on day~6), the average losses in the diaphragms amounted to \SI{0.7}{mA}, or 6\% of the initial beam current, corresponding to an average power of \SI{657}{W} with constant steerer~4 settings. 
Losses in the RFQ remained stable and below the IPHI design value, with less than 3\% loss (\textit{i.e.}, RFQ transmission=97\%).

Throughout the experimental campaign, losses in the Experimental Line (depicted in green) were consistently observed. The average beam loss measured was \SI{1.37}{mA}. In the initial days of the experiment, it was noted that these losses increased with beam power (considering that the peak current was kept constant): starting from \SI{0.8}{mA} at an 8\% duty cycle (\SI{1.9}{kW}), rising to \SI{1.3}{mA} at a 15\% duty cycle (\SI{5.85}{kW}), and eventually stabilizing at \SI{1.0}{mA} at the nominal duty cycle after ten days of operation.

From day~11 onward (indicated by the vertical dashed line), losses in the Experimental Line significantly decreased, reaching the milliampere range. This can be attributed to different source and LEBT settings, better RFQ stabilization leading to an improved RFQ voltage law, and a higher proportion of protons being accelerated to the correct energy of \SI{3}{MeV} by the RFQ. 
With fewer un-accelerated particles, the losses could be permanently better controlled.
As the un-accelerated particles have an energy of \SI{95}{keV}, the power associated with these losses is not \SI{1}{kW} but around \SI{33}{W}. This significantly reduces the risks of activation and degradation of the vacuum chambers.
This assessment is confirmed by temperature measurements taken along the beamline (using thermocouples regularly located along the beam pipe, particularly at the entrance and exit of the dipole), which remained stable throughout the entire experimental campaign.

\subsection{Neutron monitoring}

\begin{figure}[h!]
    \centering    
    \includegraphics[width=1.\linewidth]{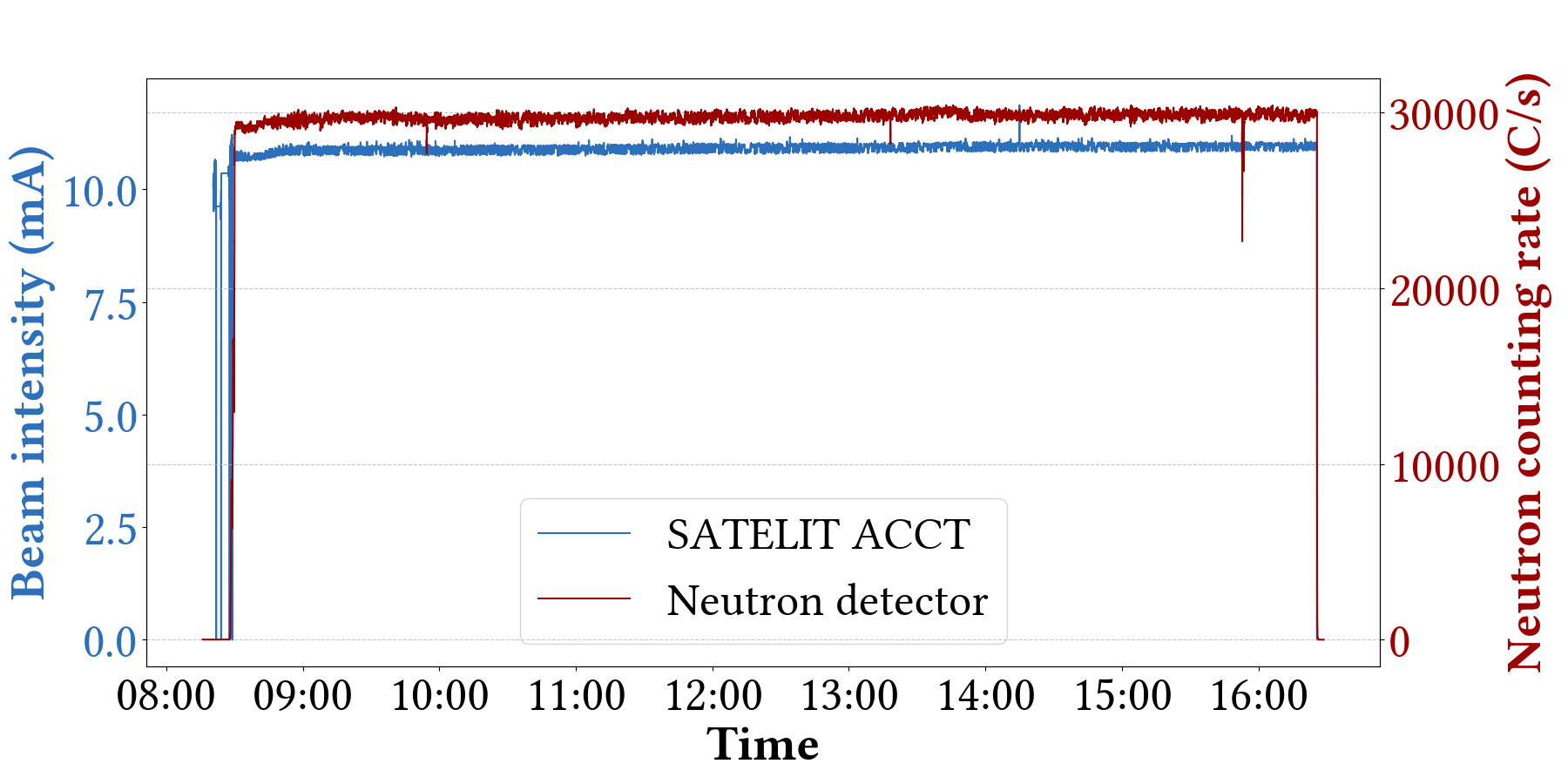}
    \caption{Peak beam current on target (in blue, left axis) and neutron counting rate (in red, right axis) during the last day of experiment. No beam trips or interlocks occurred on that day. The beam delivery to the lithium target was maintained continuously for the full eight-hour experimental session.}
    \label{fig:ACCT_Neutron_Detector}
\end{figure}

The online measurement of the proton current provides indirect access to the neutron flux, assuming that the lithium flows through the nozzle under nominal conditions. While this method is always applicable, it may not be reliable if the liquid lithium proton-to-neutron converter stops circulating or if the lithium film at the interaction point becomes too thin to stop the protons. Therefore, correlating the proton current given by the ACCTs and the neutron flux measured by neutron detectors is necessary. A loss of neutron signal while still detecting a proton signal may indicate a loss of the lithium film in the \satelit nozzle. Consequently, the absence of a neutron signal while detecting a proton signal should stop the experiment for safety reasons; otherwise, the \SI{10}{kW} proton beam could melt the nozzle, jeopardizing the experiment's integrity.
For this purpose, we used fission chambers designed by CEA, namely CF8Rgr chambers with an \SI{8}{mm} diameter containing \SI{5}{g} of enriched \isotope[235]{U} and depleted \isotope[234]{U}. The fission chambers operate in pulse mode, providing responses in counts per second. Two chambers were used sequentially in the neutron beam and demonstrated excellent proportionality to the proton current with a 5\% accuracy. They were calibrated absolutely using activation foil measurements (see Section~\ref{subsec:fluxMeasurements}).
An example of neutron flux measurements well correlated and proportional to the proton beam current delivered to the lithium target during a full day of experiment is shown in Figure~\ref{fig:ACCT_Neutron_Detector}.

\section{Saclay target with liquid lithium}
\label{sec:satelit}

\begin{figure}[h!]
    \centering    
    \includegraphics[width=1.\linewidth]{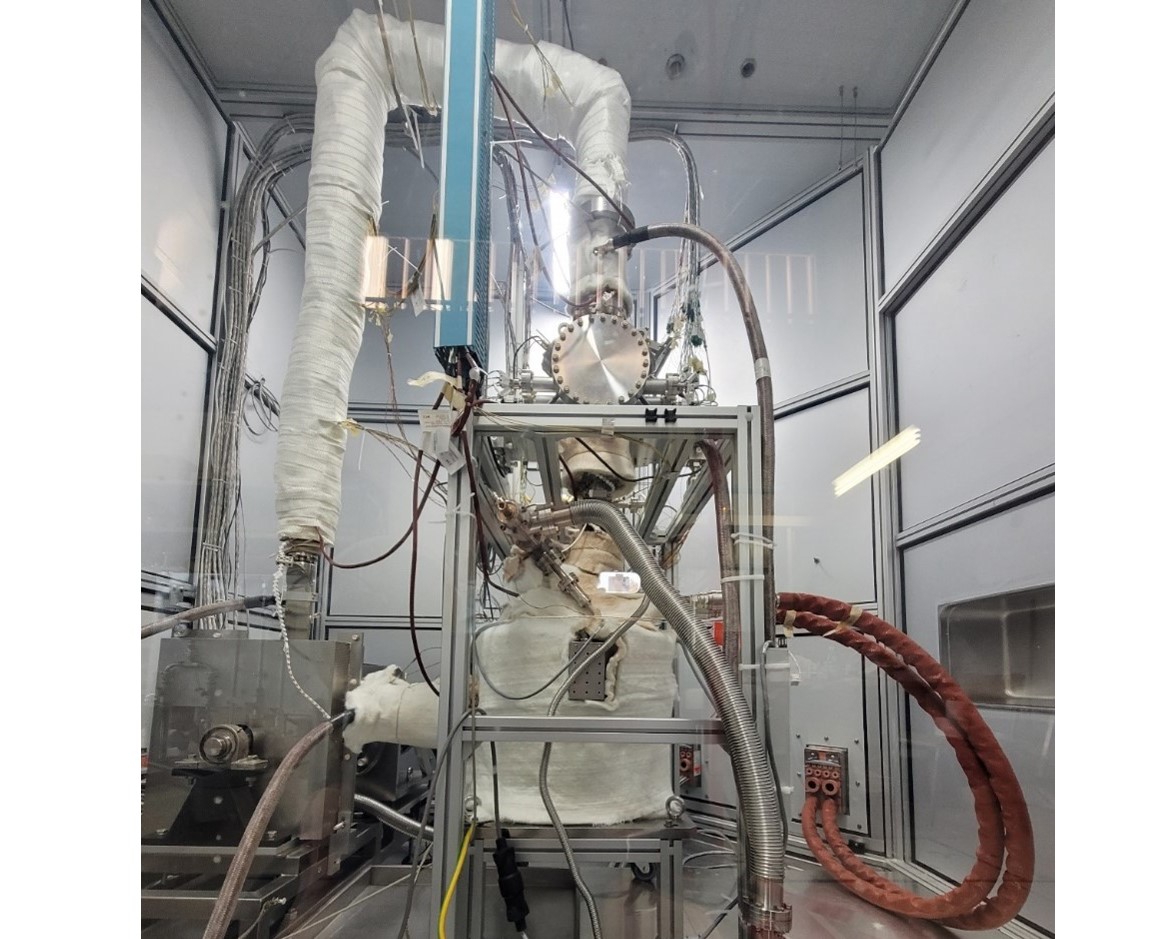}
    \caption{Picture of the \satelit liquid lithium loop taken during Phase~0 of the project.}
    \label{fig:photo_boucle1}
\end{figure}

The \satelit target consists of a mechanical assembly designed to circulate liquid lithium in a closed loop, producing a two-millimeter-thick liquid lithium film within a nozzle. This nozzle is located in a vacuum interaction chamber, open to the accelerator vacuum, where the proton beam impinges on the lithium film to produce neutrons. The overall dimensions of the lithium loop are \SI{2}{m} in height and \SI{1.3}{m} in width (Figure~\ref{fig:photo_boucle1}). Its main components include a lithium storage tank, an electromagnetic pump to circulate the metallic lithium, a nozzle to create the two-millimeter-thick film, and connecting pipes. This system requires \SI{22}{\liter} (\SI{11}{kg}) of liquid lithium at 220~\degree.
The development of this lithium loop for the HiCANS facility has been carried out in two phases. Phase 0, initiated in 2020, focused on circulating liquid lithium without neutron production to avoid radiation protection concerns. This phase aimed to validate the loop's performance, particularly the vacuum integrity, the argon inerting system surrounding the loop, the temperature control system, and the liquid metal circulation. In 2024-2025, Phase 1 aimed to integrate this lithium loop within the \iphi accelerator to produce and efficiently extract neutrons.

In the following sections, we will first discuss the critical compatibility between lithium and the materials used, and then detail the various components of the liquid lithium loop.

\begin{figure}[h!]
    \centering    
    \includegraphics[width=1.\linewidth]{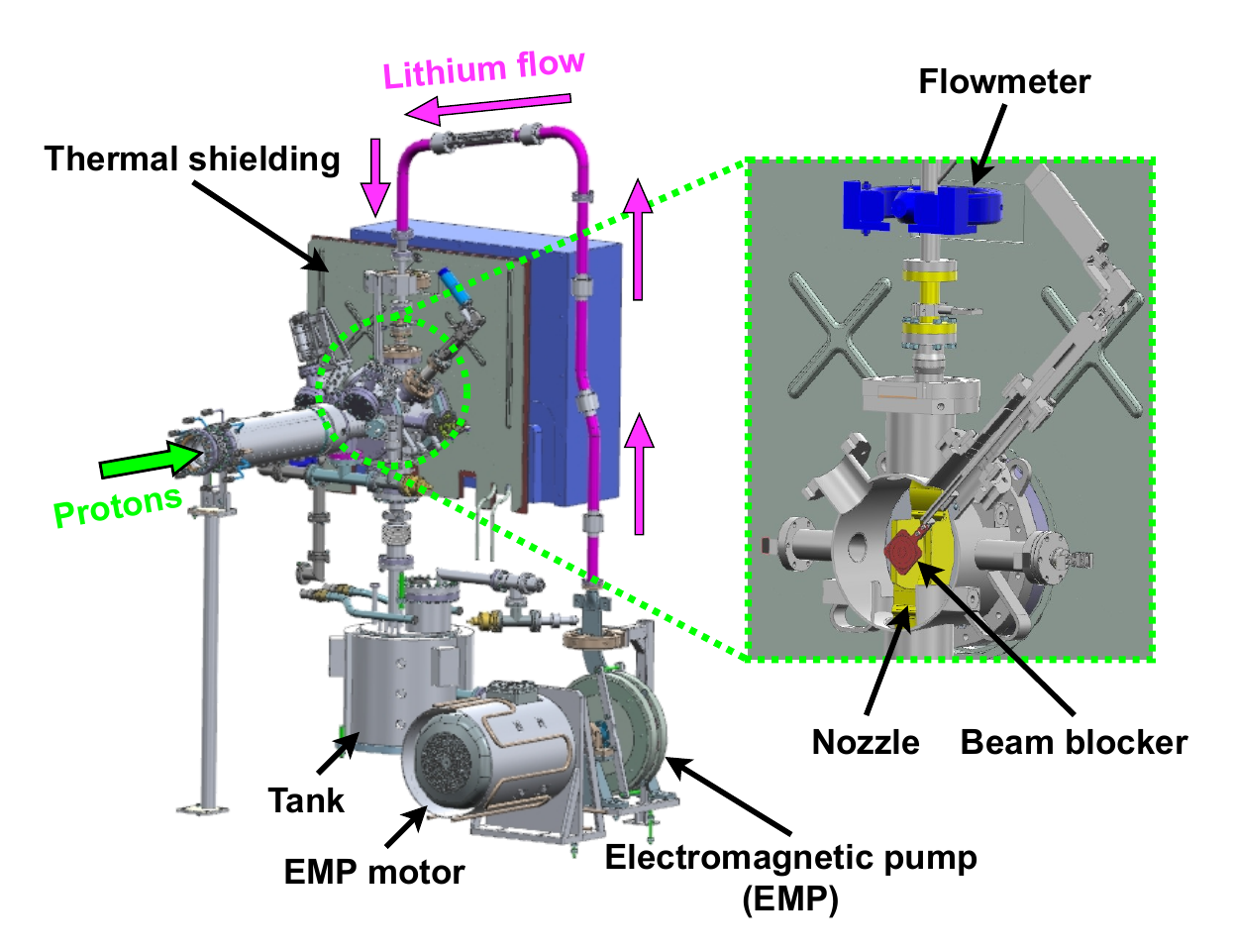}
    \caption{3D CAD model of \satelit showing its main components.}
    \label{fig:loop1}
\end{figure}

\subsection{Lithium compatibility with materials}

During the development of the \satelit liquid lithium target, special attention was given to the compatibility between lithium and various materials, as our goal is to develop a HiCANS that can operate with minimal maintenance. Consequently, lithium corrosion studies were conducted on 316L stainless steel, which is the material used for the \satelit lithium loop in this Phase 1, as well as on several titanium grades (ASME Grade 2 and Grade 4) and alloys (Grade 5) considered for future phases to minimize material neutron activation. Here, we present the studies performed on 316L stainless steel for this \satelit Phase~1, followed by a discussion on the selection of seals for lithium and vacuum tightness.

\subsubsection{Corrosion of structural materials}

In this work, no system for monitoring or purifying impurities was implemented, as such systems are not commercially available and would add excessive complexity to the circuits~\cite{KONDO2016}. Therefore, from the beginning, the main source of contamination responsible for corrosion, namely nitrogen, was limited to its maximum extent. This was achieved starting from the handling of the lithium ingots, through the filling of the \satelit tank, and throughout the entire operation of the lithium loop.
Since the operating temperature was expected to remain below 300~\degree, the corrosion was assessed as low to negligible, even with relatively high nitrogen content, such as several hundred parts per million by weight (ppmw). This is in comparison to the 10~ppmw limit generally agreed upon for higher temperatures~\cite{FURUKAWA2014}.

When the lithium ingots were melted in the tank, all potential contamination due to air entrance was carefully minimized through the use of a vacuum or argon as a cover gas. Given that the nitrogen solubility is around 4000~ppmw at 250~\degree in liquid lithium, it is estimated that the risk of lithium nitride buildup within the system is low. This is important because lithium nitride can eventually cause flow disturbances, heat transfer resistance, or even clogs.
The corrosion process proceeds via the homogeneous dissolution of the surface layer of the steel and depends on the temperature, nitrogen concentration, and liquid lithium flow rate. The dissolved nitrogen is hypothesized to participate in complexation reactions involving the formation of Li-Fe/Cr-N compounds in the liquid metal solution~\cite{borgstedt1987,MEDDEB2023,Chopra1986}, whose kinetics are generally limited by mass transfer through the boundary layer. However, no specific data were available for both low temperatures and high nitrogen concentrations. Therefore, it was supposed that operating at low temperatures with a service lifetime restricted to 1500 hours (one year typical operation time) in a perfectly sealed system would be sufficient to ensure low corrosion of the structure.

Other potential damaging phenomena, such as liquid metal embrittlement and erosion, were similarly assessed as manageable. Additive manufacturing 316L was also tested in liquid lithium to assess its corrosion resistance but was discarded upon further examination due to crack initiations.

\begin{figure}[h!]
    \centering    
    \includegraphics[width=1.\linewidth]{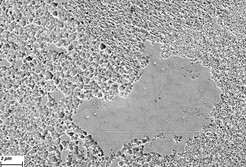}
    \caption{Scanning Electron Microscope (SEM) imaging of a 316L(N) sample immersed in liquid lithium at 300~\degree for 1015 hours with high nitrogen content ($\geq$ 200~ppmw) shows dissolution patterns as microporosities according to the crystallographic orientation, as well as the unaffected and possibly unwetted initial surface. 
    }
    \label{fig:SEM_316LN}
    
    \centering    
    \includegraphics[width=1.\linewidth]{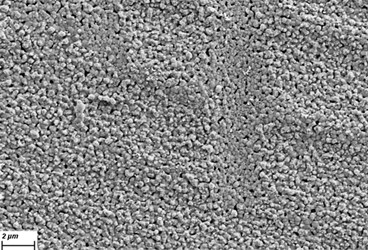}
    \caption{Scanning Electron Microscope (SEM) imaging of the 316L surface of the Omega-shaped piece used during the \satelit Phase~0 lithium loop operation for 36 hours at 250 -- 300~\degree shows microporosity patterns caused by dissolution in the liquid lithium.
    }
    \label{fig:Omega_316L}
\end{figure}

To support these assumptions, corrosion tests with specimen of 316L steel and pure iron were done in static conditions as a function of the temperature and nitrogen content in the liquid metal corrosion laboratory at CEA-Saclay~\cite{MEDDEB2023}. 
The potential evolution of the mechanical properties was assessed by comparing tensile tests conducted under argon and liquid lithium with those of reference specimens and aged specimens in liquid lithium, while the effects of temperature and nitrogen concentrations were assessed in static tests in controlled conditions. Indeed, the nitrogen was controlled through high-temperature purification using zirconium foils (getter) for a duration of 7 to 28 days at 650~\degree, along with a controlled addition of Li$_3$N powder, followed by a dissolution step at 600~\degree for 3 days. As illustrated in Figures~\ref{fig:SEM_316LN} and~\ref{fig:Omega_316L}, homogeneous and nearly negligible dissolution is effectively observed, aligning well with the Phase~0 observations (less than 1~$\mu$m in all cases). No mechanical evolution is detected either. However, porosities could create sites where beryllium compound particles, resulting from the \isotope[7][]{Li}(p,n)\isotope[7][]{Be} nuclear reaction, may be deposited (see Section~\ref{sec:7Be_buildup}). When compared to 316L steel, pure iron exhibits a slightly higher sensitivity to intergranular corrosion over an extremely limited range ($\leq$1~$\mu$m), which is attributed to a potentially higher reactivity at the grain boundaries, although this phenomenon is not yet fully understood.

\subsubsection{Seals for lithium and vacuum tightness}

During the assembly of \satelit, special attention was given to ensuring that all connections between the different parts were both lithium and vacuum tight. To achieve this, each connection was equipped with a Conflat (CF) type seal, which provides a mixed seal for lithium and vacuum. The leak tightness of the metal-to-metal seal depends on the smoothness of the seal's surface, its deformation when compressed between the knife edges, and its thermal stability in response to temperature variations and aging during operation.

The material of the seal is chosen for its low hardness and low oxidation in air, preventing damage to the knife-edge of the flanges during tightening. Therefore, the seals near high temperatures and close to the chamber are made of copper, while those in areas below 200~\degree and not in contact with lithium are made of aluminum. However, when there is potential contact with lithium, soft \armco iron seals were systematically installed at these locations, as they are corrosion-resistant and have low solubility in liquid lithium.
On soft \armco iron seals, specific corrosion tests were conducted under more severe conditions than anticipated in the metal liquid loop (450~\degree, 500~ppmw of nitrogen, for 4000 hours) to verify that the corrosion remained globally homogeneous and negligible.
During Phase~1 of the project, \armco iron seals were purchased from various suppliers, resulting in inconsistent leak tightness performance due to variations in surface quality across different batches. Some batches exhibited oxidized surfaces, hindering proper contact between the knife-edge and the seal, and highlighting deficiencies in surface chemical etching and quality control. Additionally, mechanical and thermal treatments were not consistently applied, and some batches had higher hardness than expected, which compromised sealing efficiency and damaged the flange knives. To ensure reliable performance, we systematically applied chemical and vacuum heat treatments to all batches of seals.

\subsection{Description of the \satelit target components}
\label{subsec:Satelit_Components}

\subsubsection{Overview of the monitoring}

\begin{figure}[h!]
    \centering    
    \includegraphics[width=1.\linewidth]{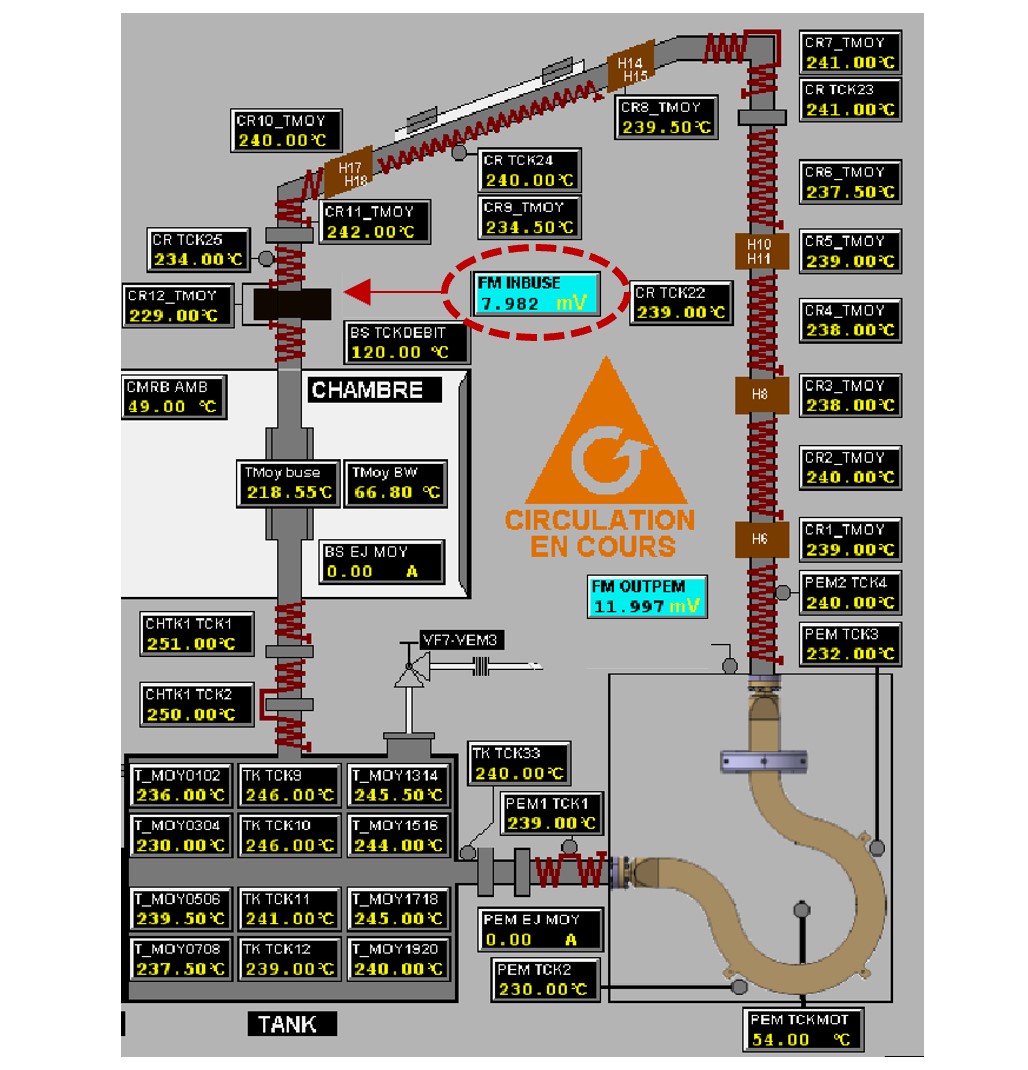}
    \caption{Screenshot of the MUSCADE\copright temperature monitoring.}    
    \label{fig:muscade}
\end{figure}

To avoid the solidification of lithium, which could block its flow, the lithium loop is maintained at a temperature around 220~\degree -- 250~\degree using heating cords and wrapped with an insulation layer made of glass wool. The temperature is monitored with type K thermocouples, as shown in Figure~\ref{fig:photo_boucle1}. Additionally, the pipes and the tank are wrapped with a beaded wire system that allows for the detection of lithium leaks: when lithium envelops the wire, an electrical contact with the structure is established, shutting down the circuit.
All measurements from these various instruments are supervised by a custom SCADA (Supervisory Control And Data Acquisition) system named MUSCADE\copright. It manages alarms, sends messages, and stores relevant data throughout the experiment, as illustrated in Figure~\ref{fig:muscade} for temperature supervision.

\subsubsection{Nozzle design and performances}

\begin{figure}[h!]
    \centering    
    \includegraphics[scale=0.7]{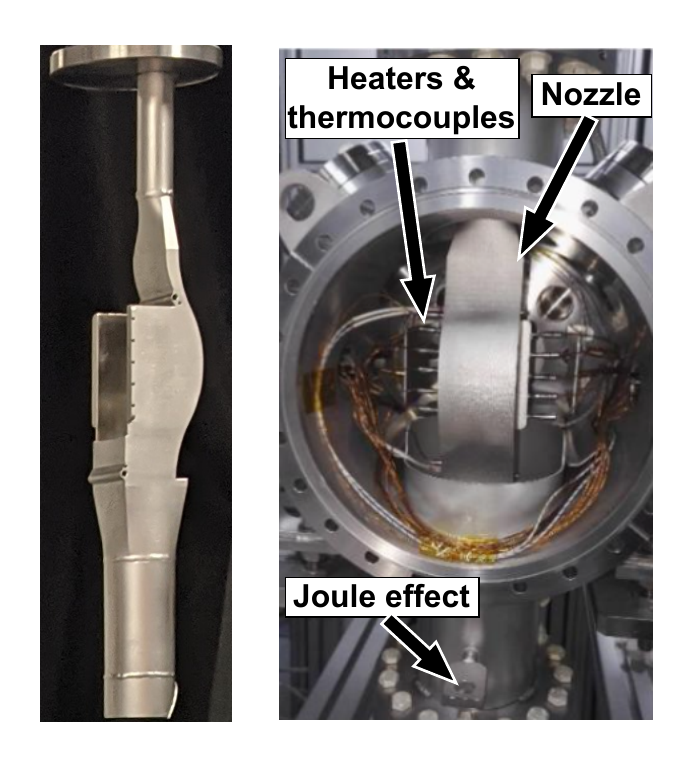}
    \caption{Left: Pictures of the nozzle with the protons coming from the left side. Right: rear view of this nozzle in the interaction chamber. The lithium flow from the top to the bottom in the direction of the tank.}
    \label{fig:nozzle1}
\end{figure}

The nozzle, made of 316L stainless steel, is the crucial part of the system since it produces the two-millimeter-thick lithium film that interacts with the proton beam. It is located in the interaction chamber, made of grade~2 titanium to minimize neutron activation, and is under vacuum, as shown in Figure~\ref{fig:nozzle1}.

\begin{figure}[h!]
    \centering    
    \includegraphics[width=1.\linewidth]{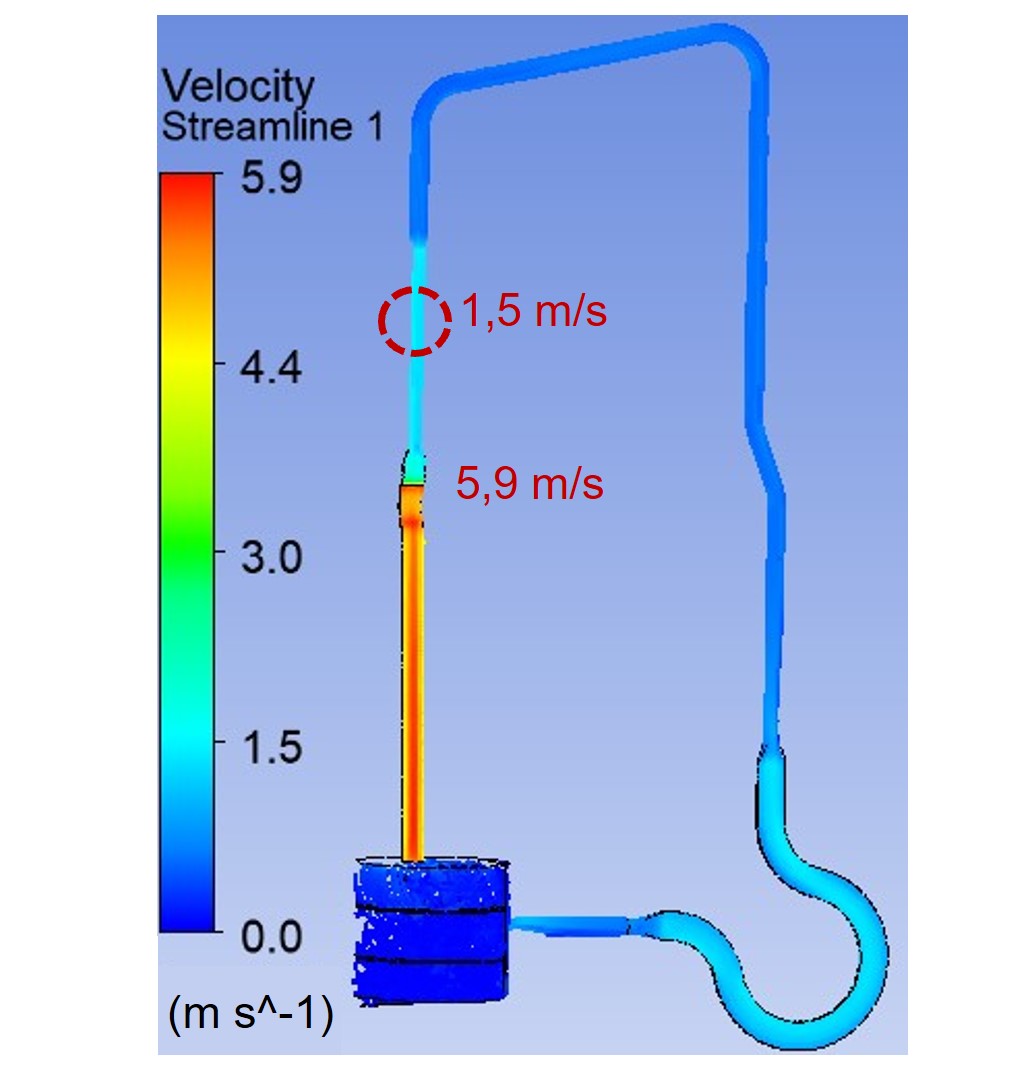}
    \caption{Velocity field of the liquid lithium inside the \satelit's loop, simulated with a computational fluid dynamics model performed with the ANSYS software\copright.}    
    \label{fig:CFD}
\end{figure}

During the project, hydrodynamic and magnetohydrodynamics simulations of the loop, performed with the ANSYS-CFX\copright software, provided crucial information such as pressure and velocity fields, as shown in Figure~\ref{fig:CFD}. This allowed for continuous improvement of the lithium flow by modifying the tank and the Omega-shaped pipe (see Section~\ref{subsubsec:Tank_EMP}) throughout the project from Phase~0 to Phase~1.

The thickness of the lithium film is determined by the dimensions of the nozzle's rectangular cross-section, which is \SI{46}{mm} wide and \SI{2}{mm} thick. The lithium enters the nozzle at a speed of 1.5~m.s$^{-1}$, is accelerated by the curved geometry of the nozzle to over 5~m.s$^{-1}$, and then returns to the tank, as shown in Figures~\ref{fig:nozzle1} and \ref{fig:CFD}. 

In operation, a camera placed on the interaction chamber allows for the live visualization of the lithium film flow. Additionally, the flow of lithium is measured using two electromagnetic flowmeters positioned above the nozzle (see Figure~\ref{fig:loop1}) and at the outlet of the Omega-shaped pipe. Each flowmeter consists of two permanent magnets with a magnetic field strength around \SI{0.5}{T} and has been calibrated at the Institute of Physics of the University of Latvia (IPUL) using an InGaSn eutectic alloy. 
During the experiment, the flowmeter located above the nozzle measured a velocity of 1.5~m.s$^{-1}$, in good agreement with the simulations shown in Figure~\ref{fig:CFD}. This confirms that the behavior of the loop is under control.

It should be noted that in the current configuration, the maximum allowed velocity is around 5~m.s$^{-1}$. Beyond this speed, cavitation phenomena occur, characterized by noise and saturation of the lithium flow velocity despite an increase in motor rotation speed. To avoid this, a higher lithium flow must be achieved, for example by lengthening the loop to achieve a higher Net Positive Suction Head (NPSH) for pump priming~\cite{BREKIS2023113919,Kravalis2022}. However, a lithium flow of 5~m.s$^{-1}$ was sufficient in this work to demonstrate our ability to produce neutrons with a 10~kW beam power.

Before the lithium circulation, the nozzle is heated by the Joule effect and heaters inserted on the nozzle, as depicted in Figure~\ref{fig:nozzle1}. If, by accident, the proton beam deteriorates and passes through the nozzle, a thin 0.5~mm isolated aluminum foil is placed behind it to detect the protons and stop the experiment.

Upstream of the nozzle, in the proton beam line, a cold trap composed of 6 washers intercepts and solidifies lithium vapors coming from the nozzle upon contact, limiting the contamination of the proton beam line. The CFD simulations also allowed for sensitivity studies to assess the maximum allowed lithium evaporation rate as a function of the lithium flow rate and proton beam characteristics. The maximum evaporation criterion has been set to around 10~mg.h$^{-1}$, which is a commonly agreed value~\cite{Silverman2020} that has to be consolidated with a long-term experiment. In fact, beyond this value, despite the presence of the cold trap positioned in the interaction chamber, there is a risk of lithium in the beam line and accumulation in the experimental setup, which could prevent its operation by disturbing electronic systems such as ACCTs in the beam line.

\subsubsection{Tank and electromagnetic pump}
\label{subsubsec:Tank_EMP}

\begin{figure}[h!]
    \centering    
    \includegraphics[scale=0.5]{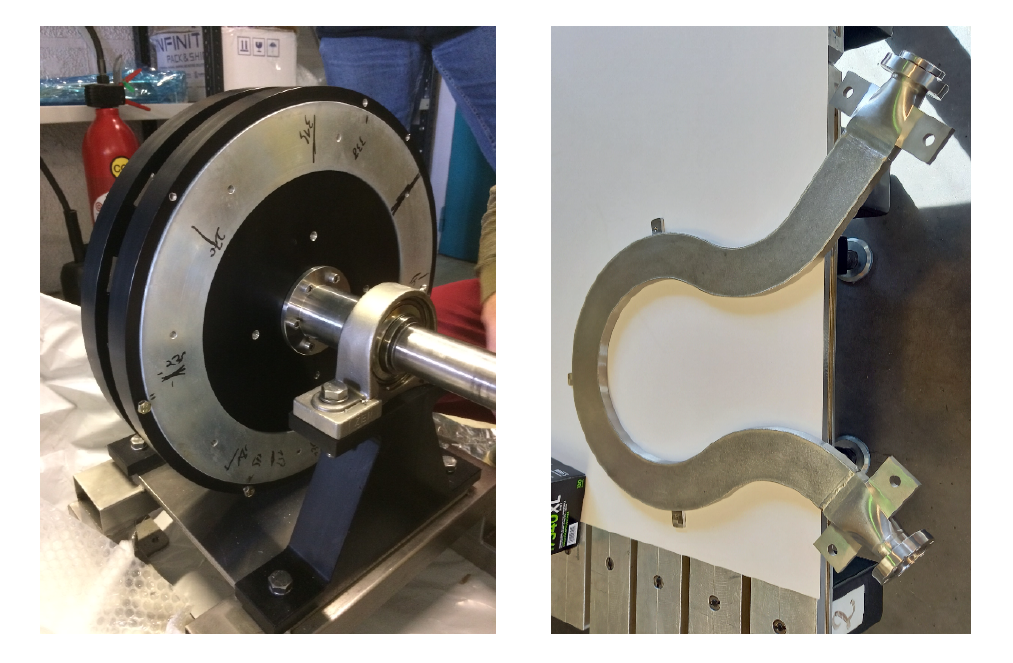}
    \caption{Pictures of the two discs supporting the 64 Sm$_2$Co$_{17}$ permanent magnets of the electromagnetic pump (left) and of the Omega-shaped pipe placed in between the two magnetic discs (right).}
    \label{fig:EMP}
\end{figure}

The lithium tank is located directly beneath the interaction chamber, as shown in Figure~\ref{fig:loop1}. Its primary role is to liquefy the lithium (melting temperature of 180~\degree) and maintain the operational temperatures between 220~\degree and 250~\degree, either by heating or cooling. The latter is necessary to remove the 10~kW beam-deposited power or to speed-up lithium solidification. The tank is completely insulated with glass wool and is heated with 3~kW external ceramic heaters and an internal oil circuit. All these operations are controlled by a ToolTemp\copright TT390/2 thermo regulator, able to extract up to 90~kW. Furthermore, in the tank, the speed of the lithium coming from the nozzle decreases which reduces the turbulence before suction by the pump.
Its volume is designed to accommodate the circulation of lithium throughout the loop, functioning as a buffer volume and priming for the electromagnetic pump (EMP)~\cite{bucenieks2000perspectives,bucenieks2011efficiency}.
\newline
\indent
The EMP consists of two discs containing 64 Sm$_2$Co$_{17}$ permanent magnets, each with a field of \SI{0.5}{T}, sandwiching an Omega-shaped pipe that connects the tank to the rest of the lithium loop. The two separate parts are shown in Figure~\ref{fig:EMP}. The magnets are positioned such that opposite poles are alternated. When the two discs rotate, driven by a motor, an alternating magnetic field is generated, inducing a current in the lithium and creating a Lorentz force that propels the lithium through the pipes. It should be noted that the Omega-shaped pipe is heated by the Joule effect with electrical contacts at its ends, allowing it to expand during thermal transients without touching the pump or its frame.

The tank and the electromagnetic pump are shielded by a sarcophagus made of 5~cm thick antimony-free lead integrated into a box with a 5~mm thickness of S355 steel. This shielding is designed to mitigate the radiological issues arising from \isotope[7][]{Be}, produced by the \isotope[7][]{Li}(p,n)\isotope[7][]{Be} nuclear reaction and dissolved in the lithium.

\subsubsection{Integration of the loop in its environment}

\begin{figure}[h!]
    \centering    
    \includegraphics[width=1.\linewidth]{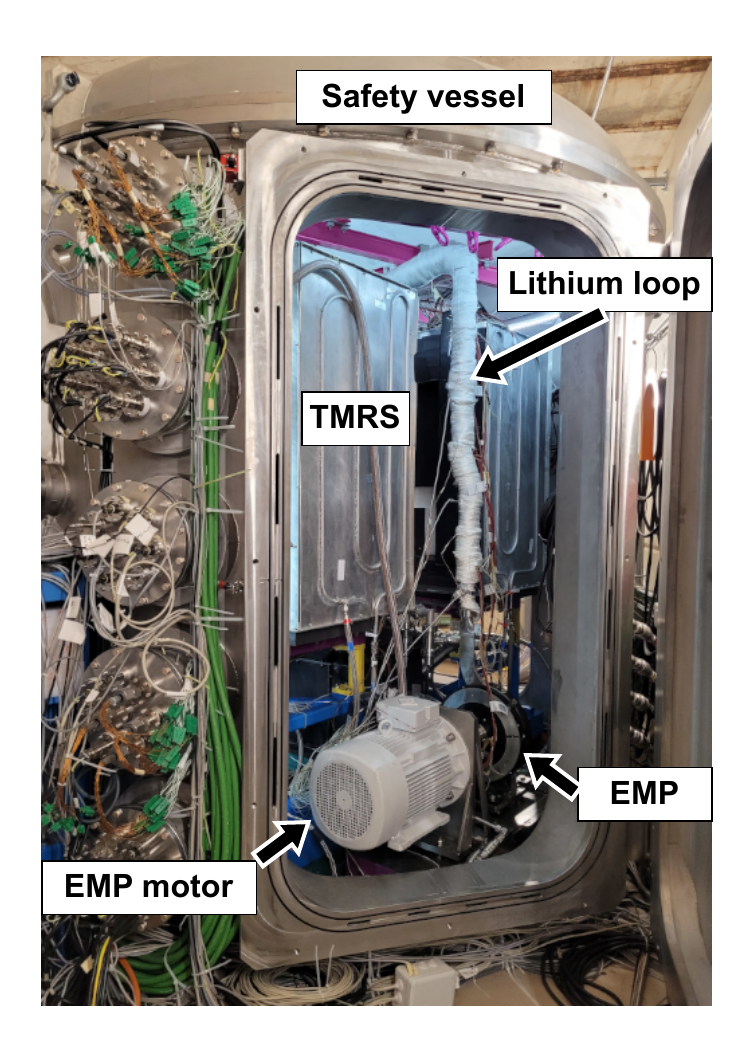}
    \caption{Picture of the \satelit lithium loop along with the target-moderator-reflector-shielding (TMRS) in the safety vessel. The TMRS is divided into two parts and can slide on its specific frame to create a 600~mm gap along the beam axis, facilitating maintenance of the lithium loop.}
    \label{fig:enclosure11}
\end{figure}

The lithium loop is contained within a cylindrical safety vessel, 3~m in diameter and height, featuring two independent nested walls to minimize the risk of lithium ignition from air contact, as illustrated in Figure~\ref{fig:enclosure11}. Each vessel has its own vacuum seal, providing two safety barriers. The inner vessel is pressurized to \SI{1.05}{bar} with argon to prevent lithium ignition, while the outer vessel is slightly under-pressurized at approximately 0.9~bar to contain potential contaminants.

Due to the lithium loop's operating temperature around 250~\degree, the equipment within the double wall vessel experiences significant heat. Therefore, cooling the components of the target-moderator-reflector-shielding (TMRS) assembly (see Section~\ref{sec:cmrbDesign}) and instrumentation is crucial. Practically, this is handled by the Eurodifroid\copright cooling system. Attention must be given to the maximum allowable temperature of the various materials, minimizing thermal losses in the lithium loop to prevent cold spots, and addressing thermomechanical constraints to ensure equipment longevity and functionality.

\begin{figure}[h!]
    \centering    
    \includegraphics[width=1.\linewidth]{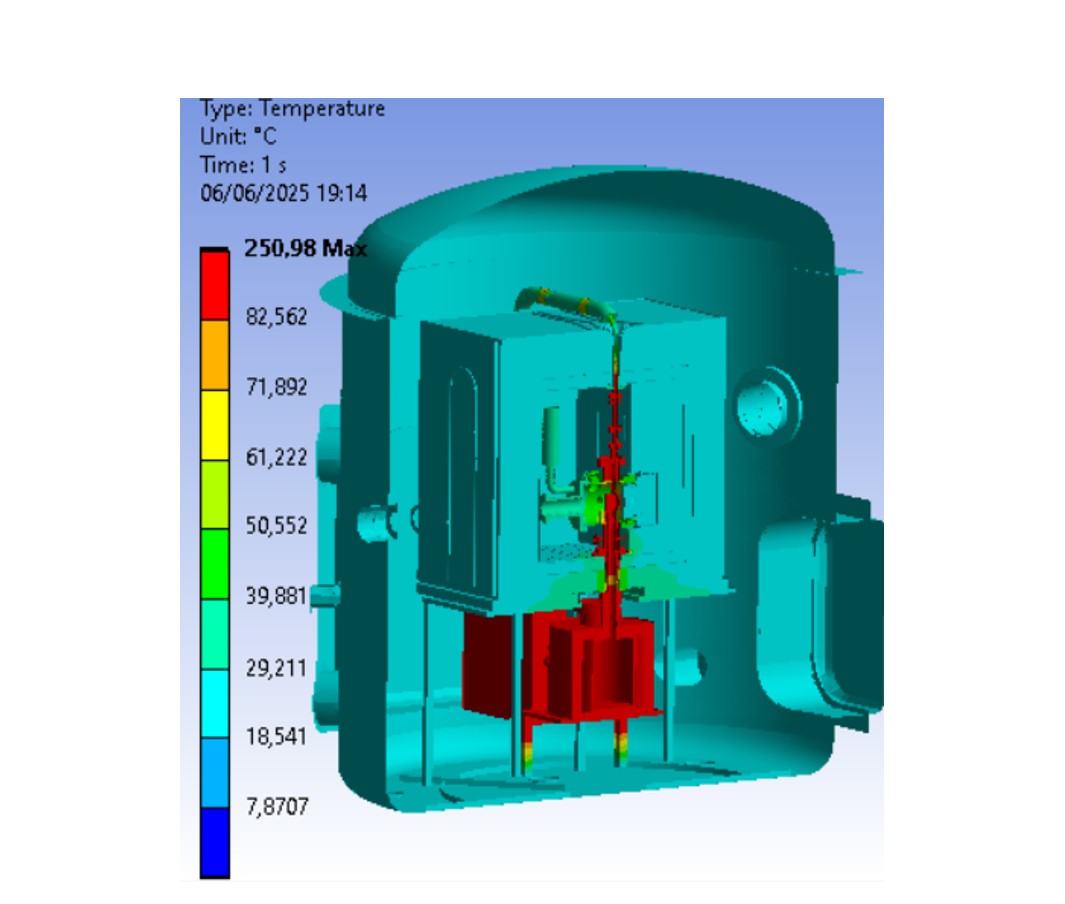}
    \caption{Temperature field of \satelit simulated with a computational fluid dynamics model performed with the ANSYS software\copright with all the heating and cooling devices at work.}
    \label{fig:thermique2}
\end{figure}

To tackle these thermal challenges, a comprehensive computational fluid dynamics (CFD) model was developed using ANSYS\copright software to size the cooling systems, including thermal radiation screens and a cooling oil system. The temperature distribution of the entire assembly, shown in Figure~\ref{fig:thermique2}, reflects the effectiveness of all cooling methods. A significant temperature contrast is observed, with the lithium loop at 250~\degree and surrounding areas around 50~\degree. These predictions were validated by measurements taken during the experiment with various temperature monitors.

Throughout the experimental campaign, before the loop operation, the temperature was increased from 180~\degree to 250~\degree in the tank as well as in all the components of the loop. Then the electromagnetic pump was turned on, and the lithium flow was set to the defined velocity for safety purpose. At the end of the day, the temperature was decreased and maintained throughout the night at 180~\degree (the lithium melting temperature) to be ready for the next day. Overall, the circulation of the liquid metal and control of the temperature proceeded smoothly, enabling a focus on neutron production and characterization.

\section{Description of the target - moderator - shielding assembly} 
\label{sec:cmrbDesign}

The objective of this experiment was to demonstrate the feasibility of delivering thermal neutron beams suitable for experiments at a HiCANS facility utilizing a liquid lithium loop. Neutrons produced by the \isotope[7][]{Li}(p,n) reaction at keV to MeV energies must be slowed down to $\sim$25~meV (thermal energies) using a material known as a moderator. In this study, the moderator is made of polyethylene.

\begin{figure}[h!]
    \centering
    \includegraphics[width=1.\linewidth]{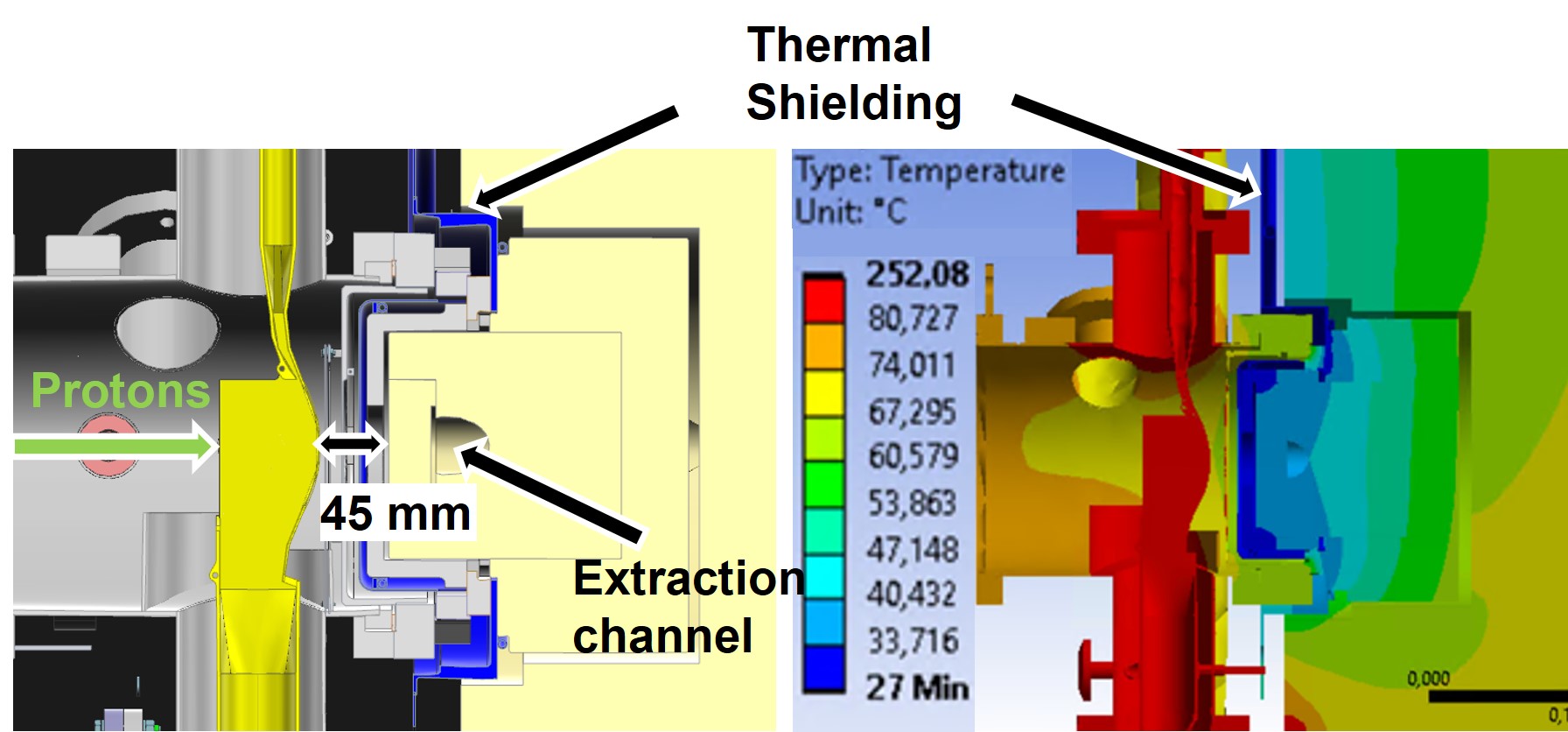}
    \caption{CAD drawing of the close coupling between the \satelit nozzle and moderator (left) with the associated temperature field simulated with a computational fluid dynamics model performed with the ANSYS software\copright with all the heating and cooling devices at work (right).}
    \label{fig:CMRB5}
\end{figure}

\begin{figure}[h!]
    \centering
    \includegraphics[width=1.\linewidth]{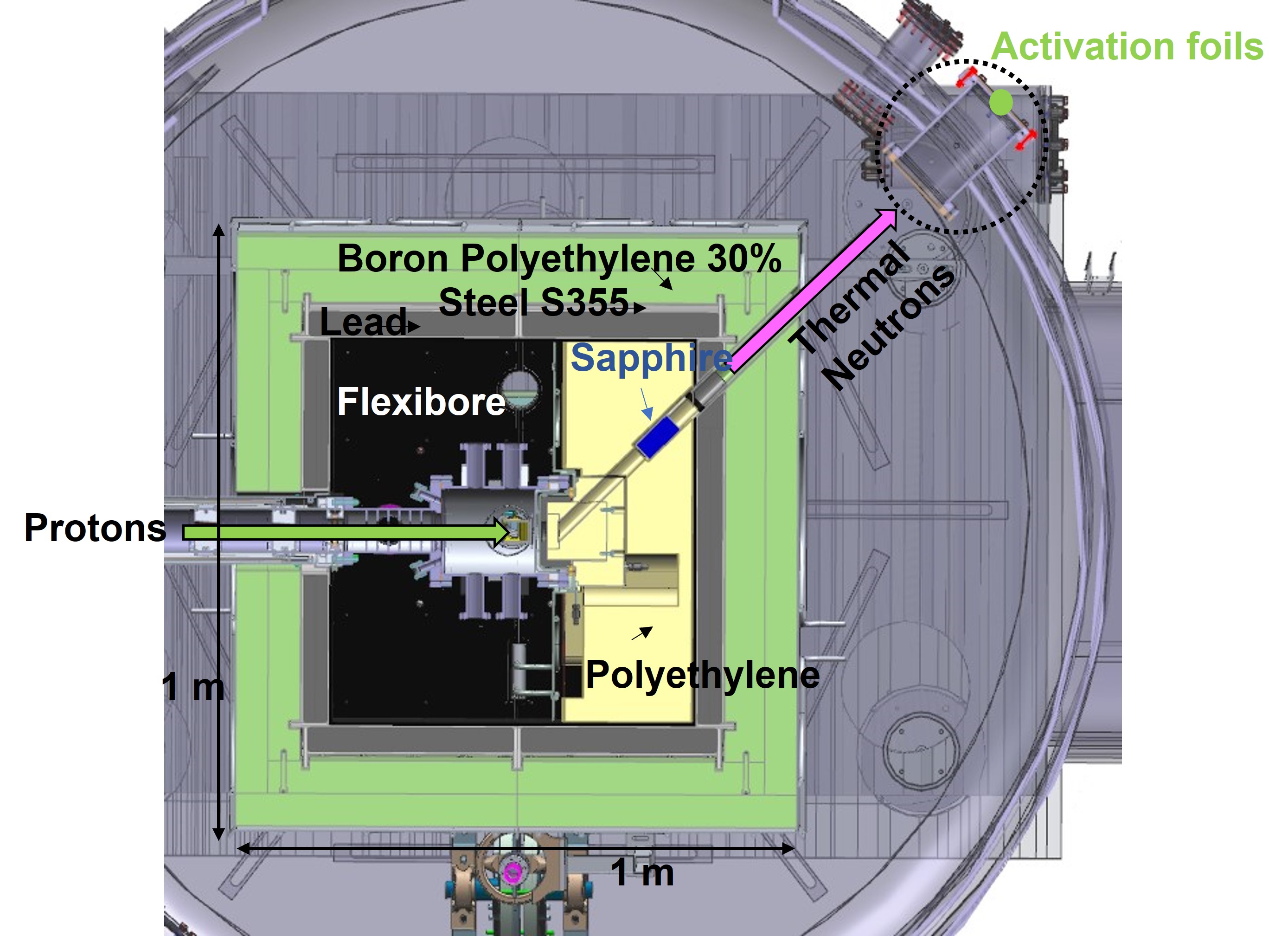}
    \caption{Top sectional view of the moderator and radiological shielding integrated to the \satelit target, inside the safety vessel.}
    \label{fig:CMRB1}
\end{figure}

To maximize the extracted thermal neutron flux, the moderator is positioned as close as possible to the nozzle, specifically 45~mm away, as illustrated in Figure~\ref{fig:CMRB5}. The overall dimensions of the moderator are 800$\times$800$\times$300 mm, as shown in Figure~\ref{fig:CMRB1}. Given that the polyethylene moderator is located near the nozzle, which operates at 250~\degree, a thermal screen cooled by oil is placed between the nozzle and the moderator to maintain the moderator's temperature at 50~\degree, thereby preserving its performance.

The moderator is housed within a radiation shielding structure, which consists of several layers as depicted in Figure~\ref{fig:CMRB1}. From the inside to the outside, the shielding includes a 5~mm layer of B$_4$C-loaded carpet (54\% B$_4$C) to absorb escaping thermal neutrons, followed by 50~mm of antimony-free lead sandwiched between a few millimeters of S355 steel plates for mechanical and thermal resistance against gamma rays. Finally, a 150~mm layer of borated polyethylene loaded with 30\% boron absorbs the remaining neutrons.

A conical hole, starting with a diameter of 35~mm and angled at 44\degree from the beam axis, is drilled through the various components of the TMRS to extract thermal neutrons. This extraction channel begins 25~mm from the start of the moderator to optimize thermal neutron collection. Additionally, it features a rod for inserting or removing a sapphire neutron filter (see Section~\ref{subsec:simulation}), which has a diameter of 5~cm and a length of 10~cm. The neutrons are then directed toward two 2017A aluminum portholes in the safety vessel, each with a thickness of 3~mm and a diameter of 154~mm.

\section{Neutron transport simulation for design and optimization}
\label{sec:simulation}

\subsection{Primary neutrons from a thick target \isotope[7][]{Li}(p,n) source}
\label{subsec:primaryNeutron}

When a 3~MeV proton impinges on the liquid target, it slows down \textit{via} mainly electromagnetic interactions until it either stops or undergoes a nuclear reaction with a \isotope[7][]{Li} isotope, representing an atomic fraction $f_{\text{7Li}}$ = 0.924 of the naturally abundant lithium (\isotope[6][]{Li} represents $f_{\text{6Li}}$ = 0.076).
The slowing down and, consequently, the energy loss of the proton in the target are computed using the \srim software, which provides stopping power curves $S(E_{\text{p}}^{\text{lab}}) = \frac{dE_{\text{p}}^{\text{lab}}}{dx}$~\cite{ZIEGLER2010}. For 3~MeV protons, the computed projected stopping range is 320~$\mu$m. Therefore, the thickness of the liquid lithium must be greater than this value to prevent damage to the nozzle from the proton beam.
At a 3~MeV proton incident energy ($E_{\text{p}}$), the nuclear reaction of interest producing a neutron is \isotope[7][]{Li}(p,n), which has two channels. The first channel involves the production of \isotope[7][]{Be} in its ground state and has a Q-value of -1.644~MeV, translating to an energy threshold of $E_{\text{th}}^{\text{GS}}$ = 1.88~MeV. In the second channel, the first excited state of \isotope[7][]{Be} at 0.429~MeV is populated, with a Q-value of -2.075~MeV and an energy threshold of $E_{\text{th}}^{\text{1st}}$ = 2.31~MeV.

\begin{figure}[!h]
    \centering
    \includegraphics[width=1\linewidth]{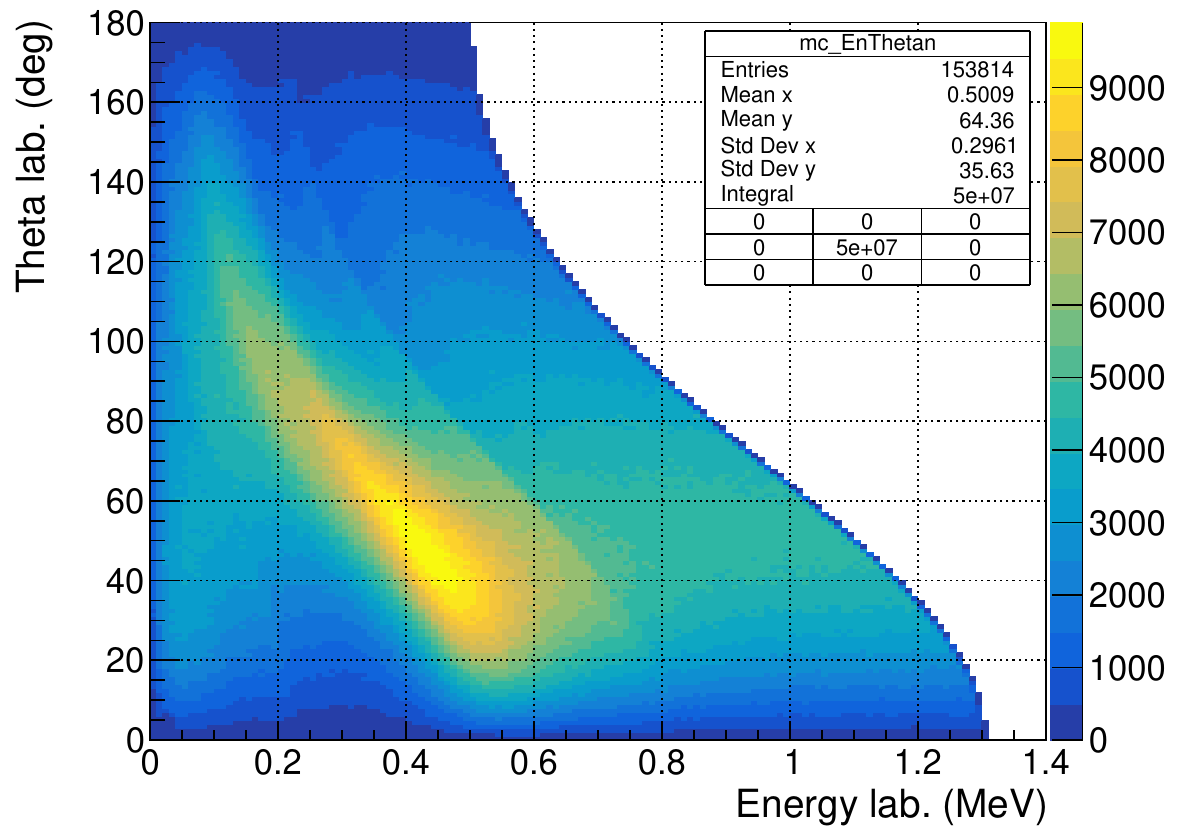}
    \caption{Emitted neutron number as a function of the laboratory emission angle and energy.}
    \label{fig:PDF_angle_energy_primaryNeutrons}
\end{figure}

The methodology used to build the neutron source term is described in~\cite{Ritchie_1976} and yields the neutron production as:

\begin{equation}\label{eq:7Li(p,n)}
    \frac{d^2N_{\text{n}}}{dE_{\text{n}}^{\text{lab}} d\Omega_{\text{n}}^{\text{lab}}} = 
    \frac{I_{\text{beam}}}{q}
    \frac{f_{\text{7Li}} \rho_{\text{Li}} N_{\text{Av.}} }{ M_{\text{Li}} }
    \frac{1}{S(E_{\text{p}}^{lab})}\frac{d\sigma}{d\Omega_{\text{n}}^{\text{CM}}}
    \frac{d\Omega_{\text{n}}^{\text{CM}}}{d\Omega_{\text{n}}^{\text{lab}}} 
    \frac{dE_{\text{p}}^{\text{lab}}}{dE_{\text{n}}^{\text{lab}}}
\end{equation}
\noindent
For $E_{\text{p}} \geq$ 1.95~MeV, the Liskien and Paulsen evaluation~\cite{LISKIEN197557} provides a parametrization of the differential cross section ($\frac{d\sigma}{d\Omega_n^{\text{CM}}}$) in the center-of-mass frame using Legendre polynomials. 
For $E_{\text{p}} \leq$ 1.95~MeV down to $E_{\text{th}}^{\text{GS}}$, the cross-section in the center-of-mass frame is described analytically by a Breit-Weigner single resonance whose parameters are those of an S-wave resonance. The partial width of this resonance is assumed to be large compared to the resonance energy~\cite{LEE19991,HERRERA2014}.
The code for generating neutron yields as a function of solid angle and energy of emitted neutrons has been validated at low energy using experimental data from Kokonov et al.~\cite{Kononov1977}.
The 2D plot representing the probability density as a function of the laboratory emission angle and energy is presented in Figure~\ref{fig:PDF_angle_energy_primaryNeutrons}. From this plot, the primary neutron characteristics can be sampled during the Monte-Carlo simulation. The neutron yield associated with this reaction, used to normalize the simulation, is 2.4$\times10^{-4}$~n/p~\cite{Porges1970}. This yields to the production of 5$\times$10$^{12}$~n$_{\text{fast}}$.s$^{-1}$ at 10~kW.

\subsection{Primary gamma rays from a thick target p+\isotope[7][]{Li} source}

In addition to the primary neutron sources, it is crucial to have a good estimate of the primary gamma-ray production in \isotope[7][]{Li}(p,$\gamma$) reactions to assess its contribution to any experimental setup and to design radiological shielding. These reactions can be decomposed into \isotope[7][]{Li}(p,p'$\gamma$)\isotope[7][]{Li}, \isotope[7][]{Li}(p,n$\gamma$)\isotope[7][]{Be}, and \isotope[7][]{Li}(p,$\gamma$)\isotope[8][]{Be}$^{\ast}$ reactions.
The first reaction leads to the emission of a 0.478~MeV gamma ray from the first excited state of \isotope[7][]{Li}. The second excited state at 4.630~MeV cannot be reached with 3~MeV protons. The yield for the 0.478~MeV gamma ray is $Y_{0.478~\text{MeV}} = 1.0 \times 10^{-4}$~$\gamma$/p.

The second reaction occurs when, after neutron emission, the first excited state of \isotope[7][]{Be} is populated, leading to the emission of a 0.429~MeV gamma ray (see Section~\ref{subsec:primaryNeutron}). The computed yield with the cross section from~\cite{BACHILLERPEREA2017} for 3~MeV protons is $Y_{0.429~\text{MeV}} = 1.2 \times 10^{-5}$~$\gamma$/p.

The third reaction produces a \isotope[8][]{Be} compound nucleus with a proton separation energy of 17.25~MeV, leading to a potential maximum available excited energy of 20.25~MeV. Then, the \isotope[8][]{Be} de-excites by emitting high-energy gamma rays of 17.6~MeV and 14.6~MeV~\cite{Zahnow1995}. A good understanding of the production of these high-energy gamma rays is of primary importance since they are difficult to shield. Zahnow et al.~\cite{Zahnow1995} have determined the production cross sections up to $E_{\text{p}}$ = 1.5~MeV. It is then assumed that the cross section is constant from 1.5~MeV to 3~MeV. Under this assumption, the estimated yields are $Y_{17.64~\text{MeV}} = 4.3 \times 10^{-8}$~$\gamma$/p and $Y_{14.61~\text{MeV}} = 7.7 \times 10^{-8}$~$\gamma$/p. Apart from being difficult to shield, these gamma rays could be used to perform photo-nuclear reaction studies.
New measurements would be needed to further understand the production of these primary gamma rays.

\subsection{Neutron transport simulation with \toucans}
\label{subsec:simulation}

The open-source Monte-Carlo \geant toolkit is developed by an international collaboration with the aim of supporting high-energy physics applications as well as other domains such as space science, medical physics, nuclear engineering and physics. Its Neutron-HP package (where -HP stands for High Precision) allows for the accurate transport of neutrons with energies below 20~MeV. This package has undergone numerous developments and benchmarks over the years, placing \geant on par with reference neutron transport codes such as \mcnp~\cite{Werner2018} and \tripoli~\cite{Brun2015} in terms of physics~\cite{Mendoza2014,Mendoza2018,Tran2018,Thulliez2022,Zmeskal2023,Zmeskal2024}.

\begin{figure}[!h]
    \centering
    \includegraphics[width=1\linewidth]{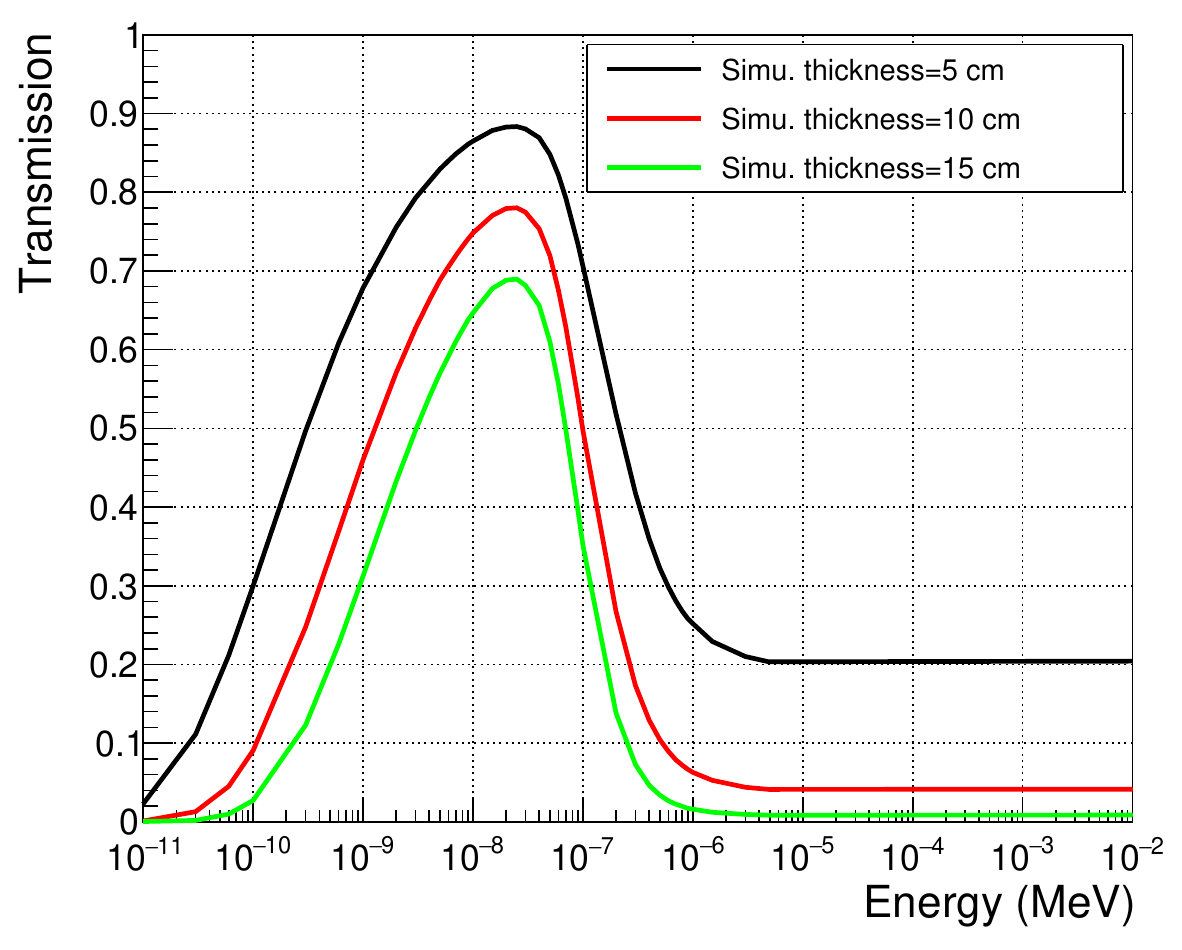} 
    \caption{Neutron transmission through sapphire single crystals computed with \ncrystal library.}
    \label{fig:neutronFilters}
\end{figure}

The \toucans code has been developed to provide an intuitive key/value user interface that hides the complexity of \geant. Therefore, it can be used to perform multi-objective optimization and build a Pareto front to choose the best solution. It can also be coupled with external softwares, such as \promethee, to leverage advanced algorithm-based optimization, such as the Kriging method~\cite{Mom2022}.
In addition to the well-benchmarked \geant toolkit, regarding neutron transport below 20~MeV, \toucans has been successfully validated on three different beryllium target experimental configurations~\cite{Tran2018,Tran2020,Thulliez2020,Schwindling2022}. This gives us confidence in its predictions and use for neutron experiment design.
Furthermore, particular attention has been paid to the description of neutron-crystal interactions since single crystals are used as neutron filters in numerous neutron experimental setups to eliminate fast neutrons and gamma rays while conserving thermal ones. In fact, its neutron transmission curve, reflecting the underlying cross section~\cite{Squires_2012}, as seen in Figure~\ref{fig:neutronFilters}, shows a transmission peak at thermal energies and drops around 0.5~eV. To simulate this behavior, either \ncrystal~\cite{caiKittelmann2020,KittelmannCai2021} or the JEFF3-3 library can be used, both yielding similar results to the percent level~\cite{MomThese2023}. To evaluate the impact of such single crystals on HiCANS performance, sapphire single crystals were characterized at the ILL nuclear research reactor on the TENIS neutron beamline using a CCD camera with a \isotope[6][]{Li}F:ZnS screen along with gold and indium activation foils. The data validated the \ncrystal and JEFF3-3 library; further details are available in~\cite{MomThese2023}.
\newline
\indent
One other key feature of \geant in such projects is its ability to transport particles in CAD files thanks to the \cadmesh library~\cite{poole2012acad,poole2012fast}. In fact, it is crucial in designing and optimizing a CANS to be able to constantly switch between the Monte-Carlo simulation and the CAD file of the mechanical structure, whether regarding the geometry or the choice of materials, to maximize the extracted thermal neutron flux while minimizing the neutron activation of the structure.

\section{Study of the \isotope[7][]{Be} build-up}
\label{sec:7Be_buildup}

The 3~MeV proton beam directed toward the lithium target produces neutrons through \isotope[7][]{Li}(p,n)\isotope[7][]{Be} reactions.
While neutrons are the quantity of interest, the same reaction unavoidably produces \isotope[7][]{Be} radionuclides, which are assumed to dissolve homogeneously in the liquid lithium volume and accumulate with integrated beam power. 
The \isotope[7][]{Be} nuclei decay back to \isotope[7][]{Li} by electron capture ($Q_{EC} = 861.9$ keV) with a half-life of 53.22 days.
In 10.4\% of the decays, the \isotope[7][]{Li} daughter nucleus is formed in its 477.6~keV first excited state, yielding a gamma ray.
The radiation protection issue arises because external exposure can range from $10^{-7}$ to $10^{-4}$~$\mu$Sv.h$^{-1}$.Bq$^{-1}$ depending on the geometrical configuration~\cite{Delacroix2006}. At the end of the experiment, \textit{i.e.} after nearly 100~h of irradiation by a 10~kW proton beam, the total activity of \isotope[7][]{Be} diluted in the \SI{22}{\liter} of lithium is around 200~GBq, which could lead to significant radiological exposure.
This radiation protection issue is exacerbated by the inevitable solidification of a fraction of the liquid lithium on the inner pipes of the loop when the experiment is stopped, as well as within low-flow stagnation zones such as in the tank and Omega-shaped pipe. That is why the latter two are shielded with 5~cm of lead (see Section~\ref{subsubsec:Tank_EMP}). However, shielding is very difficult to implement for the \satelit pipes since they are already instrumented with heating devices and insulating layers (see Section~\ref{sec:satelit}). Therefore, it is crucial to understand how the liquid lithium, and thus the dissolved \isotope[7][]{Be}, freezes on the inner pipe walls, since the lithium crust thickness will directly impact the dose that a person will receive upon contact with the pipes.

\begin{figure}[!h]
    \centering
    \includegraphics[width=1\linewidth]{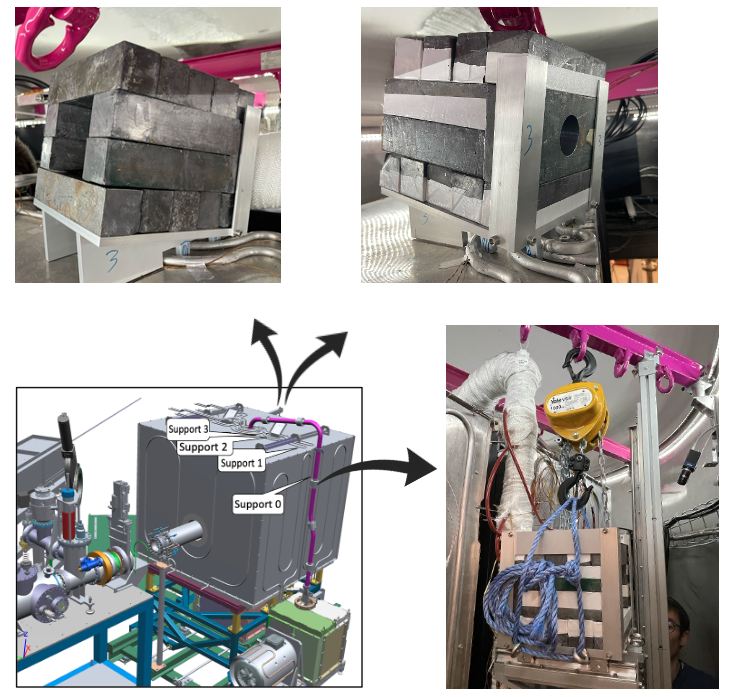}
    \caption{
    Experimental arrangement for \isotope[7]{Be} spectroscopy and hot-spot localization along the \satelit liquid-lithium loop. A 3''$\times$3'' NaI(Tl) scintillation detector is mounted behind a 50~mm-thick lead collimator with a 50~mm circular aperture. The assembly can be positioned at three horizontal stations (Support~1 to 3) and translated to three heights via a winch-driven vertical stage (Support~0). The two upper photographs show front- and rear-side views of the collimator at the central horizontal station (Support~2) before the detector is inserted, while the image on the right highlights the winch mechanism.
    }
    \label{fig:BeDevice}
\end{figure}

\begin{figure}[!h]
    \centering
    \includegraphics[scale=0.5]{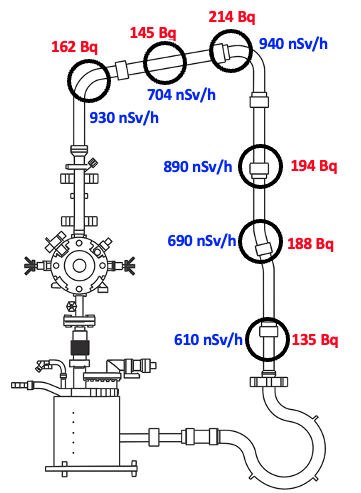}
    \caption{ 
    Energy-integrated count rates obtained with the collimated 3'' × 3'' NaI(Tl) detector at six survey points along the SATELIT lithium loop (red figures). Ambient dose-rate readings collected with an unshielded handheld 6150~AD probe at the same six points (blue figures). Both data sets correspond to an integrated beam power of 330~W.h.
    }
    \label{fig:BeResults}
\end{figure}

Two complementary $\gamma$-ray diagnostics were deployed for the \isotope[7]{Be} monitoring.

First, a 3'' × 3'' NaI(Tl) detector placed in a 50~mm thick lead shielding with a circular aperture of 50~mm diameter can be positioned at six well-defined positions along the external pipe of the loop (three horizontal stations "Support~1–3" and three elevations on a winch-driven stage "Support 0"), as shown in Figure~\ref{fig:BeDevice}.

The linear response and the energy resolution of the detector were determined beforehand using \isotope[133]{Ba} (356~keV), \isotope[137]{Cs} (662~keV), and \isotope[60]{Co} (1173~keV, 1333~keV), enabling a good understanding of the detector response at 478~keV.

At an integrated beam power of 330~W.h, a \SI{900}{\second} spectrum and an energy-integrated count rate were recorded at each station; the latter results are plotted as red symbols in the left panel of Figure~\ref{fig:BeResults}.

\begin{figure}[!h]
    \centering
    \includegraphics[width=1\linewidth]{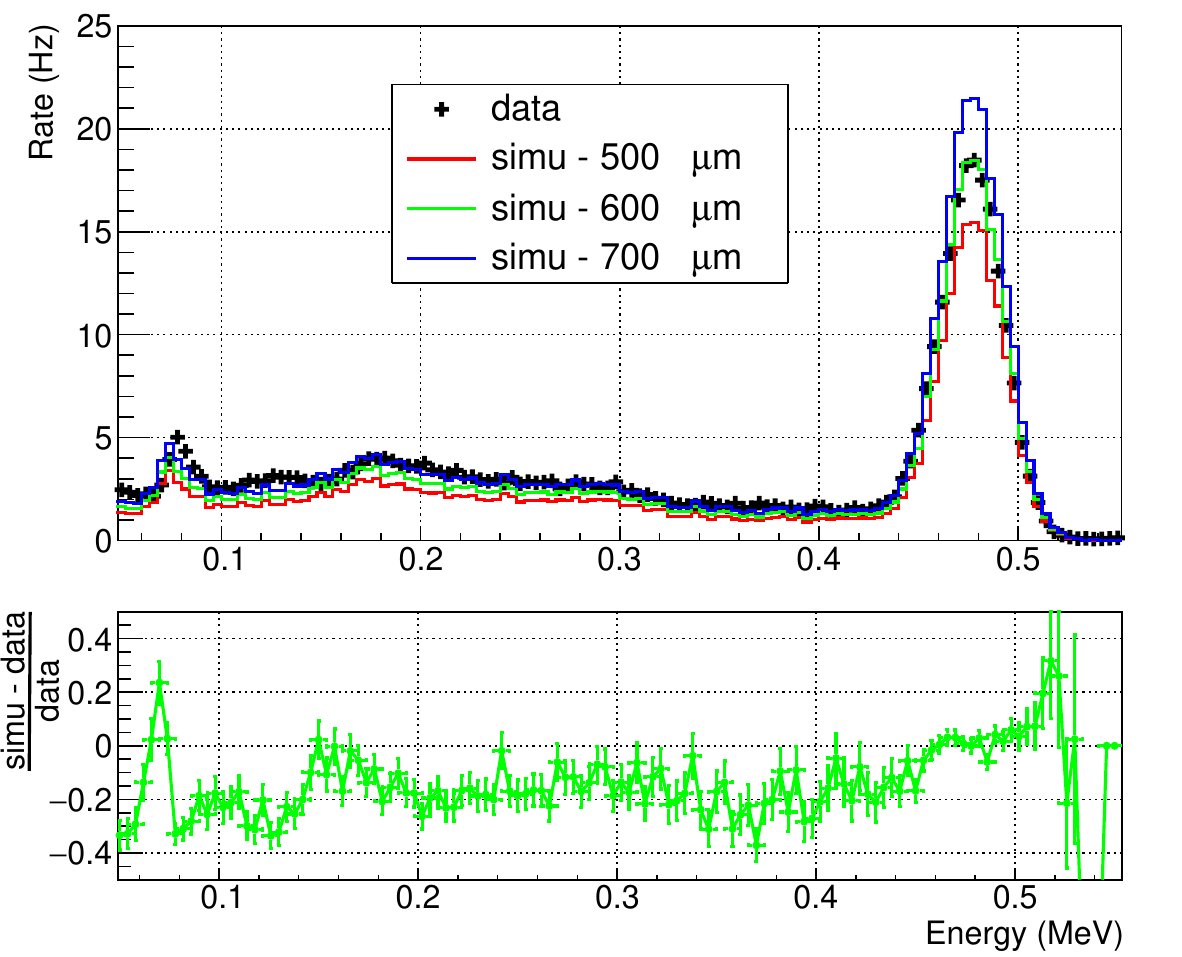}
    \caption{
   Energy spectrum of \isotope[7]{Be}, diluted homogeneously in the lithium liquid, at the central position of the vertical pipe measured with a NaI detector (black) compared to simulations with different lithium film thicknesses. The 600~$\mu$m thickness best describes the data (green).
    }
    \label{fig:BeSpectrum}
\end{figure}

The spectrum obtained at the central vertical position (Support~0) is shown in Figure~\ref{fig:BeSpectrum}; apart from a Compton continuum, it contains a single full-energy peak at \SI{478}{\kilo\electronvolt}, unambiguously assigning the measured activity to \isotope[7]{Be}.

\begin{figure}[!h]
    \centering
    \includegraphics[width=1\linewidth]{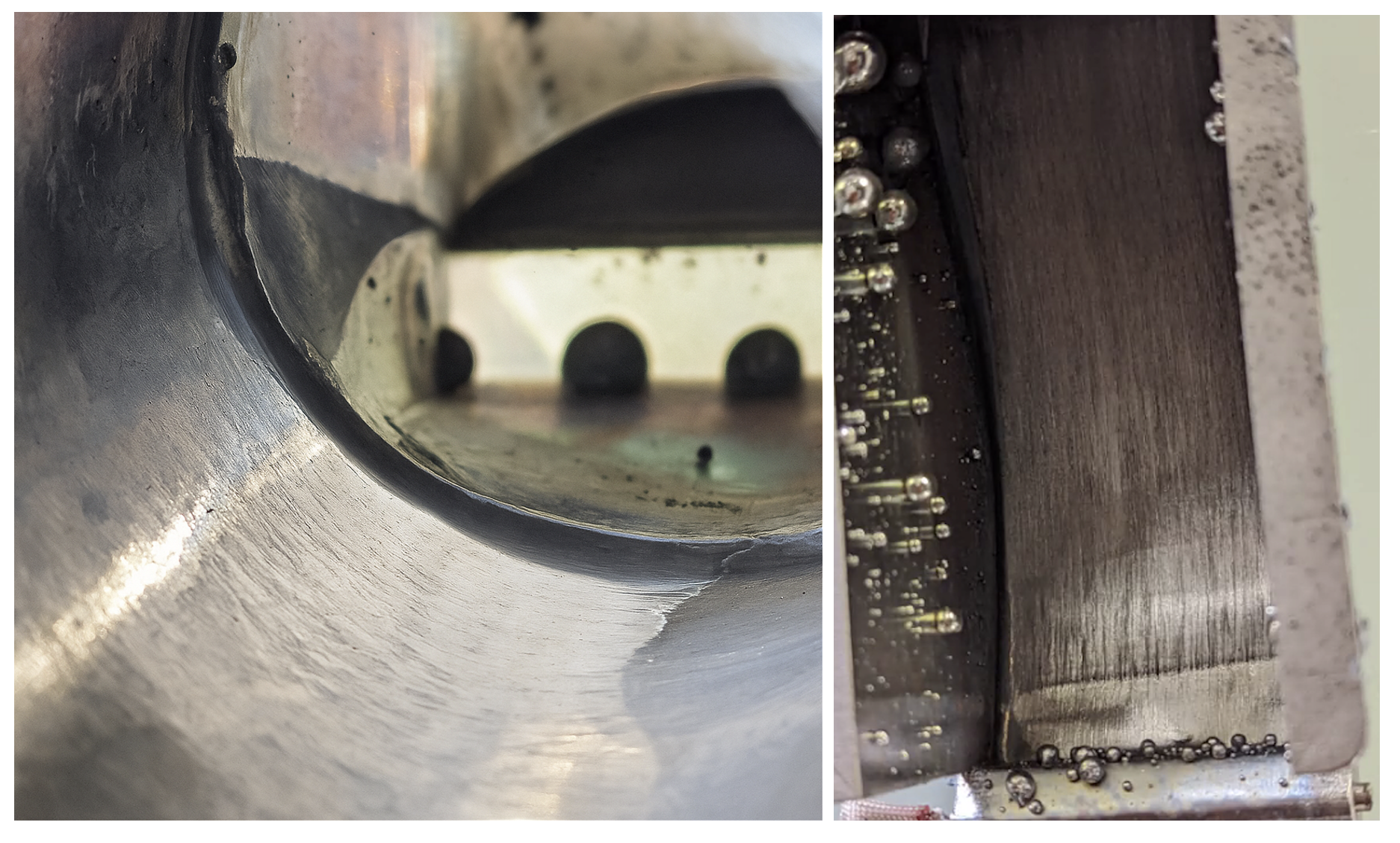}
    \caption{
    Visual evidence of lithium splash in \satelit Phase~0. Left photograph: the diffuse grey veil indicates a $\approx0.5$~mm frozen lithium film on the inner wall of the loop. Right photograph: numerous spherical lithium droplets produced by spray (aspersion) are visible on the inner wall of the nozzle. Localized bright droplets highlight spots in \satelit Phase~1 where the \isotope[7]{Be} activity exceeds the mean level, accounting for the strong dose-rate non-uniformity observed with the handheld survey meter.
    }
    \label{fig:BePhoto}
\end{figure}

A dedicated \geant model of a \SI{30}{\centi\metre} pipe segment with a 3~mm effective thickness (1~mm is added to the real 2~mm to take into account the impact of the instrumentation around the pipe), including the TMRS, the NaI detector and its shielding allowed to simulate the detector response and to estimate the film thickness which is a variable in the simulation normalization factor written as:

\begin{equation}
    \frac{P_{\text{cumul}}}{E_{\text{p}}}Q_{\text{mA}}Y_{\text{n/p}}\lambda_{\text{$^7$Be}} BR_{\text{$^7$Li$^{\ast}$}}\frac{L_{\text{pipe}}\pi \left( r^{2}_{\text{pipe-in}} - (r_{\text{pipe-in}}-x_{\text{Li}})^{2}\right)}{V_{\text{Li}}^{\text{tot}}}
\end{equation}
\noindent
with $P_{\text{cumul}} = 0.330$~kW.h as the cumulative deposited power on the target, $E_{\text{p}} = 3$~MeV as the proton energy, $Q_{\text{mA}} = 6.243 \times 10^{15}$ as the number of charges per mA, $Y_{\text{n/p}} = 2.4 \times 10^{-4}$ as the neutron yield, $\lambda_{\text{$^7$Be}} = 1.5074 \times 10^{-7}$~s$^{-1}$ as the \isotope[7][]{Be} decay constant, $BR_{\text{$^7$Li$^{\ast}$}} = 0.1044$ as the branching ratio, $L_{\text{pipe}} = 30$~cm as the length of the pipe under consideration, $r_{\text{pipe-in}} = 1.7$~mm as the inner pipe radius, $x_{\text{Li}}$ as the thickness of the lithium film, assumed to be constant, on the pipe inner wall which needs to be determined, and $V_{\text{Li}}^{\text{tot}} = \SI{22}{\liter}$ as the total volume of lithium in the system.
At the central position of the vertical pipe, there is good agreement between the data and the simulation for a 600~$\mu$m thick lithium crust on the inner pipe, as shown in Figure~\ref{fig:BeSpectrum}. In fact, the full energy peak is well described since the relative difference between the data and the simulation is close to zero, and the left part of the spectrum (Compton, back-scattering, etc.), which is sensitive to the detailed environment around the pipe, agrees well to the 20\% level.
This result is consistent with the lithium film observed optically in Figure~\ref{fig:BePhoto}. To confirm this, it would be necessary to investigate the inner part of the pipes.

\begin{figure}[!h]
    \centering
    \includegraphics[width=1\linewidth]{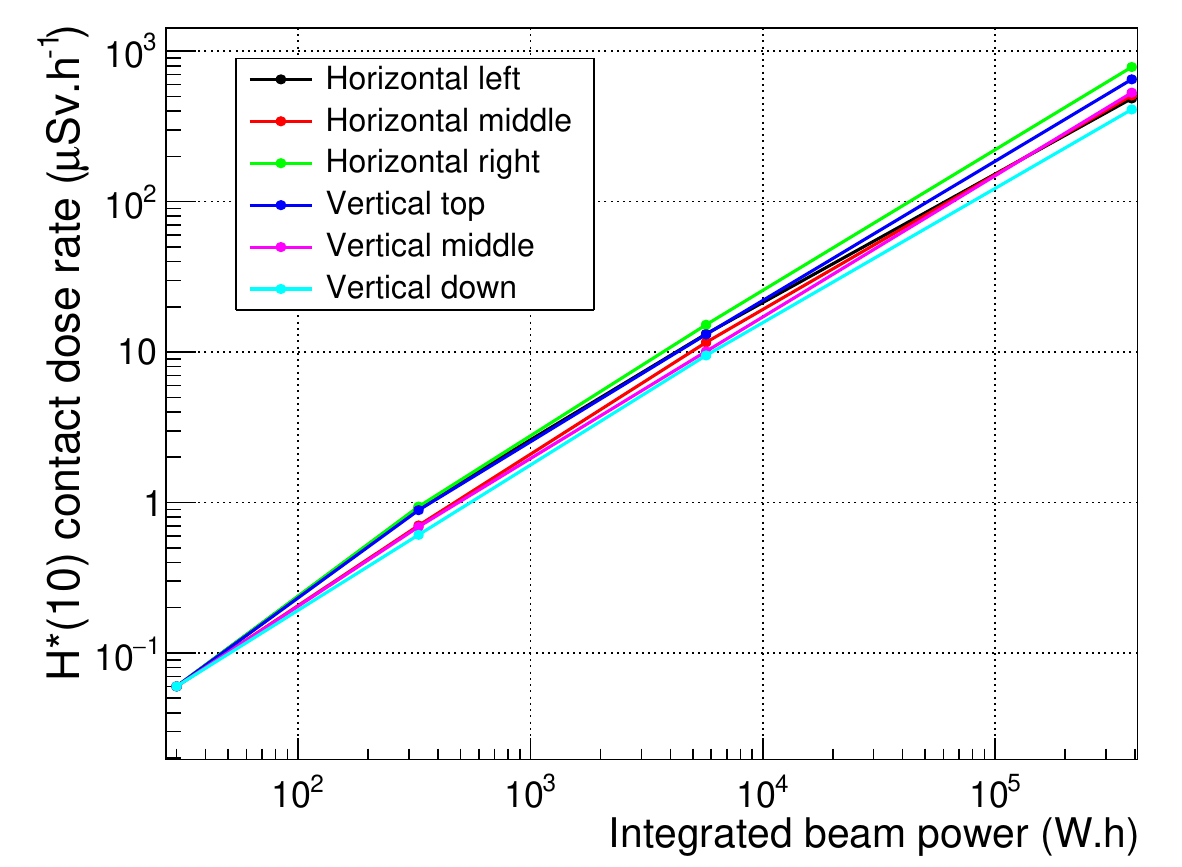}
    \caption{
    $\text{H}^{\ast}(10)$ contact dose rate as a function of integrated beam power, for the 6 different positions indicated in Figure~\ref{fig:BeResults}.
    }
    \label{fig:DoseVsPower}
\end{figure}

The second diagnostic was a 6150~AD 3''$\times$3'' plastic scintillator doped with ZnS survey meter~\cite{plastic_SPRE} operated without shielding for real-time radiological supervision.
Ambient dose-equivalent rates $\text{H}^{\ast}(10)$ were measured at the same six positions; the blue symbols in Figure~\ref{fig:BeResults} correspond to the same 330~W.h beam charge.
Although less selective, this probe enabled continuous dose monitoring during Phase~1 operation. Figure~\ref{fig:DoseVsPower} shows that the contact dose rate $\text{H}^{\ast}(10)$ increases linearly as a function of the time-integrated beam power, over more than four orders of magnitude. This linearity is expected since the test durations were relatively short in comparison to the \isotope[7]{Be} half-life.
When fitting the function:
$ \text{H}^{\ast}(10) \simeq \text{k} \, \text{P}_{\text{beam}}$
to the data, the best-fit slopes vary by a factor of two between locations:
\begin{equation}
    \text{k}\simeq(1–2)\times 10^{-3} (\mu\text{Sv}.\text{h}^{-1})/(\text{W.h}),
\end{equation}
\noindent
with the highest values observed at the Horizontal-Right station—the same hot-spot already identified by the collimated probe in Figure~\ref{fig:BeResults}.
It is evident that long-term operation beyond 100 hours could lead to radiological issues if short-term maintenance has to be performed on the loop pipes. Therefore, if HiCANS aims to leverage the target lithium loop, the \isotope[7][]{Be} buildup must be mitigated. Some ideas, which will require R\&D, include having a working cold trap in the lithium tank (currently, none has proven to be effective), remote maintenance with robotic arms, chemical distillation of \isotope[7][]{Be} from \isotope[7][]{Li} (which could be possible due to their very different fusion temperatures), or rinsing the pipes with \isotope[7][]{Be}-free lithium~\cite{Bechtold1980}.

%
\section{Experimental characterization of the neutron beam}
\label{sec:measurements}
The experimental campaign allowed to produce a neutron beam for nearly 100~hours with a 10~kW proton beam impinging on SATELIT. During the experiment, the extracted neutron beam was carefully studied.

\subsection{Flux measurements with activation foils}
\label{subsec:fluxMeasurements}

\begin{table*}
\centering
\begin{tabular}{|c|c|c|c|c|}
\hline
\multirow{2}{*}{\textbf{Gold foil activation}} & \multicolumn{2}{c|}{WITH sapphire filter} & \multicolumn{2}{c|}{WITHOUT sapphire filter} \\
\cline{2-5}
 & Center & Center+4~cm  & Center & Center+4~cm  \\
\hline

\multirow{3}{*}{\parbox{4cm}{\centering Diam. [mm] / Thick. [mm] \\ \centering Mass [mg]  \\ \centering Activity (bare) [Bq.mg$^{-1}$]}} 
 & 10.00 / 0.10 & 6.00 / 0.22 & 10.00 / 0.10 & 6.00 / 0.22 \\
 & 139.724 & 112.604 & 143.807 & 112.708 \\
 & (1.540 $\pm$0.019) 10$^{1}$ & (1.364 $\pm$0.018) 10$^{1}$ & (2.207 $\pm$0.026) 10$^{1}$ & (1.857 $\pm$0.024) 10$^{1}$  \\

\hline

\multirow{3}{*}{\parbox{4cm}{\centering Diam. [mm] / Thick. [mm] \\ \centering Mass [mg]  \\ \centering Activity (Cd)  [Bq.mg$^{-1}$]}} 
 & 10.00 / 0.10 & 6.00 / 0.22  & 10.00 / 0.10 & 6.00 / 0.22 \\
& 137.344 & 112.948 & 144.971 & 112.966 \\
& (6.574 $\pm$0.079) 10$^{-1}$ & (4.638 $\pm$0.065) 10$^{-1}$ & 4.683 $\pm$0.056 & 3.213 $\pm$0.042  \\

\hline

R$_{\text{Cd}}$=A$_{\text{bare}}$/A$_{\text{Cd}}$ & (2.343 $\pm$0.040) 10$^{1}$ & (2.941 $\pm$0.056) 10$^{1}$ & 4.713 $\pm$ 0.080 & 5.780 $\pm$0.106  \\
\hline
$\phi_{\text{thermal}}$ [n.cm$^{-2}$.s$^{-1}$.kW$^{-1}$] & (8.214 $\pm$0.269) 10$^{4}$ & (7.599 $\pm$0.251) 10$^{4}$ & (1.154 $\pm$0.039) 10$^{5}$ & (1.019 $\pm$0.035) 10$^{5}$  \\
\hline
$\phi_{\text{epithermal}}^{@4.90~\text{eV}}$ [n.cm$^{-2}$.s$^{-1}$.kW$^{-1}$]  & (9.986 $\pm$1.023) 10$^{2}$ & (9.825 $\pm$0.972) 10$^{2}$ & (8.513 $\pm$0.876) 10$^{3}$ & (5.833 $\pm$0.602) 10$^{3}$  \\
\hline
\end{tabular}
\caption{Gold foil activation measurements. The results have been obtained for an irradiation by a 10~kW proton beam. The uncertainty is taken to 1$\sigma$ confidence level.}
\label{tab:goldActivation}
\end{table*}

\begin{table*}
\centering
\begin{tabular}{|c|c|c|}
\hline
\multirow{2}{*}{\textbf{Indium foil activation}} & \multicolumn{1}{c|}{WITH sapphire filter} & \multicolumn{1}{c|}{WITHOUT sapphire filter} \\
\cline{2-3}
 & Center & Center  \\
\hline
\multirow{3}{*}{\parbox{4cm}{\centering Diam. [mm] / Thick. [mm] \\ Mass [mg]  \\ \centering Activity (bare) [Bq.mg$^{-1}$]}}  
&  14.0 / 0.1        &   14.0 / 0.1       \\
& 129.392 & 127.857 \\
& (6.206 $\pm$0.310) 10$^{2}$ & (8.797 $\pm$0.440) 10$^{2}$  \\

\hline

\multirow{2}{*}{\parbox{4cm}{\centering Mass [mg]  \\ \centering Activity (Cd) [Bq.mg$^{-1}$] }} 
& 14.0 / 0.1 & 14.0 / 0.1         \\
& 126.134  & 126.021  \\
& (2.149 $\pm$0.193) 10$^{2}$ & (1.466 $\pm$0.132) 10$^{2}$ \\

\hline

R$_{\text{Cd}}$=A$_{\text{bare}}$/A$_{\text{Cd}}$ & (2.888 $\pm$0.297) 10$^{1}$  & 6.002 $\pm$0.618  \\
\hline
$\phi_{\text{thermal}}$  [n.cm$^{-2}$.s$^{-1}$.kW$^{-1}$] & (7.832 $\pm$0.481) 10$^{4}$ & (1.244 $\pm$0.088) 10$^{5}$ \\
\hline
$\phi_{\text{epithermal}}^{@1.46~\text{eV}}$  [n.cm$^{-2}$.s$^{-1}$.kW$^{-1}$]  & (8.503 $\pm$1.703) 10$^{2}$ & (7.551 $\pm$1.535) 10$^{3}$  \\
\hline
\end{tabular}
\caption{Indium foil activation measurements. The results have been obtained for an irradiation by a 10~kW proton beam. The uncertainty is taken to 1$\sigma$ confidence level.}
\label{tab:indiumActivation}
\end{table*}

The neutron fluxes were measured with gold and indium activation foils outside of the safety vessel at its exit port, as shown in Figure~\ref{fig:CMRB1}, \textit{i.e.} 140~cm from the beginning of the neutron extraction channel. When the foils are placed under a 1~mm cadmium layer, they are only sensitive to neutrons above 512~meV (cadmium cutoff) and more precisely mainly at 1.46~eV and 4.90~eV, respectively, for \isotope[115][]{In} and \isotope[197][]{Au} due to huge resonances in their cross sections that dominate the reaction rates. Measurements with bare and cadmium-covered foils allow for the determination of the thermal neutron flux below 512~meV. After the irradiation, the activity of gold (T$_{1/2}^{\isotope[198][]{Au}}$ = 2.694~d) foils was measured to the 1\% level at the Madere platform at CEA-Cadarache using an HPGe detector. The indium foils (T$_{1/2}^{\isotope[116m][]{In}}$ = 54.29~min) were measured to the 10\% level at LNHB at CEA-Saclay.
The analysis assumes that the neutrons have an energy of 25~meV and takes into account the self-shielding correction factors related to the geometry of the detectors~\cite{international1970iaea}.
During the experiments, the irradiation was performed with a 10~kW proton beam.

\begin{figure}[!h]
    \centering
    \includegraphics[width=1\linewidth]{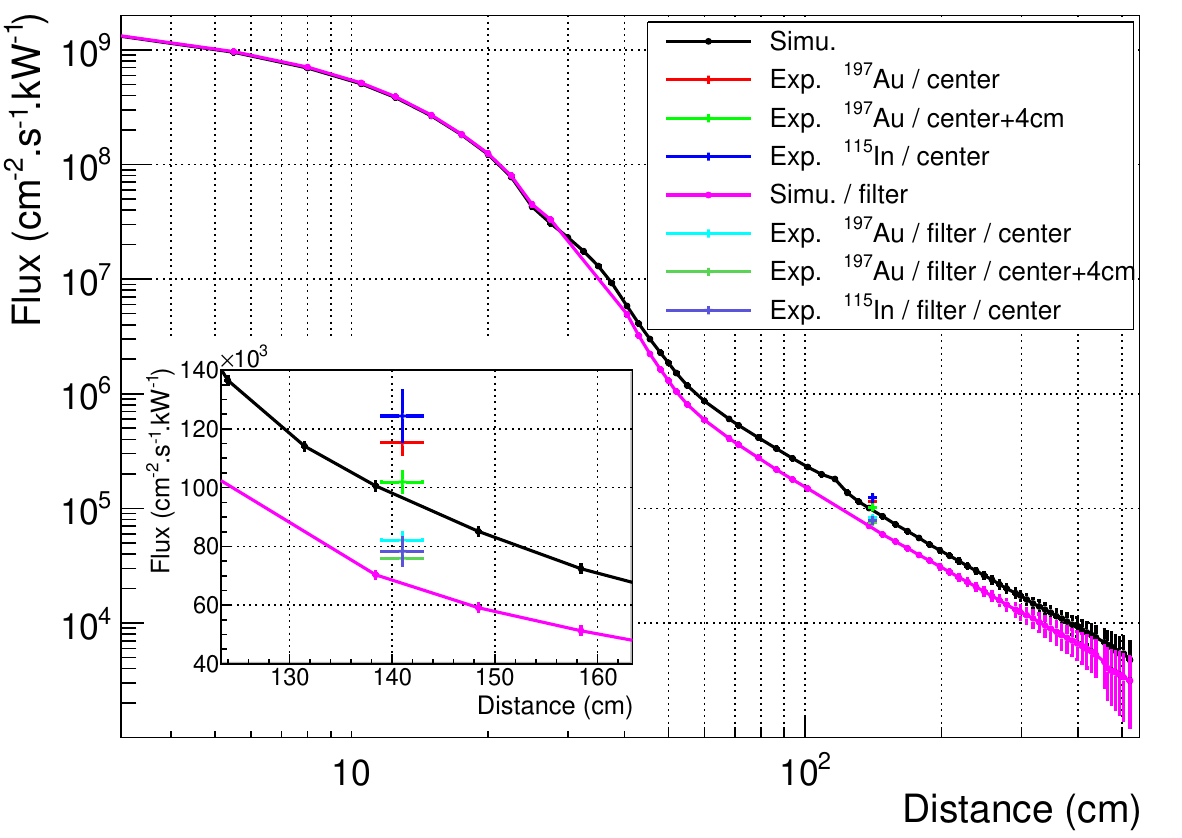}
    \caption{Thermal neutron fluxes (E$_{\text{n}}\leq$~512~meV) as a function of the distance from the start of the neutron extractor, without and with sapphire filter. Comparison between simulation and experimental results obtained from activation foil measurements. }
    \label{fig:thermalFlux_comparison}
\end{figure}

First of all, the results presented for gold and indium activation, respectively in Table~\ref{tab:goldActivation} and~\ref{tab:indiumActivation}, are consistent with each other.
The two datasets show that without the sapphire filter, for a 10~kW proton beam, the thermal neutron flux is above 10$^6$~n.cm$^{-2}$·s$^{-1}$. It should be noted that this value is sufficient for a variety of applications~\cite{ZAKALEK2025104163}. Tables~\ref{tab:goldActivation} and~\ref{tab:indiumActivation} also show that placing a sapphire filter in the neutron path reduces the thermal flux by approximately 30\% while drastically reducing the epithermal and fast neutron flux by a factor of 10. The cadmium ratio (R$_{\text{Cd}}$), often used as a criterion to characterize the neutron energy spectrum in neutron applications, is around~5 without the sapphire filter and increases up to 25 with the filter. 
Figure~\ref{fig:thermalFlux_comparison} shows that the simulation predictions are in good agreement with the experimental results. The underestimation of the predictions by 30\% could be explained by the discrepancies between the simulated (and simplified) geometry and the real geometry. In fact, around the \satelit nozzle and in the near moderator, there are many instruments (thermocouples, etc.) that allow monitoring of the behavior of the different components. Knowing, for example, that changing the distance between the nozzle and the start of the near moderator by 1~mm affects the neutron flux by 10\%, it is not surprising to have a 30\% discrepancy at the end.

\subsection{Geometric beam properties with neutron imaging}

\subsubsection{Neutron imaging tools}
The main characteristics of the beam we want to evaluate here are its spatial distribution, angular divergence, and beam purity (fast neutrons/gamma rays versus thermal neutrons).
To achieve this, we naturally used imaging tools commonly employed for qualifying neutron radiography facilities~\cite{international2008iaea,Anderson2009,Kardjilov2017}.
As imaging detectors, we used the standard radiographic film technique~\cite{CHANKOW2012}. This technique consists of a 20~$\mu$m gadolinium (Gd) layer deposited on a flat glass pane, serving as a neutron converter thanks to gadolinium's high neutron capture cross-section at thermal energies. The radiative capture reaction primarily results in the emission of conversion electrons at 29~keV~\cite{crha2019}. An X-ray sensitive radiographic film, also sensitive to the conversion electrons, is placed flat on the glass pane.
The entire assembly is housed in a vacuum casket with a thin (1.5~mm) aluminum entrance window, applying uniform pressure on the film over an active area of 35$\times$35~cm$^{2}$.
The active side of the film faces the gadolinium layer. Thus, the neutrons pass through the window, go through the film, and are captured in the gadolinium layer. The conversion electrons and X-rays emitted backward imprint the silver-halide emulsion layer of the film. We selected an AGFA STRUCTURIX\copright D3 SC (where D3 characterizes the sensitivity and SC means single layer~\cite{dutra2021,lanier2014}), known for its excellent intrinsic spatial resolution, less than 50~$\mu$m~\cite{alvarado2024}.

\begin{figure}[!h]
    \centering
    \includegraphics[width=1\linewidth]{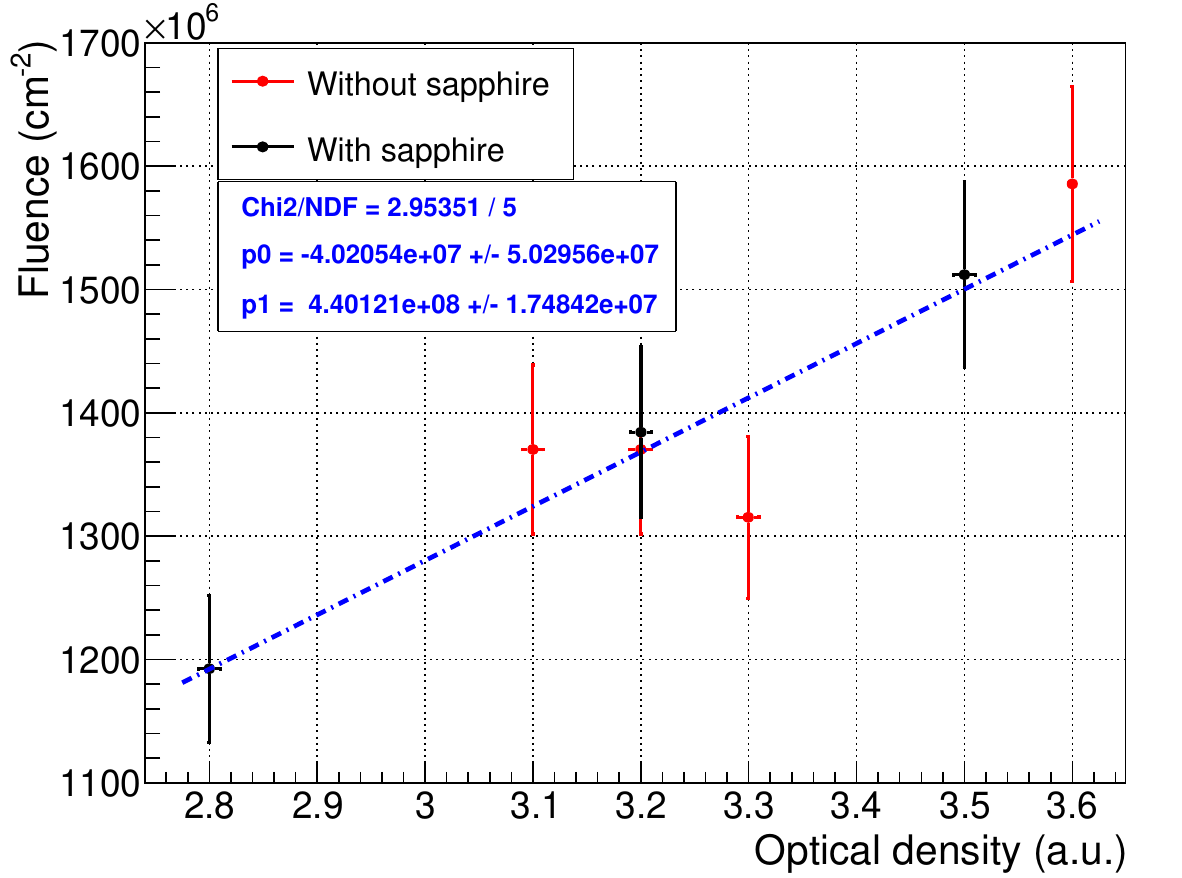}
    \caption{Neutron fluence as a function of the neutron image optical density.}
    \label{fig:fluence_vs_dose}
\end{figure}

\begin{figure}[!h]
    \centering
    \includegraphics[scale=0.49]{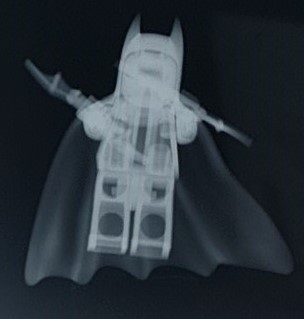}
    \caption{Neutron image of a \lego Batman figure.}
    \label{fig:batman}
\end{figure}

After the irradiation, the photographic film is developed using a STRUCTURIX NOVA developing machine from GE Inspection Technologies, with a development time set to 8~min and the temperatures of the developer and fixer chemical baths set to 28~\degree.
Once the photographic film is developed, a densitometer is used to read the optical density (OD) at a given point of the image.
The grey scale calibration of the developing machine is checked before each development to ensure the stability of its parameters using Process Monitoring Control (PMC) strips. This ensures a good linearity between the optical density of the image as a function of the exposure/neutron fluence (\textit{i.e.} the neutron flux integrated over the irradiation time), as shown in Figure~\ref{fig:fluence_vs_dose}. Typically, an exposure of 10$^8$ -- 10$^9$ n.cm$^{-2}$ is needed to print the photographic film~\cite{alvarado2024}, with an optimal optical density around 3, in the middle of its linear response regime. The exposure time of the photographic plate depends on its distance to the target, as seen in Figure~\ref{fig:thermalFlux_comparison}. For example, this setup allows for taking a picture of a hydrogenated figure, such as the one presented in Figure~\ref{fig:batman}.
The setup is placed on a lifting table on wheels, enabling the variation of the L/D parameter (L varies while D is fixed), and has been aligned beforehand with \satelit by a geometrician to ensure its good positioning within the frame of the facility when changing the position of the table.

\subsubsection{Spatial extension and divergence of the beam}

\begin{figure}[!h]
     \centering
     \includegraphics[width=1\linewidth]{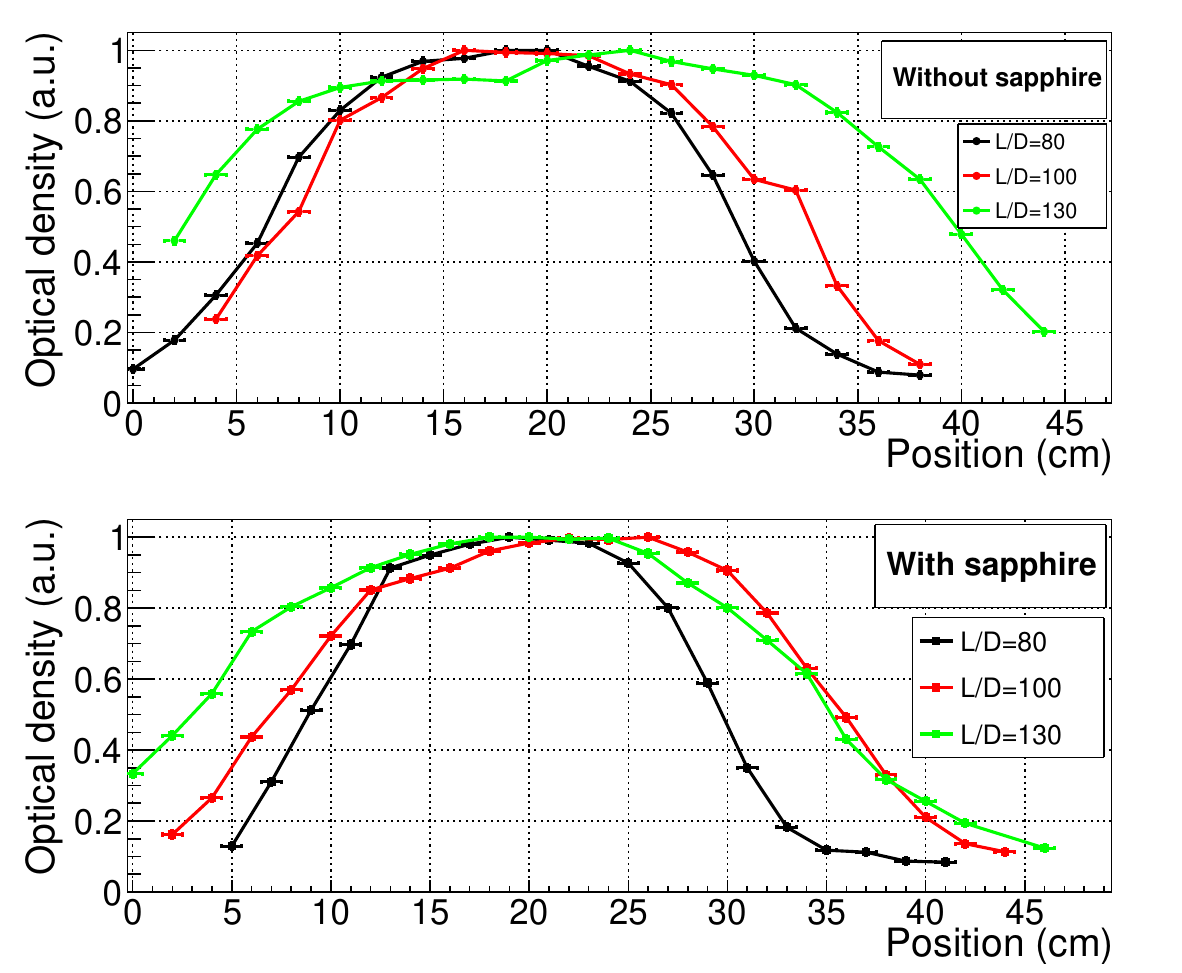}
     \caption{Spatial distribution of the neutron beam as a function of L/D, \textit{i.e.} the distance, without and with the sapphire filter. The optical density is normalized to one to not depend on the exposure time.}
     \label{fig:neutronBeam_spatial_distribution}
 \end{figure}

\begin{figure}[!h]
     \centering
     \includegraphics[width=1\linewidth]{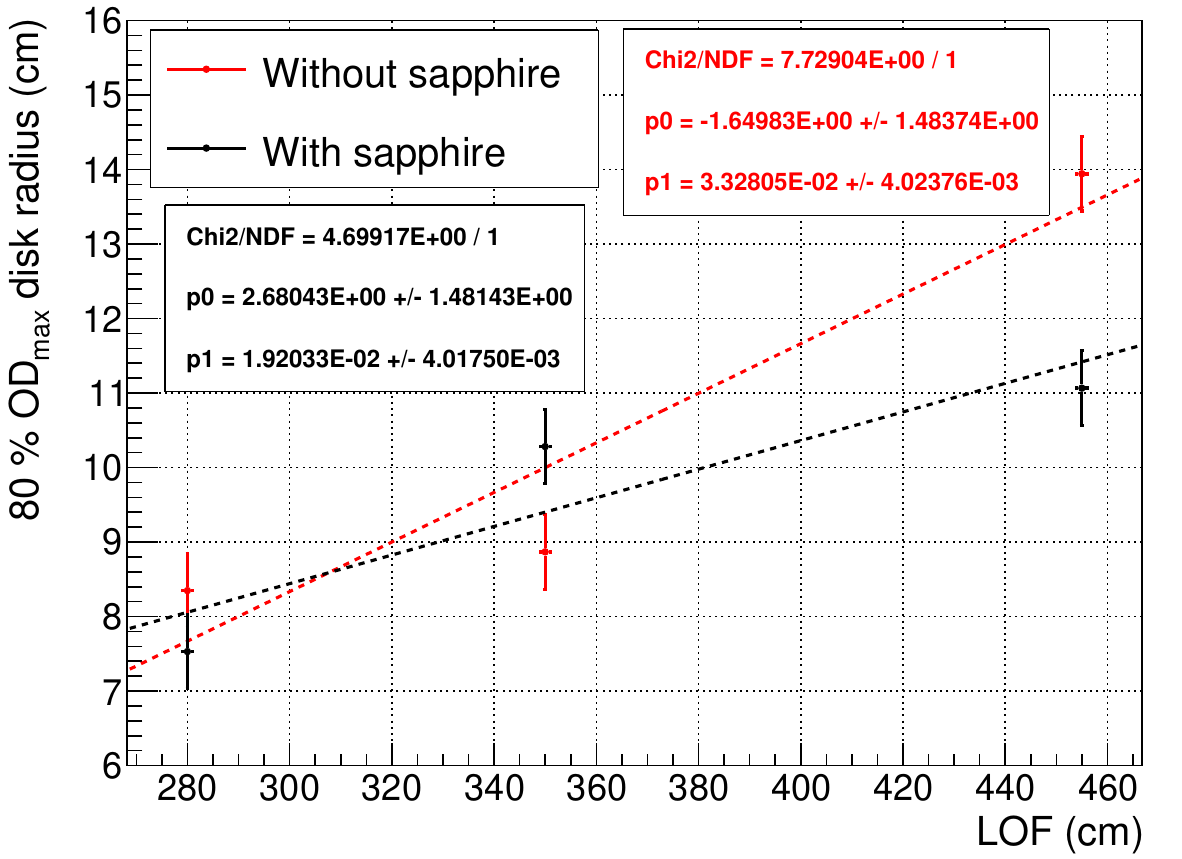}
     \caption{Radius of the disk representing the limits of 80~\% of the maximal optical density (OD$_{\text{max}}$) 80~\% as a function of the neutron length of flight (LOF) from the beginning of the extraction channel.}
     \label{fig:neutronBeam_divergence}
 \end{figure}

To determine the spatial extension of the neutron beam, the optical density is measured with the densitometer along the diagonal of the photographic film. Its shape evolution as a function of the distance is studied at three different distances: 280~cm (L/D=80), 350~cm (L/D=100), and 455~cm (L/D=130). Figure~\ref{fig:neutronBeam_spatial_distribution} shows that, as expected, the farther the detector is, the wider the neutron beam becomes, since the neutrons have a divergence of roughly 2 degrees. At L/D = 130, the spatial uniformity of the beam is better than 20\% on a disk surface with a diameter of D$_{\text{0.8}}$ $\sim$ 20 -- 30~cm and decreases to D$_{\text{0.8}}$ $\sim$ 15~cm at L/D = 80. The beam divergence can be deduced using the formula L $\times$ tan$\theta$ = D$_{\text{0.8}}$/2, where $\theta$ is the beam divergence and is obtained by a linear fit, as shown in Figure~\ref{fig:neutronBeam_divergence}. As a result, the beam divergences are 1.91 $\pm$ 0.23 degrees without and 1.10 $\pm$ 0.23 degrees with the sapphire filter in place. The difference between the two configurations could be due to the fact that the sapphire diameter does not fully fit the neutron extractor channel: gaps appear, modifying the neutron beam profile. These divergence values are close to the expected value of 2 degrees.

\subsubsection{Study of the spatial resolution}

\begin{figure}[!h]
    \centering
    \includegraphics[width=0.7\linewidth]{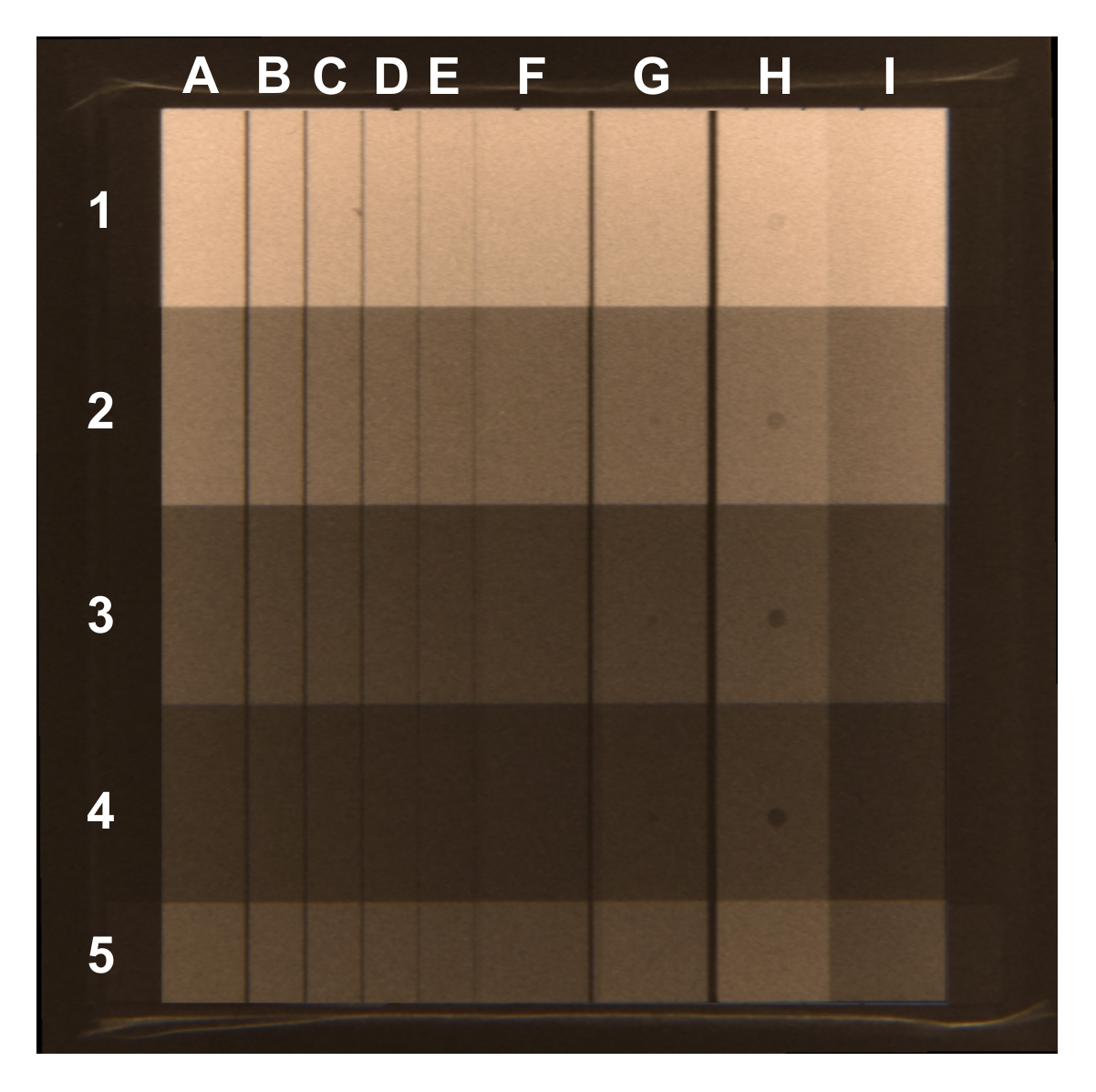}
    \caption{Neutron image of an Image Quality Indicators (IQI).}
    \label{fig:IQI_images_neutron}
\end{figure}

\begin{figure}[!h]
    \centering
    \includegraphics[width=1\linewidth]{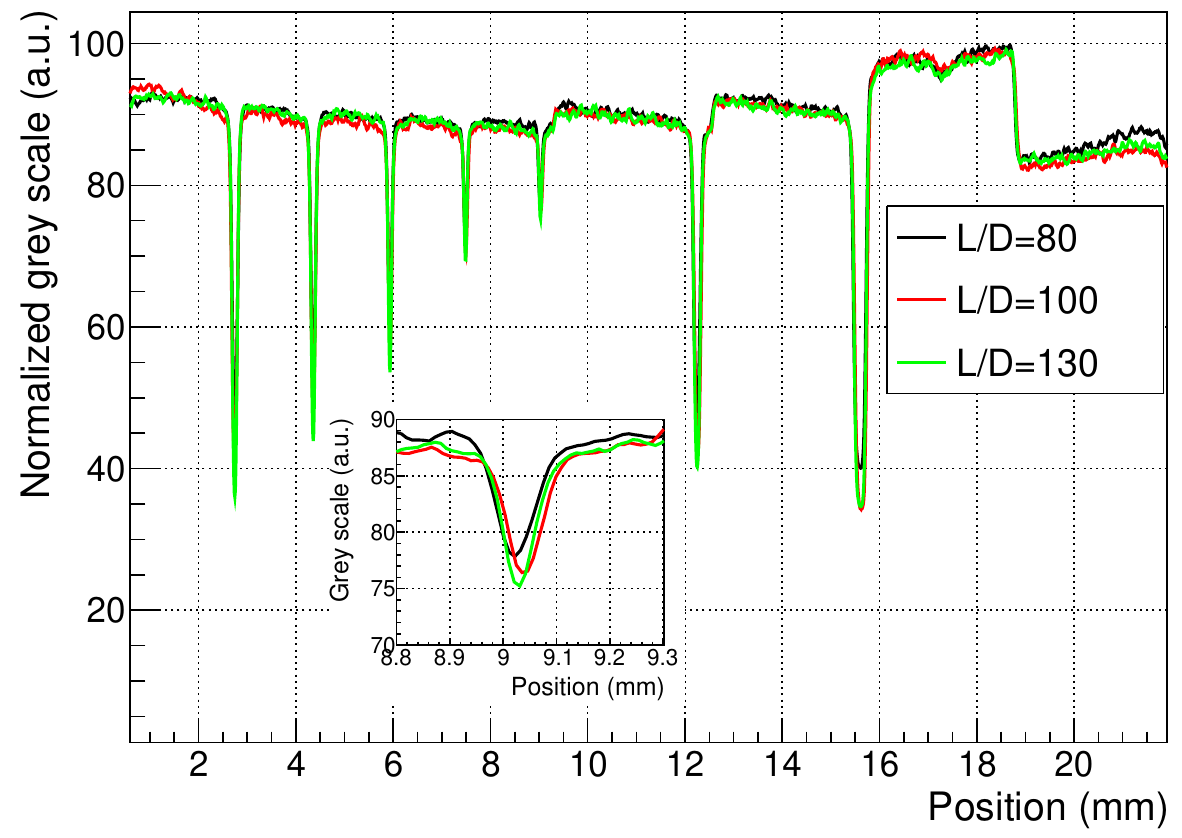}
    \caption{IQI profiles taken along row 2 for different L/D.}
    \label{fig:IQI_profile}
\end{figure}

\begin{figure}[!h]
    \centering
    \includegraphics[width=1\linewidth]{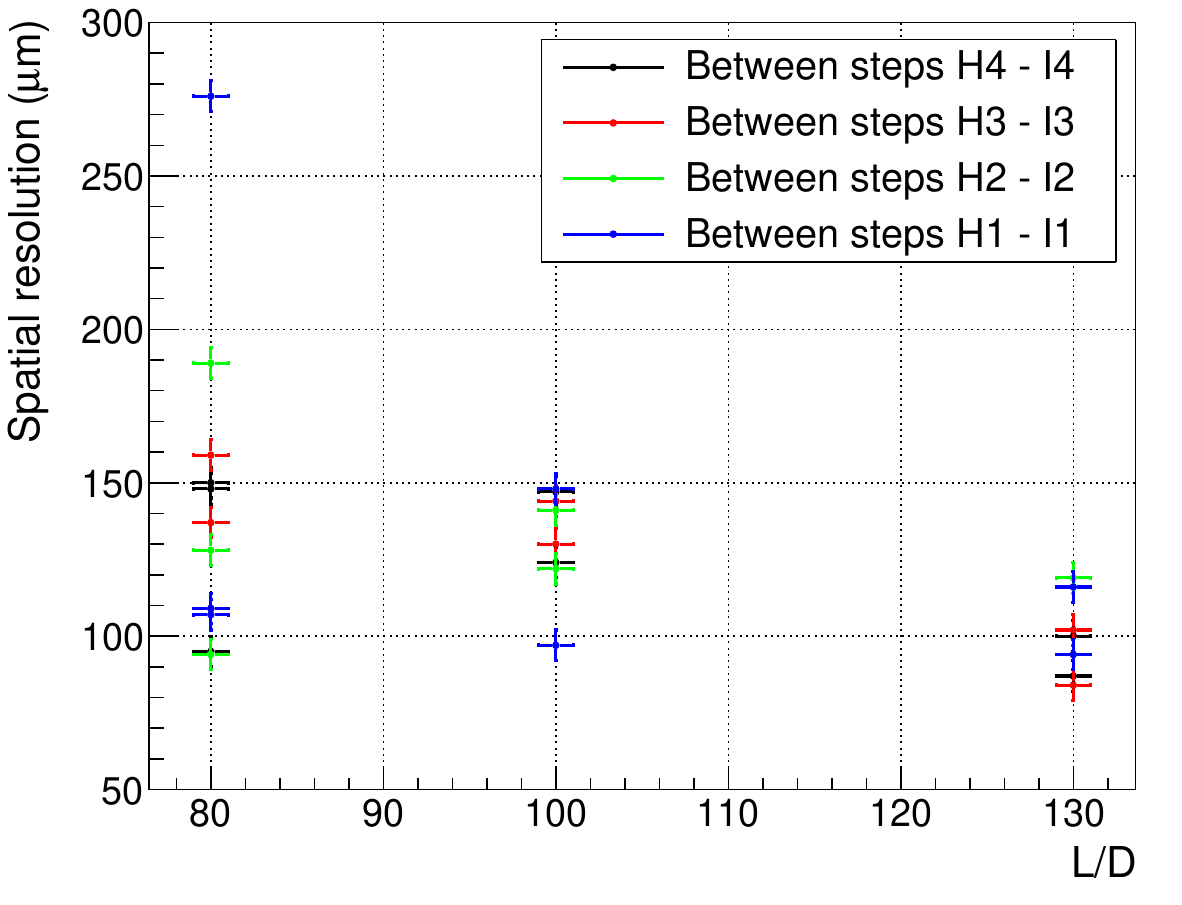}
    \caption{Spatial resolution as a function of L/D.}
    \label{fig:resolution_fct_LoverD}
\end{figure}

With this neutron beam and imaging apparatus, we also examine the spatial resolution as a by-product of the previous characterization.
The spatial resolution of an image has two contributions. The first, often neglected compared to the other, is the intrinsic detector spatial resolution, better than 50~$\mu$m with D3SC films. The second contribution is due to the beam divergence and the geometry of the detection setup, commonly referred to as the unsharpness ($U_{\text{g}}$) defined as:
\begin{equation}
    U_{\text{g}} = \frac{d_{\text{obj}\rightarrow\text{Gd-conv}}}{L/D}
    \label{eq:unsharpness}
\end{equation}

\noindent
To assess the value of this parameter, a standard Image Quality Indicator (IQI) (its characteristics are given by ASTM E-2003~\cite{ASTME2023} and E-545~\cite{ASTME545}) has been neutron-imaged.
The developed photographic film has been digitized with a Canon D5100 camera using a 250~mm objective in macro mode, allowing us to obtain the image shown in Figure~\ref{fig:IQI_images_neutron} with a pixel size of $\sim$10~$\mu$m.
Firstly, by eye, on the full-size neutron image, all the aluminum shims between polymethacrylate (PMMA) steps and seven holes are clearly visible. According to these values, this would place the experimental setup in category I for neutron radiography~\cite{ASTME545}. The ImageJ software~\cite{ImageJ} is also used to investigate how well the shims can be detected, as seen in Figure~\ref{fig:IQI_profile}.
The spatial resolution is also studied by selecting and profiling the transition between the steps in the same row (from 1 to 5) of a given thickness between two columns (here H and I) and then fitting this profile with an error function defined as:
\begin{equation}
  OD = a+b \times \text{erf}((x - c)/d)  
\end{equation}

\noindent
where erf is the error function, while $a$, $b$, $c$ and $d$ are free parameters, with 1.8$\times d$ being the searched for spatial resolution. The evolution of the spatial resolution as a function of the L/D parameter is presented in Figure~\ref{fig:resolution_fct_LoverD}.

It has to be noticed that a Beam Purity Indicator did not show any gamma contribution impacting the image.

\section{Conclusion}

A new High-Current Accelerator-driven Neutron Source (HiCANS) has been developed at CEA-Saclay, based on the 3~MeV high current proton accelerator \iphi and the lithium liquid proton-to-neutron converter \satelit. The facility was operated with great stability and reliability for nearly 100~h with a deposited beam power of 10~kW on the lithium target, for a total beam power accumulation of 840~kW.h, which is a world premiere at this proton beam energy. Moreover, on two days, the facility was continuously operated over 11~h for a total deposited power of 110~kW.h each.
The integration of an optimized neutron moderator around the target allowed for the provision of a stable thermal neutron flux of the order of 10$^{6}$~n.cm$^{-2}$·s$^{-1}$ at the exit of the \satelit safety vessel, in agreement with the simulation performed with the \toucans code based on \geant. Moreover, the beam purity was determined based on the cadmium ratio, whose values are between 5 and 25, respectively, without and with a sapphire neutron filter. The spatial extension of the beam was neutron-imaged using photographic films, which also allowed for the study of the evolution of the spatial resolution with an IQI object as a function of the L/D parameter, giving a resolution of around 100~$\mu$m at L/D=130.
The next step to use this facility on a long-term basis is to find a way to mitigate the \isotope[7][]{Be} build-up inside the \satelit pipes.
Overall, this work demonstrates that by coupling a target liquid lithium loop with a high-current accelerator, we are able to fill the gap in terms of medium neutron sources in the neutron source landscape.

\begin{acknowledgements}
The authors are grateful for the technical and administrative support of V.~Aquilina, M.~Authier, E.~Bougamont and E.~Petit.
Regarding the preparation and operation of the \iphi accelerator we especially acknowledge the support of K.~Aouij, B.~Bolzon, A.~Boufker, A.-C.~Chauveau, M.-P.~Combet, T.~Dalla-Foglia, A.~Dubois, Y.~Gauthier, T.~Hamelin, F.~Hassane, P.~Legou, M.~Oublaid, G.~Perreu, O.~Piquet and E.~Stanojevic.
For their work on radiological assessment, we particularly thank S.~Come, G.~Gigante, Q.~Guillon, K.~Hanssens, F.~Jasserand, L.~Joyeux, L.~Pages and B.~Rannou.
For their work in the validation of the thermal treatment of \armco iron seals we particularly thanks T. Alpettaz, S. Chatain, N. Lochet and J. Fortas.

\end{acknowledgements}


\bibliographystyle{unsrtnat}
\bibliography{bibliography}

\begin{thebibliography}{65}
\providecommand{\natexlab}[1]{#1}
\providecommand{\url}[1]{\texttt{#1}}
\expandafter\ifx\csname urlstyle\endcsname\relax
  \providecommand{\doi}[1]{doi: #1}\else
  \providecommand{\doi}{doi: \begingroup \urlstyle{rm}\Url}\fi

\bibitem[Zakalek et~al.(2025)Zakalek, Gutberlet, and Brückel]{ZAKALEK2025104163}
P.~Zakalek, T.~Gutberlet, and Th. Brückel.
\newblock {Neutron sources for large scale user facilities: The potential of high current accelerator-driven neutron sources}.
\newblock \emph{Prog. Part. Nucl. Phys.}, 142:\penalty0 104163, 2025.
\newblock \doi{https://doi.org/10.1016/j.ppnp.2025.104163}.
\newblock URL \url{https://www.sciencedirect.com/science/article/pii/S0146641025000109}.

\bibitem[ELENA Association()]{ELENA}
ELENA Association.
\newblock URL \url{www.ELENA-neutron.eu}.

\bibitem[HBS()]{HBS}
{High Brilliance neutron Source}.
\newblock URL \url{https://www.fz-juelich.de/en/jcns/jcns-2/expertise/high-brilliance-neutron-source/conceptual-and-technical-design}.

\bibitem[Baggemann et~al.(2024)Baggemann, Gutberlet, Zakalek, Wolters, Rücker, Mauerhofer, Li, Ding, Loewenhoff, Dorow-Gerspach, Bessler, and Brückel]{BAGGEMANN2024169912}
J.~Baggemann, T.~Gutberlet, P.~Zakalek, J.~Wolters, U.~Rücker, E.~Mauerhofer, J.~Li, Q.~Ding, Th. Loewenhoff, D.~Dorow-Gerspach, Y.~Bessler, and Th. Brückel.
\newblock High power target for the high brilliance neutron source.
\newblock \emph{Nucl. Instrum. Methods A}, 1069:\penalty0 169912, 2024.
\newblock \doi{https://doi.org/10.1016/j.nima.2024.169912}.
\newblock URL \url{https://www.sciencedirect.com/science/article/pii/S0168900224008386}.

\bibitem[Pérez et~al.(2020)Pérez, Sordo, Bustinduy, Muñoz, and Villacorta]{Perez01102020}
M.~Pérez, F.~Sordo, I.~Bustinduy, J.L. Muñoz, and F.J. Villacorta.
\newblock {ARGITU compact accelerator neutron source: A unique infrastructure fostering R\&D ecosystem in Euskadi}.
\newblock \emph{Neutron News}, 31\penalty0 (2-4):\penalty0 19--25, 2020.
\newblock \doi{10.1080/10448632.2020.1819140}.
\newblock URL \url{https://doi.org/10.1080/10448632.2020.1819140}.

\bibitem[ICO(2023)]{ICONE2023}
{ICONE White Book - Une nouvelle source de diffusion neutronique française}, 2023.
\newblock URL \url{https://2fdn.cnrs.fr/wp-content/uploads/2023/09/ICONE-digital.pdf}.

\bibitem[ICO()]{ICONE_website}
{ICONE website}.
\newblock URL \url{https://www.icone-neutron.fr}.

\bibitem[Tran et~al.(2018)Tran, Marchix, Letourneau, Chauvin, Menelle, Ott, and Schwindling]{Tran2018}
H.N. Tran, A.~Marchix, A.~Letourneau, N.~Chauvin, A.~Menelle, F.~Ott, and J.~Schwindling.
\newblock {Validation of Geant4 simulation tool for low energy proton induced reactions}.
\newblock \emph{J. Phys.: Conf. Ser.}, 1021:\penalty0 012008, may 2018.
\newblock \doi{10.1088/1742-6596/1021/1/012008}.
\newblock URL \url{https://doi.org/10.1088/1742-6596/1021/1/012008}.

\bibitem[Tran et~al.(2020)Tran, Ott, Darpentigny, Marchix, Letourneau, Chauvin, Prunes, Homatter, Annigh\"ofer, Menelle, and Schwindling]{Tran2020}
H.N. Tran, F.~Ott, J.~Darpentigny, A.~Marchix, A.~Letourneau, N.~Chauvin, F.~Prunes, B.~Homatter, B.~Annigh\"ofer, A.~Menelle, and J.~Schwindling.
\newblock {Neutrons production on the IPHI accelerator for the validation of the design of the compact neutron source SONATE}.
\newblock \emph{EPJ Web Conf.}, 231:\penalty0 01007, 2020.
\newblock \doi{10.1051/epjconf/202023101007}.
\newblock URL \url{https://doi.org/10.1051/epjconf/202023101007}.

\bibitem[Thulliez et~al.(2020)Thulliez, Letourneau, Schwindling, Chauvin, Sellami, Ott, Menelle, and Annigh\"ofer]{Thulliez2020}
L.~Thulliez, A.~Letourneau, J.~Schwindling, N.~Chauvin, N.~Sellami, F.~Ott, A.~Menelle, and B.~Annigh\"ofer.
\newblock {First steps toward the development of SONATE, a Compact Accelerator driven Neutron Source}.
\newblock \emph{EPJ Web Conf.}, 239:\penalty0 17011, 2020.
\newblock \doi{10.1051/epjconf/202023917011}.
\newblock URL \url{https://doi.org/10.1051/epjconf/202023917011}.

\bibitem[Schwindling et~al.(2022)Schwindling, Annighöfer, Chauvin, Meuriot, Mom, Ott, Sellami, and Thulliez]{Schwindling2022}
J.~Schwindling, B.~Annighöfer, N.~Chauvin, J.-L. Meuriot, B.~Mom, F.~Ott, N.~Sellami, and L.~Thulliez.
\newblock {Long term operation of a 30 kW Beryllium target at IPHI}.
\newblock \emph{{J. Neutron Res.}}, 24\penalty0 (3-4):\penalty0 289--298, 2022.
\newblock \doi{10.3233/JNR-220024}.
\newblock URL \url{https://journals.sagepub.com/doi/abs/10.3233/JNR-220024}.

\bibitem[Darpentigny and Ott(2022)]{Darpentigny2022}
J.~Darpentigny and F.~Ott.
\newblock {Neutron scattering on DIoGENE at IPHI–neutrons}.
\newblock \emph{{J. Neutron Res.}}, 24\penalty0 (3-4):\penalty0 385--393, 2022.
\newblock \doi{10.3233/JNR-220018}.

\bibitem[Porges and al.(1971)]{Porges1970}
K.~Porges and al.
\newblock Applied physics division annual report (july 1, 1970, to june 30, 1971).
\newblock Technical report, Argonne National Laboratory (ANL), 12 1971.
\newblock URL \url{https://www.osti.gov/biblio/4592825}.
\newblock {Section III-14: Thick Target Neutron Yields of Lithium and Beryllium Taregets Bombarded with Protons and Deuterons, p.361}.

\bibitem[Paul et~al.(2015)]{Paul2015}
M.~Paul et~al.
\newblock {A high-power liquid-lithium target (LiLiT) for neutron production}.
\newblock \emph{J. Radioanal. Nucl. Chem.}, 305:\penalty0 783–786, 2015.
\newblock \doi{10.3233/JNR-220024}.
\newblock URL \url{https://doi.org/10.1007/s10967-015-4027-3}.

\bibitem[{Paul, M.} et~al.(2020){Paul, M.}, {Silverman, I.}, {Halfon, S.}, {Sukoriansky, S.}, {Mikhailovich, B.}, {Palchan, T.}, {Kapusta, A.}, {Shoihet, A.}, {Kijel, D.}, {Arenshtam, A.}, and {Barami, E.}]{Silverman2020}
{Paul, M.}, {Silverman, I.}, {Halfon, S.}, {Sukoriansky, S.}, {Mikhailovich, B.}, {Palchan, T.}, {Kapusta, A.}, {Shoihet, A.}, {Kijel, D.}, {Arenshtam, A.}, and {Barami, E.}
\newblock {A 50 kW Liquid-Lithium Target for BNCT and Material-Science Applications}.
\newblock \emph{EPJ Web Conf.}, 231:\penalty0 03004, 2020.
\newblock \doi{10.1051/epjconf/202023103004}.
\newblock URL \url{https://doi.org/10.1051/epjconf/202023103004}.

\bibitem[S{\'{e}}n{\'{e}}e et~al.(2018)S{\'{e}}n{\'{e}}e, Chauvin, Chel, Gobin, Knaster, Okumura, Shinto, and Valette]{Senee2018}
F.~S{\'{e}}n{\'{e}}e, N.~Chauvin, S.~Chel, R.~Gobin, J.~Knaster, Y.~Okumura, K.~Shinto, and M.~Valette.
\newblock {Increase of IPHI Beam power at CEA Saclay}.
\newblock In \emph{Proc. of International Particle Accelerator Conference (IPAC'18)}, Vancouver, BC, Canada, 2018.
\newblock \doi{10.18429/JACoW-IPAC2018-TUPAF016}.

\bibitem[Gobin et~al.(2002)Gobin, Beauvais, Bogard, Charruau, Delferri\`ere, Menezes, France, Ferdinand, Gauthier, Harrault, Jannin, Lagniel, Leroy, Matt\'ei, Sherman, Sinanna, Ausset, Bousson, and Pottin]{Gobin_RSI2002}
R.~Gobin, P.-Y. Beauvais, D.~Bogard, G.~Charruau, O.~Delferri\`ere, D.~De Menezes, A.~France, R.~Ferdinand, Y.~Gauthier, F.~Harrault, J.-L. Jannin, J.-M. Lagniel, P.-A. Leroy, P.~Matt\'ei, J.~Sherman, A.~Sinanna, P.~Ausset, S.~Bousson, and B.~Pottin.
\newblock {High intensity ECR ion source (H$^+$, D$^+$, H$^-$) developments at CEA/Saclay}.
\newblock \emph{Rev. Sci. Instrum.}, 73\penalty0 (2):\penalty0 922--924, 2002.
\newblock \doi{10.1063/1.1428783}.

\bibitem[Piquet(2016)]{Piquet_IPAC2016}
O.~Piquet.
\newblock {RFQ} {D}evelopments at {CEA-IRFU}.
\newblock In \emph{Proc. of International Particle Accelerator Conference (IPAC'16), Busan, Korea}, 2016.
\newblock \doi{10.18429/JACoW-IPAC2016-MOOCA02}.
\newblock URL \url{http://jacow.org/ipac2016/papers/mooca02.pdf}.

\bibitem[Uriot and Pichoff(2015)]{Uriot_IPAC2015}
D.~Uriot and N.~Pichoff.
\newblock {S}tatus of {T}race{W}in {C}ode.
\newblock In \emph{Proc. 6th International Particle Accelerator Conference (IPAC'15), Richmond, VA, USA}, 2015.
\newblock \doi{10.18429/JACoW-IPAC2015-MOPWA008}.
\newblock URL \url{http://jacow.org/ipac2015/papers/mopwa008.pdf}.
\newblock https://www.dacm-logiciels.fr/tracewin.

\bibitem[Kondo et~al.(2016)Kondo, Kanemura, Furukawa, Hirakawa, Wakai, and Knaster]{KONDO2016}
H.~Kondo, T.~Kanemura, T.~Furukawa, Y.~Hirakawa, E.~Wakai, and J.~Knaster.
\newblock {Demonstration of Li target facility in IFMIF/EVEDA project: Li target stability in continuous operation of entire system}.
\newblock \emph{Fusion Eng. Des.}, 109-111:\penalty0 1759--1763, 2016.
\newblock \doi{https://doi.org/10.1016/j.fusengdes.2015.09.016}.

\bibitem[Furukawa et~al.(2014)Furukawa, Hirakawa, Kato, Iijima, Ohtaka, Kondo, Kanemura, and Wakai]{FURUKAWA2014}
T.~Furukawa, Y.~Hirakawa, S.~Kato, M.~Iijima, M.~Ohtaka, H.~Kondo, T.~Kanemura, and E.~Wakai.
\newblock {Current status of the technology development on lithium safety handling under IFMIF/EVEDA}.
\newblock \emph{Fusion Eng. Des.}, 89\penalty0 (12):\penalty0 2902--2909, 2014.
\newblock \doi{https://doi.org/10.1016/j.fusengdes.2014.07.010}.

\bibitem[Borgstedt and Mathews(1987)]{borgstedt1987}
H.U. Borgstedt and C.K. Mathews.
\newblock \emph{Applied Chemistry of the Alkali Metals}.
\newblock Springer US, 1987.
\newblock ISBN 9780306423260.

\bibitem[Meddeb et~al.(2023)Meddeb, Courouau, Rouhard, Cavaliere, and Giorgi]{MEDDEB2023}
S.~Meddeb, J.-L. Courouau, M.~Rouhard, N.~Cavaliere, and M.-L. Giorgi.
\newblock Kinetics of the complexation reaction of iron with dissolved oxygen in liquid sodium.
\newblock \emph{J. Nucl. Mater.}, 584:\penalty0 154541, 2023.
\newblock \doi{https://doi.org/10.1016/j.jnucmat.2023.154541}.

\bibitem[Chopra and Smith(1986)]{Chopra1986}
O.K. Chopra and D.L. Smith.
\newblock Influence of temperature and lithium purity on corrosion of ferrous alloys in a flowing lithium environment.
\newblock \emph{J. Nucl. Mater.}, 141-143:\penalty0 584--591, 1986.
\newblock \doi{https://doi.org/10.1016/0022-3115(86)90058-9}.

\bibitem[Brēķis et~al.(2023)Brēķis, Buligins, Bucenieks, Goldšteins, Kravalis, Lācis, Mikanovskis, and Jēkabsons]{BREKIS2023113919}
A.~Brēķis, L.~Buligins, I.~Bucenieks, L.~Goldšteins, K.~Kravalis, A.~Lācis, O.~Mikanovskis, and N.~Jēkabsons.
\newblock {Electromagnetic pump with rotating permanent magnets operation at low inlet pressures}.
\newblock \emph{Fusion Engineering and Design}, 194:\penalty0 113919, 2023.
\newblock ISSN 0920-3796.
\newblock \doi{https://doi.org/10.1016/j.fusengdes.2023.113919}.
\newblock URL \url{https://www.sciencedirect.com/science/article/pii/S092037962300501X}.

\bibitem[Kravalis et~al.(2022)Kravalis, Boix, Bucenieks, Buligins, Delonca, Goldšteins, and Stora]{Kravalis2022}
K.~Kravalis, F.~Boix, I.~Bucenieks, L.~Buligins, M.~Delonca, L.~Goldšteins, and T.~Stora.
\newblock {Experimental cavitation investigation of the electromagnetic PbBi pump with rotating permanent magnets}.
\newblock \emph{Magnetohydrodynamics}, 58:\penalty0 195--203, 02 2022.
\newblock \doi{10.22364/mhd.58.1-2.21}.

\bibitem[Bucenieks(2000)]{bucenieks2000perspectives}
I.~Bucenieks.
\newblock {Perspectives of using rotating permanent magnets for electromagnetic induction pump design}.
\newblock \emph{Magnetohydrodynamics}, 36\penalty0 (2):\penalty0 151--156, 2000.

\bibitem[Bucenieks and Kravalis(2011)]{bucenieks2011efficiency}
I.~Bucenieks and K.~Kravalis.
\newblock {Efficiency of EM induction pumps with permanent magnets}.
\newblock \emph{Magnetohydrodynamics}, 47\penalty0 (1):\penalty0 89--96, 2011.

\bibitem[Ziegler et~al.(2010)Ziegler, Ziegler, and Biersack]{ZIEGLER2010}
J.F. Ziegler, M.D. Ziegler, and J.P. Biersack.
\newblock {SRIM – The stopping and range of ions in matter (2010)}.
\newblock \emph{Nucl. Instrum. Methods B}, 268\penalty0 (11):\penalty0 1818--1823, 2010.
\newblock \doi{https://doi.org/10.1016/j.nimb.2010.02.091}.
\newblock URL \url{https://www.sciencedirect.com/science/article/pii/S0168583X10001862}.
\newblock 19th International Conference on Ion Beam Analysis.

\bibitem[Ritchie(1976)]{Ritchie_1976}
A.I.M. Ritchie.
\newblock {Neutron yields and energy spectra from the thick target Li(p,n) source}.
\newblock \emph{J. Phys. D: Appl. Phys.}, 9\penalty0 (1):\penalty0 15, jan 1976.
\newblock \doi{10.1088/0022-3727/9/1/008}.
\newblock URL \url{https://dx.doi.org/10.1088/0022-3727/9/1/008}.

\bibitem[Liskien and Paulsen(1975)]{LISKIEN197557}
H.~Liskien and A.~Paulsen.
\newblock {Neutron production cross sections and energies for the reactions 7Li(p,n)7Be and 7Li(p,n)7Be$^\ast$}.
\newblock \emph{Atomic Data and Nuclear Data Tables}, 15\penalty0 (1):\penalty0 57--84, 1975.
\newblock \doi{https://doi.org/10.1016/0092-640X(75)90004-2}.
\newblock URL \url{https://www.sciencedirect.com/science/article/pii/0092640X75900042}.

\bibitem[Lee and Zhou(1999)]{LEE19991}
C.L. Lee and X.-L. Zhou.
\newblock {Thick target neutron yields for the 7Li(p,n)7Be reaction near threshold}.
\newblock \emph{Nucl. Instrum. Methods B}, 152\penalty0 (1):\penalty0 1--11, 1999.
\newblock \doi{https://doi.org/10.1016/S0168-583X(99)00026-9}.
\newblock URL \url{https://www.sciencedirect.com/science/article/pii/S0168583X99000269}.

\bibitem[Herrera et~al.(2014)Herrera, Moreno, and Kreiner]{HERRERA2014}
M.S. Herrera, G.A. Moreno, and A.J. Kreiner.
\newblock {Revisiting the 7Li(p,n)7Be reaction near threshold}.
\newblock \emph{Appl. Radiat. Isotopes}, 88:\penalty0 243--246, 2014.
\newblock \doi{https://doi.org/10.1016/j.apradiso.2013.11.042}.
\newblock URL \url{https://www.sciencedirect.com/science/article/pii/S0969804313004880}.
\newblock 15th International Congress on Neutron Capture Therapy Impact of a new radiotherapy against cancer.

\bibitem[Kononov et~al.(1977)Kononov, Poletaev, and Yurlov]{Kononov1977}
V.N. Kononov, E.D. Poletaev, and B.D. Yurlov.
\newblock {Absolute yield and spectrum of neutrons from the reaction 7Li(p,n)7Be}.
\newblock \emph{At. {\'E}nerg}, 43\penalty0 (4):\penalty0 303--305, 1977.

\bibitem[{Bachiller Perea} et~al.(2017){Bachiller Perea}, Corvisiero, {Jiménez Rey}, Joco, {Maira Vidal}, {Muñoz Martin}, and Zucchiatti]{BACHILLERPEREA2017}
D.~{Bachiller Perea}, P.~Corvisiero, D.~{Jiménez Rey}, V.~Joco, A.~{Maira Vidal}, A.~{Muñoz Martin}, and A.~Zucchiatti.
\newblock {Measurement of gamma-ray production cross sections in Li and F induced by protons from 810 to 3700keV}.
\newblock \emph{{Nucl. Instrum. Methods B}}, 406:\penalty0 161--166, 2017.
\newblock \doi{https://doi.org/10.1016/j.nimb.2017.02.017}.
\newblock URL \url{https://www.sciencedirect.com/science/article/pii/S0168583X17301283}.
\newblock Proceedings of the 12th European Conference on Accelerators in Applied Research and Technology (ECAART12).

\bibitem[Zahnow et~al.(1995)Zahnow, Angulo, Rolfs, Schmidt, Schulte, and Somorjai]{Zahnow1995}
D.~Zahnow, C.~Angulo, C.~Rolfs, S.~Schmidt, W.H. Schulte, and E.~Somorjai.
\newblock {The S(E) factor of $^7$Li(p, $\gamma$)$^8$Be and consequences for S(E) extrapolation in $^7$Be(p, $\gamma_0$)8B}.
\newblock \emph{Zeitschrift für Physik A Hadrons and Nuclei}, 351:\penalty0 229–236, 1995.
\newblock \doi{10.1007/BF01289534}.
\newblock URL \url{https://doi.org/10.1007/BF01289534}.

\bibitem[et~al.(2018)]{Werner2018}
C.J.~Werner et~al.
\newblock {MCNP Version 6.2 Release Notes}.
\newblock Technical report, Los Alamos National Laboratory (LANL), 2 2018.

\bibitem[et~al.(2015)]{Brun2015}
E.~Brun et~al.
\newblock {TRIPOLI-4®, CEA, EDF and AREVA reference Monte Carlo code}.
\newblock \emph{Ann. Nucl. Energy}, 82:\penalty0 151--160, 8 2015.
\newblock \doi{10.1016/j.anucene.2014.07.053}.

\bibitem[et~al.(2014)]{Mendoza2014}
E.~Mendoza et~al.
\newblock {New standard evaluated neutron cross section libraries for the GEANT4 code and first verification}.
\newblock \emph{IEEE Trans. Nucl. Sci.}, 61:\penalty0 2357--2364, 2014.
\newblock \doi{10.1109/TNS.2014.2335538}.

\bibitem[Mendoza et~al.(2018)Mendoza, Cano-Ott, and Capote]{Mendoza2018}
E.~Mendoza, D.~Cano-Ott, and R.~Capote.
\newblock {Update of the Evaluated Neutron Cross Section Libraries for the Geant4 Code, IAEA technical report INDC(NDS)-0758 (releases JEFF-3.3, JEFF-3.2, ENDF/B-VIII.0, ENDF/B-VII.1, BROND-3.1 and JENDL-4.0u)}.
\newblock Technical report, CIEMAT, Vienna, 2018.
\newblock URL \url{https://www-nds.iaea.org/geant4/figures/G4\_10.04.p01\_VS\_MCNP6\_ENDF80.pdf}.

\bibitem[Thulliez et~al.(2022)Thulliez, Jouanne, and Dumonteil]{Thulliez2022}
L.~Thulliez, C.~Jouanne, and E.~Dumonteil.
\newblock {Improvement of Geant4 Neutron-HP package: From methodology to evaluated nuclear data library}.
\newblock \emph{Nucl. Instrum. Methods A}, 1027:\penalty0 166187, 3 2022.
\newblock \doi{10.1016/j.nima.2021.166187}.

\bibitem[Zmeškal et~al.(2023)Zmeškal, Thulliez, and Dumonteil]{Zmeskal2023}
M.~Zmeškal, L.~Thulliez, and E.~Dumonteil.
\newblock {Improvement of Geant4 Neutron-HP package: Doppler broadening of the neutron elastic scattering kernel}.
\newblock \emph{Ann. Nucl. Energy}, 192:\penalty0 109949, 11 2023.
\newblock \doi{10.1016/j.anucene.2023.109949}.

\bibitem[Zmeškal et~al.(2025)Zmeškal, Thulliez, Tamagno, and Dumonteil]{Zmeskal2024}
M.~Zmeškal, L.~Thulliez, P.~Tamagno, and E.~Dumonteil.
\newblock {Improvement of Geant4 Neutron-HP package: Unresolved resonance region description with probability tables}.
\newblock \emph{Ann. Nucl. Energy}, 211:\penalty0 110914, 2025.
\newblock \doi{https://doi.org/10.1016/j.anucene.2024.110914}.
\newblock URL \url{https://www.sciencedirect.com/science/article/pii/S0306454924005772}.

\bibitem[Mom et~al.(2022)Mom, Thulliez, Dumonteil, Binois, Richet, Schwindling, and Drouart]{Mom2022}
B.~Mom, L.~Thulliez, E.~Dumonteil, M.~Binois, Y.~Richet, J.~Schwindling, and A.~Drouart.
\newblock {Simulation and design of an IPHI-based neutron source, first steps toward SONATE}.
\newblock \emph{J. Neutron Res.}, 24\penalty0 (3-4):\penalty0 337--345, 2022.
\newblock \doi{10.3233/JNR-220027}.
\newblock URL \url{https://journals.sagepub.com/doi/abs/10.3233/JNR-220027}.

\bibitem[Squires(2012)]{Squires_2012}
G.~L. Squires.
\newblock \emph{Introduction to the Theory of Thermal Neutron Scattering}.
\newblock Cambridge University Press, 3 edition, 2012.

\bibitem[Cai and Kittelmann(2020)]{caiKittelmann2020}
X.-X. Cai and T.~Kittelmann.
\newblock {NCrystal: A library for thermal neutron transport}.
\newblock \emph{Comput. Phys. Commun.}, 246:\penalty0 106851, 2020.
\newblock \doi{https://doi.org/10.1016/j.cpc.2019.07.015}.

\bibitem[Kittelmann and Cai(2021)]{KittelmannCai2021}
T.~Kittelmann and X.-X. Cai.
\newblock {Elastic neutron scattering models for NCrystal}.
\newblock \emph{Comput. Phys. Commun.}, 267:\penalty0 108082, 2021.
\newblock \doi{https://doi.org/10.1016/j.cpc.2021.108082}.

\bibitem[Mom(2023)]{MomThese2023}
B.~Mom.
\newblock \emph{Modélisation et tests d'une source compacte de neutrons basée sur l'accélérateur IPHI}.
\newblock PhD thesis, 2023.
\newblock URL \url{http://www.theses.fr/2023UPASP129}.
\newblock Thèse de doctorat dirigée par Drouart, Antoine et Thulliez, Loïc Physique nucléaire université Paris-Saclay 2023.

\bibitem[Poole et~al.(2012{\natexlab{a}})Poole, Cornelius, Trapp, and Langton]{poole2012acad}
C.M. Poole, I.~Cornelius, J.~V. Trapp, and C.M. Langton.
\newblock {A CAD Interface for GEANT4}.
\newblock \emph{Australasian Physical \& Engineering Science in Medicine}, September 2012{\natexlab{a}}.
\newblock \doi{10.1007/s13246-012-0159-8}.
\newblock URL \url{https://link.springer.com/article/10.1007/s13246-012-0159-8}.

\bibitem[Poole et~al.(2012{\natexlab{b}})Poole, Cornelius, Trapp, and Langton]{poole2012fast}
C.M. Poole, I.~Cornelius, J.~Trapp, and C.M. Langton.
\newblock {Fast Tessellated Solid Navigation in GEANT4}.
\newblock \emph{IEEE Trans. Nucl. Sci.}, 99:\penalty0 1--7, 2012{\natexlab{b}}.

\bibitem[Delacroix et~al.(2006)Delacroix, Guerre, and Leblanc]{Delacroix2006}
D.~Delacroix, J.-P. Guerre, and P.~Leblanc.
\newblock \emph{{Radionucléides \& radioprotection : guide pratique ; manuel pour la manipulation de substances radioactives dans les laboratoires de faible et moyenne activité}}.
\newblock {EDP sciences}, 2006.
\newblock ISBN 2-86883-864-2.

\bibitem[{Gamma monitor 6150 AD}()]{plastic_SPRE}
{Gamma monitor 6150 AD}.
\newblock URL \url{https://www.bertin-technologies.fr/produits/sondes-ad-pour-moniteurs-6150-ad-5-et-6/}.
\newblock {Gamma monitor 3'' × 3'' plastic scintillator Adb coupled to a Geiger Müller 6150 AD 6 device}.

\bibitem[Bechtold(1980)]{Bechtold1980}
R.A. Bechtold.
\newblock {Deposition and control of 7Be in liquid lithium}.
\newblock Hanford Engineering Development Lab., Richland, WA (USA), 01 1980.
\newblock URL \url{https://www.osti.gov/biblio/5384169}.

\bibitem[int(1970)]{international1970iaea}
\emph{Neutron Fluence Measurements}.
\newblock Number 107 in Tech. Rep. Ser. IAEA, Vienna, 1970.
\newblock ISBN 92-0-135070-8.
\newblock URL \url{https://www.iaea.org/publications/1206/neutron-fluence-measurements}.

\bibitem[int(2008)]{international2008iaea}
\emph{{Neutron Imaging: A Non-destructive Tool for Materials Testing}}.
\newblock Number 1604 in TECDOC Series. IAEA, Vienna, 2008.
\newblock ISBN 978-92-0-110308-6.
\newblock URL \url{https://www.iaea.org/publications/7923/neutron-imaging-a-non-destructive-tool-for-materials-testing}.

\bibitem[Anderson et~al.(2009)Anderson, McGreevy, and Bilheux]{Anderson2009}
I.S. Anderson, R.L. McGreevy, and H.Z. Bilheux.
\newblock \emph{{Neutron Imaging and Applications}}.
\newblock 04 2009.
\newblock ISBN 978-1-4419-4619-5.
\newblock \doi{10.1007/978-0-387-78693-3}.
\newblock URL \url{https://doi.org/10.1007/978-0-387-78693-3}.

\bibitem[Kardjilov et~al.(2017)Kardjilov, Lehmann, Strobl, Woracek, and Manke]{Kardjilov2017}
N.~Kardjilov, E.~Lehmann, M.~Strobl, R.~Woracek, and I.~Manke.
\newblock \emph{{Neutron Imaging}}, pages 329--349.
\newblock Springer International Publishing, 2017.
\newblock ISBN 978-3-319-33163-8.
\newblock \doi{10.1007/978-3-319-33163-8_16}.
\newblock URL \url{https://doi.org/10.1007/978-3-319-33163-8_16}.

\bibitem[Chakow(2012)]{CHANKOW2012}
N.~Chakow.
\newblock \emph{{Neutron radiography}}.
\newblock IntechOpen, 2012.
\newblock ISBN 978-953-51-0108-6.
\newblock \doi{10.5772/2227}.

\bibitem[Crha et~al.(2019)Crha, Vila-Comamala, Lehmann, David, and Trtik]{crha2019}
J.~Crha, J.~Vila-Comamala, E.~Lehmann, C.~David, and P.~Trtik.
\newblock Light yield enhancement of 157-gadolinium oxysulfide scintillator screens for the high-resolution neutron imaging.
\newblock \emph{MethodsX}, 6:\penalty0 107--114, 2019.

\bibitem[Dutra et~al.(2021)Dutra, Cowan, Cunningham, Durand, Emig, Heeter, Knauer, Knight, Lara, Perry, et~al.]{dutra2021}
E.C. Dutra, J.~Cowan, T.~Cunningham, A.M. Durand, J.~Emig, R.F. Heeter, J.~Knauer, R.A. Knight, R.~Lara, T.S. Perry, et~al.
\newblock {Characterization of Agfa Structurix series D4 and D3sc x-ray films in the 0.7--4.6 keV energy range}.
\newblock \emph{Rev. Sci. Instrum.}, 92\penalty0 (7), 2021.

\bibitem[Lanier and Cowan(2014)]{lanier2014}
N.E. Lanier and J.S. Cowan.
\newblock Absolute calibration of the agfa structurix series films at energies between 2.7 and 6.2 kev.
\newblock \emph{Rev. Sci. Instrum.}, 85\penalty0 (11), 2014.

\bibitem[Alvarado et~al.(2024)Alvarado, Drouart, and Ott]{alvarado2024}
K.~Alvarado, A.~Drouart, and F.~Ott.
\newblock {Comparison of Resolution and Contrast Performances of Silver Film, Imaging Plate and Scintillator Images in Neutron Radiography}.
\newblock volume 298, page 02002. EDP Sciences, 2024.

\bibitem[of~ASTM~Standards(2000{\natexlab{a}})]{ASTME2023}
Annual~Book of~ASTM~Standards.
\newblock \emph{{ASTM E 2023 – 99. Standard Practice for Fabrication of Neutron Radiographic Sensitivity Indicators}}, volume 03.03.
\newblock 2000{\natexlab{a}}.

\bibitem[of~ASTM~Standards(2000{\natexlab{b}})]{ASTME545}
Annual~Book of~ASTM~Standards.
\newblock \emph{{ASTM E 545 – 99. Determining Image Quality in Direct Thermal Neutron Radiographic Examination}}, volume 03.03.
\newblock 2000{\natexlab{b}}.

\bibitem[ImageJ()]{ImageJ}
ImageJ.
\newblock URL \url{https://imagej.net/}.
\newblock Open source software for processing and analyzing scientific images.

\end{thebibliography}

\end{document}